\documentclass[12pt]{article}
\usepackage{jheppub} 
\pdfoutput=1 
\usepackage{amsfonts,amsmath,amssymb,epsf,physics}
\usepackage{amsmath,braket}
\usepackage{subcaption}
\graphicspath{{figures/} }
\usepackage{graphicx,hyperref,color}
\hypersetup{
    bookmarks=true,         
    unicode=false,          
    pdftoolbar=true,        
    pdfmenubar=true,        
    linktocpage=true,       
    pdffitwindow=false,     
    pdfstartview={FitH},    
    pdfnewwindow=true,      
    colorlinks=true,       
    linkcolor=blue,          
    citecolor=blue,        
    filecolor=blue,      
    urlcolor=blue           
}
\usepackage{tensor}     
\usepackage{romannum}
\usepackage[dvipsnames]{xcolor}

\numberwithin{equation}{section}									

\usepackage{comment}

\newcommand{\de}{\partial}
\newcommand{\be}{\begin{equation}}
\newcommand{\ba}{\begin{eqnarray}}
\newcommand{\ea}{\end{eqnarray}}
\newcommand{\ee}{\end{equation}}

\newcommand{\s}{\sqrt}

\newcommand{\ti}{\tilde}

\newcommand{\la}{\langle}
\newcommand{\lb}{\rangle}
\newcommand{\bea}{\begin{eqnarray}}
\newcommand{\eea}{\end{eqnarray}}
\newcommand{\bes}{\begin{equation*}}
\newcommand{\beas}{\begin{eqnarray*}}
\newcommand{\eeas}{\end{eqnarray*}}
\newcommand{\bas}{\begin{array*}}
\newcommand{\eas}{\end{array*}}
\newcommand{\ees}{\end{equation*}}

\newcommand{\ep}{\epsilon}

\newcommand{\eg}{{\it e.g.,}\ }
\newcommand{\ie}{{\it i.e.,}\ }
\newcommand{\viz}{{\it viz,}\ }
\newcommand{\mt}[1]{\textrm{\tiny #1}}

\newcommand{\con}{\mathrm{con}}
\newcommand{\dis}{\mathrm{dis}}
\renewcommand{\(}{\left(}
\renewcommand{\)}{\right)}
\renewcommand{\[}{\left[}
\renewcommand{\]}{\right]}

\newcommand{\GN}{G_\mt{N}}

\newcommand{\mA}{\mathcal{A}}
\newcommand{\mI}{\mathcal{I}}

\newcommand{\rmtE}{{t_{\mt{E}}}}

\newcommand{\GNN}{G_{\mt{N}}^{(2)}}

\newcommand{\mO}{\mathcal{O}}

\newcommand{\mL}{\mathcal{L}}
\newcommand{\vect}{\mathbf{t}}
\newcommand{\vecn}{\mathbf{n}}

\newcommand{\vecT}{\mathbf{T}}

\newcommand{\vecv}{\mathbf{v}}
\newcommand{\vecV}{\mathbf{V}}

\newcommand{\vecw}{\mathbf{w}}
\newcommand{\vecX}{\mathbf{X}}
\newcommand{\vecY}{\mathbf{Y}}
\newcommand{\vecZ}{\mathbf{Z}}

\newcommand{\arctanh}{\text{arctanh}}
\newcommand{\arccosh}{\text{arccosh}}

\newcommand{\const}{\text{const.}}
\newcommand{\hypgeo}[4]{ \;_2 F_1\qty(#1,#2,#3;#4)}

\usepackage{tikz}

\title{\boldmath Non-extremal Island in de Sitter Gravity}
\author[a,c]{Peng-Xiang Hao,}
\author[a]{Taishi Kawamoto,}
\author[ad]{Shan-Ming Ruan,}
\author[a,b]{Tadashi Takayanagi}

\affiliation[a]{Center for Gravitational Physics and Quantum Information, Yukawa Institute for Theoretical Physics, Kyoto University, \\
Kitashirakawa Oiwakecho, Sakyo-ku, Kyoto 606-8502, Japan}
\affiliation[b]{Inamori Research Institute for Science, \\
620 Suiginya-cho, Shimogyo-ku,Kyoto 600-8411 Japan}
\affiliation[c]{Yau Mathematical Sciences Center, Tsinghua University, \\
Haidian District, Beijing 100084, China}
\affiliation[d]{Theoretische Natuurkunde, Vrije Universiteit Brussel and The International Solvay Institutes, Pleinlaan 2, B-1050 Brussels, Belgium}

\emailAdd{pxhao@yukawa.kyoto-u.ac.jp} \emailAdd{taishi.kawamoto@yukawa.kyoto-u.ac.jp}
\emailAdd{Shan-Ming.Ruan@vub.be} \emailAdd{takayana@yukawa.kyoto-u.ac.jp}

\abstract{This paper investigates the challenges and resolutions in  computing the entanglement entropy for the quantum field theory coupled to de Sitter (dS) gravity along a timelike boundary. The conventional island formula, originally designed to calculate the fine-grained entropy for a non-gravitational system coupled to anti-de Sitter (AdS) gravity, encounters difficulties in de Sitter gravitational spacetime, failing to provide a physically plausible extremal island. To overcome these problems, we introduce a doubly holographic model by embedding a dS$_2$ braneworld in an AdS$_3$ bulk spacetime. This approach facilitates the computation of entanglement entropy through holographic correlation functions, effectively circumventing the constraints of the island formula. We demonstrate that the correct recipe for calculating entanglement entropy with dS gravity involves the non-extremal island, whose boundary is instead defined at the edge of the dS gravitational region. Our findings indicate that, during the island phase, the entanglement wedge of the non-gravitational bath includes the entire dS gravitational space. Using the second variation formula, we further show that the existence of a locally minimal surface anchored on the gravitational brane is intrinsically linked to the extrinsic curvature of the brane.}

\begin{document}
\begin{flushright}
YITP-24-91
\\
\end{flushright}
	\maketitle
	\flushbottom

\section{Introduction}

The island formula computes the entanglement entropy in the quantum field theory coupled to a gravitational theory across its boundary \cite{Penington:2019npb,Almheiri:2019psf,Almheiri:2019hni}. This formula generalizes the holographic entanglement entropy formula \cite{Ryu:2006bv,Ryu:2006ef,Hubeny:2007xt,Faulkner:2013ana,Engelhardt:2014gca} with quantum corrections in the context of the AdS/CFT correspondence \cite{Maldacena:1997re}. 
In the island prescription, the entanglement entropy for a subsystem $\mathcal{A}$ in the CFT region is calculated by positing an island region $\mI$ in the gravitational theory and extremizing the sum of the gravitational entropy of the island and the entanglement entropy in the effective field theory for the subsystem defined by the union of $\mathcal{A}$ and the island $\mI$. Notably, the island formula has been instrumental in deriving the Page curve \cite{Page:1993wv}, a crucial indicator of unitarity during black hole evaporation. This was achieved by considering a conformal field theory (CFT) on a half-space interacting with a gravitational theory containing a black hole along its timelike boundary \cite{Penington:2019npb,Almheiri:2019psf,Almheiri:2019hni}. In setups involving two-dimensional black holes, the island formula is derived directly from the gravitational path integrals of Jackiw-Teitelboim (JT) gravity, taking into account the non-perturbative contributions from replica wormholes \cite{Almheiri:2019qdq,Penington:2019kki}.

\begin{figure}[t]
    \centering
    \includegraphics[width=3in]{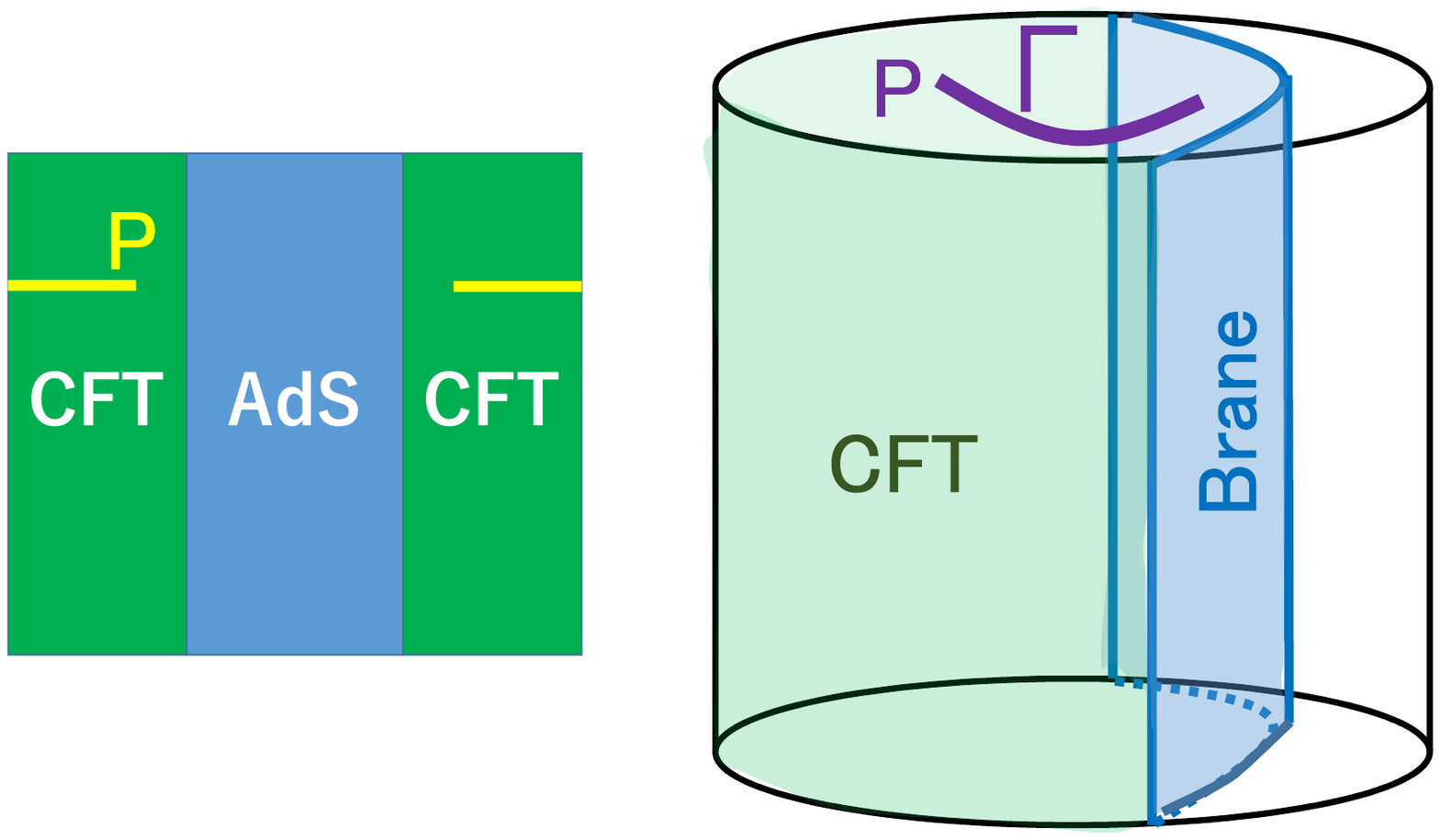}
      \includegraphics[width=3in]{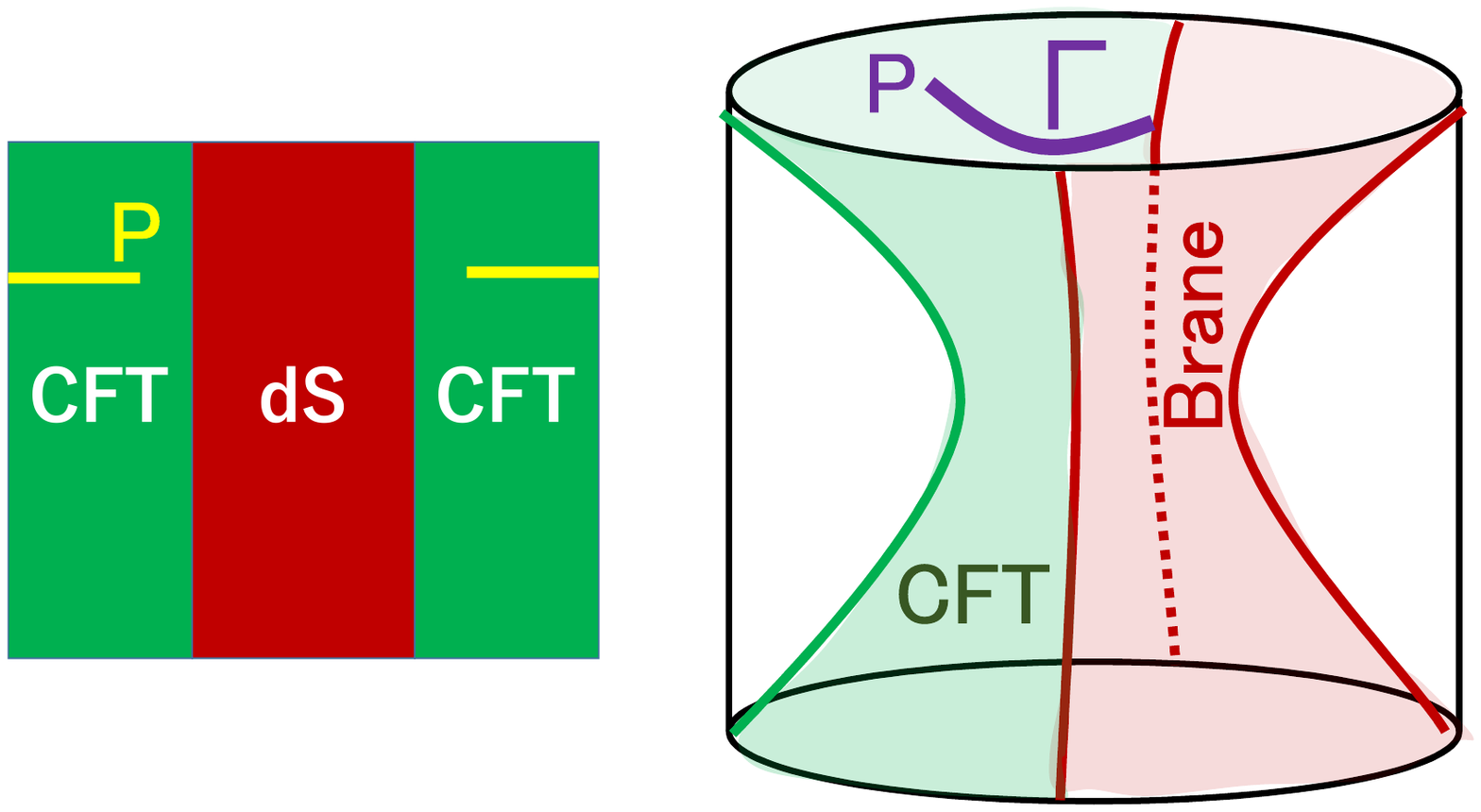}
    \caption{The two distinct setups for calculating entanglement entropy with islands are illustrated here. In the left panel, a CFT is coupled to a gravitational theory in a AdS space, whereas in the right panel, it is coupled to a gravitational theory in dS space. The subsystem $\mathcal{A}$, for which we compute the entanglement entropy, is represented by the yellow horizontal line. The doubly holographic description, employing braneworld holography, is depicted on the right side of each panel. The extremal surfaces $\Gamma$, indicated by the purple curves, each extend from a point $P$ on the AdS boundary and terminate on the EOW branes. These surfaces describe the holographic entanglement entropy in the AdS/BCFT correspondence. While the extremal surface $\Gamma$ can intersect within the AdS brane, it should only terminate at the edge of the dS brane.}
    \label{fig:setupsketch}
\end{figure}

It is also useful to note that the island formula can be understood by combining the gravity dual of boundary conformal field theories (BCFTs) \cite{Takayanagi:2011zk,Fujita:2011fp,Karch:2000gx} with braneworld holography \cite{Randall:1999ee,Randall:1999vf,Gubser:1999vj,Karch:2000ct}, a concept known as double holography \cite{Almheiri:2019hni,Suzuki:2022xwv}. In this doubly holographic approach, a $d$-dimensional CFT coupled to $d$-dimensional quantum gravity on an asymptotically AdS space is holographically described by classical gravity on a $d+1$-dimensional AdS bulk spacetime with an end-of-the-world brane (EOW brane), which is equivalent to the AdS$_{d+1}$/BCFT$_d$ model. In the classical limit of the bulk gravity, the entanglement entropy can be computed as the area of the extremal surface according to the general holographic formula \cite{Ryu:2006bv,Ryu:2006ef,Hubeny:2007xt}. In the presence of EOW branes, a notable feature is that the extremal surface can end on the EOW brane \cite{Takayanagi:2011zk,Fujita:2011fp}, necessitating extremalization over such configurations. This new twist nicely explains the emergence of island region on the EOW brane. This idea is illustrated in figure \ref{fig:setupsketch}. An important feature in such setups is the presence of a non-trivial extremal surface $\Gamma$ that departs from the AdS boundary and ends on the EOW brane. Moreover, the extremal surface $\Gamma$ is stable, as its area is minimized under spatial deformations and maximized under Lorentzian timelike direction, which is required for the consistent calculation of holographic entanglement entropy \cite{Hubeny:2007xt,Wall:2012uf}.

A pertinent question is whether the island formula retains its validity in more general spacetimes. Extending the application of the island formula to cosmological spacetimes, such as those with a positive cosmological constant, could provide important insights into the nature of holography in contexts more closely aligned with realistic settings beyond the AdS/CFT correspondence. Additionally, this investigation elucidates the structure of the Hilbert space in quantum gravity within cosmological spacetimes.  Motivated by this question, this paper aims to investigate the setup in which a non-gravitational quantum field theory is coupled to de Sitter gravity. However, in contrast to the asymptotic infinity in AdS spacetime, there is no preferred timelike boundary in global dS spacetime where the non-gravitational region could be jointed to the dS gravity region. 
In the light of the recent half dS holography proposal \cite{Kawamoto:2023nki}, which introduces a timelike boundary in dS spacetime, we are motivated to construct a bath-dS gravity system by coupling a CFT on one half of de Sitter space to the other half across this timelike boundary. Based on the half dS proposal, we are also offered a doubly holographic model, as illustrated in the right panel of figure \ref{fig:setupsketch}. To ensure analytically tractable examples, our focus is on a two-dimensional de Sitter space. Upon straightforward application of the island formula, a perplexing issue arises: there generally does not exist any physically sensible island that extremizes the generalized entropy functional in dS gravity. Although we frequently observe apparent ``fake" islands that seem to extremize the total entropy functional, these solutions are inadmissible for various reasons.

Nevertheless, the correct prescription can be determined through the framework of double holography. By analyzing holographic computations of one-point functions in the AdS$_3$/BCFT$_2$ correspondence \cite{Suzuki:2022xwv,Izumi:2022opi}, we can facilitate the computation of entanglement entropy while circumventing the constraints imposed by the island formula. Our results indicate that the boundary of {\it non-extremal island} is at the edge of the dS gravitational region. Depending on the size of the subsystem $\mathcal{A}$ in the CFT region, the island can either be an empty region or encompass the entire (gravitating) half of de Sitter space. As a result, the entanglement wedge exhibits a fascinating phase transition phenomenon.

For earlier analyses of the island formula applied to de Sitter spaces, refer to \cite{Chen:2020tes,Hartman:2020khs, Balasubramanian:2020xqf,Sybesma:2020fxg,Aguilar-Gutierrez:2021bns,Bousso:2021sji,Bousso:2022gth,Geng:2021wcq,Kames-King:2021etp,Aalsma:2021bit,Teresi:2021qff,Levine:2022wos,Aalsma:2022swk,Chang:2023gkt,Balasubramanian:2023xyd,Franken:2023pni,Shaghoulian:2023odo}, where special features unique to closed universes have been emphasized. A significant difference between the present work and earlier studies is that our dS spacetime is not closed, as it has a timelike boundary and can be directly compared with the familiar setup of islands in AdS black hole spacetimes. Another advantage of our model is that it allows us to explore a possible holographic dual of de Sitter space \cite{Strominger:2001pn,Maldacena:2002vr} (refer to \cite{Anninos:2011ui,Hikida:2021ese,Susskind:2022bia,Narovlansky:2023lfz} for explicit examples), which remains poorly understood. 
 By embedding de Sitter space into a higher-dimensional AdS/CFT via double holography, we gain more precise insights into the holographic dual of a half de Sitter space \cite{Kawamoto:2023nki}, argued to be a highly non-local field theory.

More importantly, the earlier work \cite{Shaghoulian:2021cef} explored the origin of de Sitter entropy in both three- and four-dimensional de Sitter spaces by extending holographic entanglement entropy. In this context, the author identified a surface similar to an extremal island as unphysical and suggested the relevance of a non-extremal surface. While this argument is closely related to ours, our focus in the present paper is on two-dimensional de Sitter space, which allows us to derive an analytical formula and use double holography to solve the problem in a well-controlled manner. Furthermore, our definition of the non-gravitating region, represented by a half dS, differs from that in \cite{Shaghoulian:2021cef}, where the static patch is truncated by a slice of constant radius. Our choice leads to a novel issue related to Lorentzian time evolution, which we will investigate in this paper. In addition, we would like to mention that the paper \cite{Franken:2023pni} also pointed out the potential relevance of a non-extremal surface as the correct saddle point in a similar setup.

This paper is organized as follows. In section \ref{sec:island}, we apply the island formula to a scenario in which a CFT$_2$ on one half of dS$_2$ spacetime is coupled to dS gravity on the other half, highlighting significant issues encountered when attempting to extremize the entropy functional. Section \ref{sec:island} addresses the accurate computation of entanglement entropy in the setup described in section \ref{sec:island}, utilizing the double holography model. We derive the correct prescription by examining both the holographic calculation of one-point functions and the geometric calculation of holographic entanglement entropy. In section \ref{sec:dif}, we further demonstrate that the existence of a locally minimal surface anchored on the EOW brane in AdS space is related to its extrinsic curvature, using the second variation formula for geodesic variations. Finally, section \ref{sec:dis} summarizes our conclusions and outlines directions for future research.


\section{Invalid island formula for dS gravity}\label{sec:island}

\begin{figure}[t]
	\centering
  	\includegraphics[width=3in]{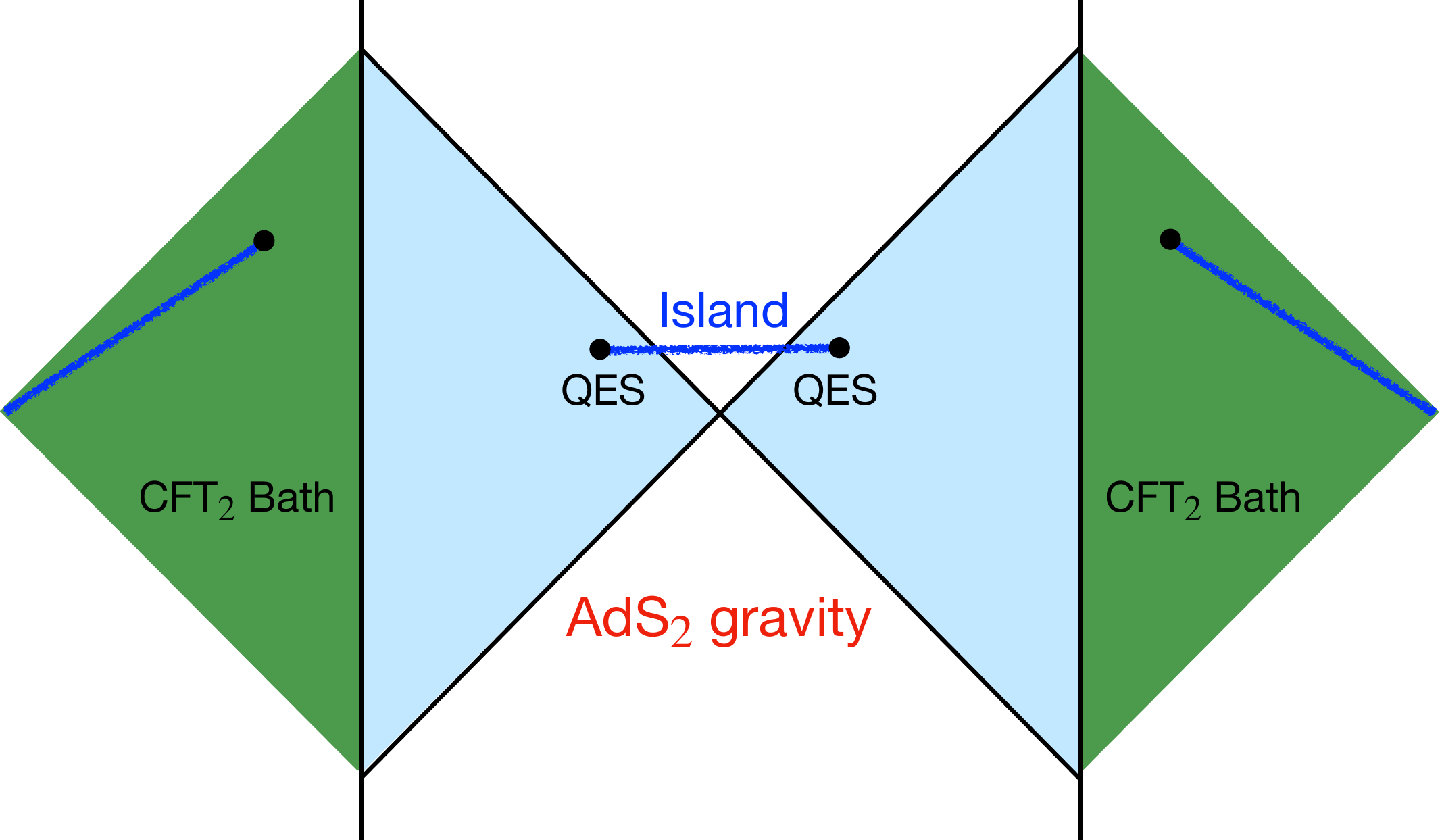}
    \quad
	\includegraphics[width=2.9in]{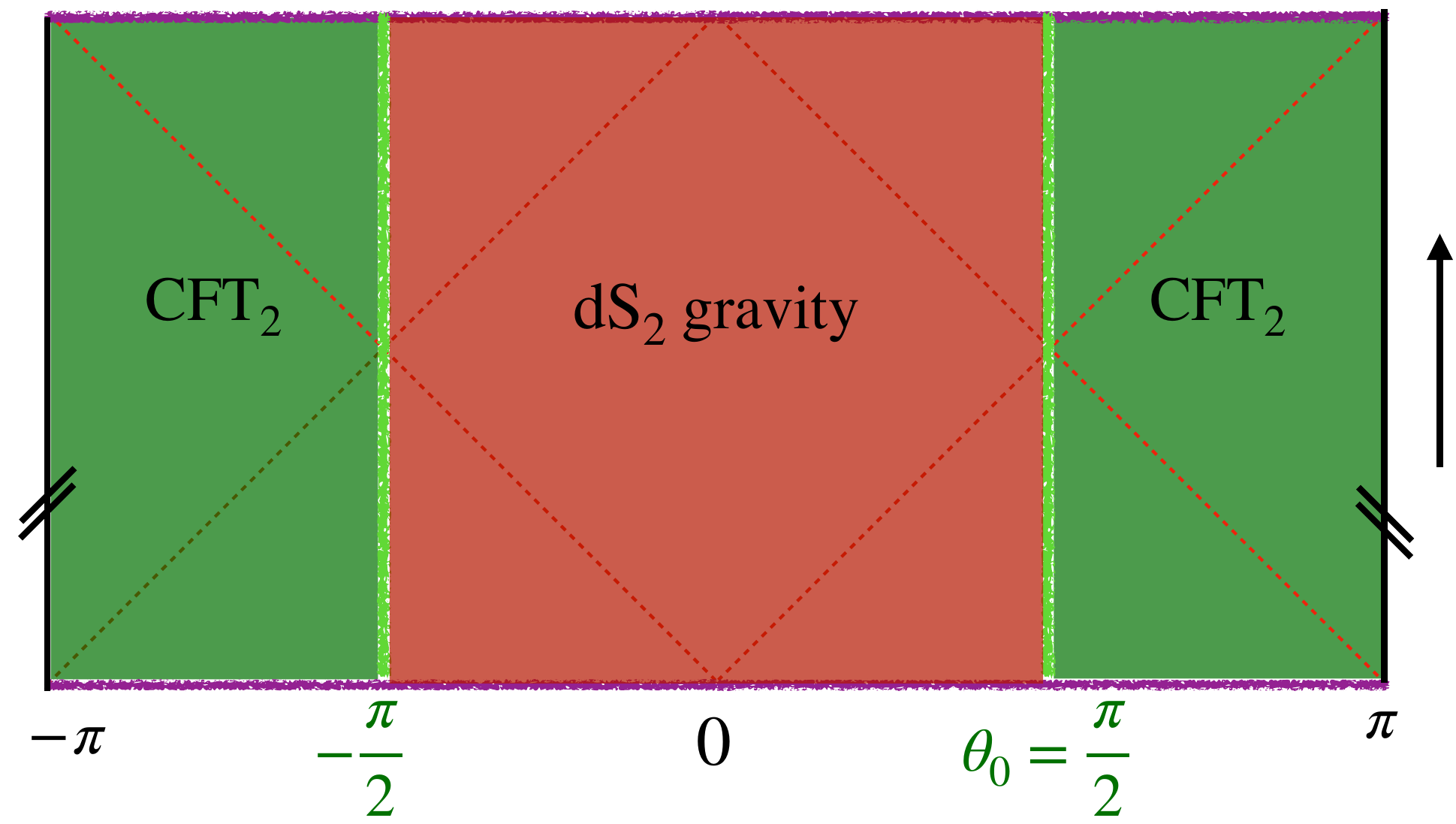}
	\caption{Left: Penrose diagram of an eternal AdS$_2$ black hole coupled to a non-gravitational thermal bath. Right: Penrose diagram of a global dS$_2$, where we identify the two north poles (indicated by the left/right black line) to glue the two copies together. One half of dS space, indicated by the red color, is gravitating.}
	\label{fig:dSPenrosedS2}
\end{figure}

As shown in figure \ref{fig:dSPenrosedS2}, our study examines a system incorporating both gravitational and non-gravitational region. For illustrative purposes, we consider a thermal bath coupled to an eternal AdS black hole, serving as a simplified model to demonstrate a Hawking-like information paradox. The island formula effectively resolves this paradox by utilizing a non-trivial quantum extremal surface, as detailed in \cite{Almheiri:2019yqk}. The primary focus of this paper is the entanglement entropy of a subregion within a non-gravitational system that is, in contrast, coupled to de Sitter gravity. We anticipate that the island formula will be essential for calculating the entanglement entropy in this context. However, our findings reveal that the island formula does not yield the dominant result, as the unique quantum extremal surface in the de Sitter gravity configuration does not represent the desired maximin surface \cite{Hubeny:2007xt,Wall:2012uf,Akers:2019lzs}.
\subsection{Half dS$_2$ gravity coupled with CFT$_2$ bath}

In this paper, we will focus on the global dS$_2$ spacetime, which can be parametrized by $(t, \theta)$ coordinates with a metric given by the following expression
\begin{equation}\label{eq:global_dS}
 \qquad ds^2 =L^2 \(  - dt^2 + \cosh^2 t  \, d \theta^2 \)   = \frac{L^2}{\cos^2 T} \(  - dT^2 + d \theta^2 \) \,. 
\end{equation}
By applying the coordinate transformation $\sinh t = \tan T$, $\cosh t = \frac{1}{\cos T}$ to the global coordinates, it is straightforward to obtain the conformal coordinates with a compact Lorentzian time $T \in (-\frac{\pi}{2}, +\frac{\pi}{2})$. Given that the spatial direction in dS$_2$ is one-dimensional, the spatial periodicity can be selected at will. For the sake of simplicity, we will consider only the canonical case, wherein the circumference of the spatial circle $S^1$ is equal to $2\pi L$, \ie  $ \theta=[-\pi, \pi]$. As illustrated in the Penrose diagram for dS$_2$, which is shown in the right panel of figure \ref{fig:dSPenrosedS2}, our choice corresponds to identifying the two poles located at  $ \theta = \pm \pi$. For later purpose, we note that dS$_2$ spacetime is conformally flat, \ie the line element can be expressed as 
\begin{equation}\label{eq:dSflat}
\quad ds^2 = \frac{L^2}{\Omega^2} dz d\bar{z} = \frac{4 L^2}{(1+ z\ \bar{z})^2} dz d\bar{z} \,,
\end{equation}
where the conformal factor is given by $\Omega = \frac{1}{2} (1+z \bar{z})$. The Lorentzian global coordinates $(z, \bar{z})$ are given as 
\begin{equation}
z = e^{-i (T -\theta)} \,, \qquad \bar{z} = e^{-i (T+\theta)}\,. 
\end{equation}

\begin{figure}[t]
	\centering
	\includegraphics[width=3in]{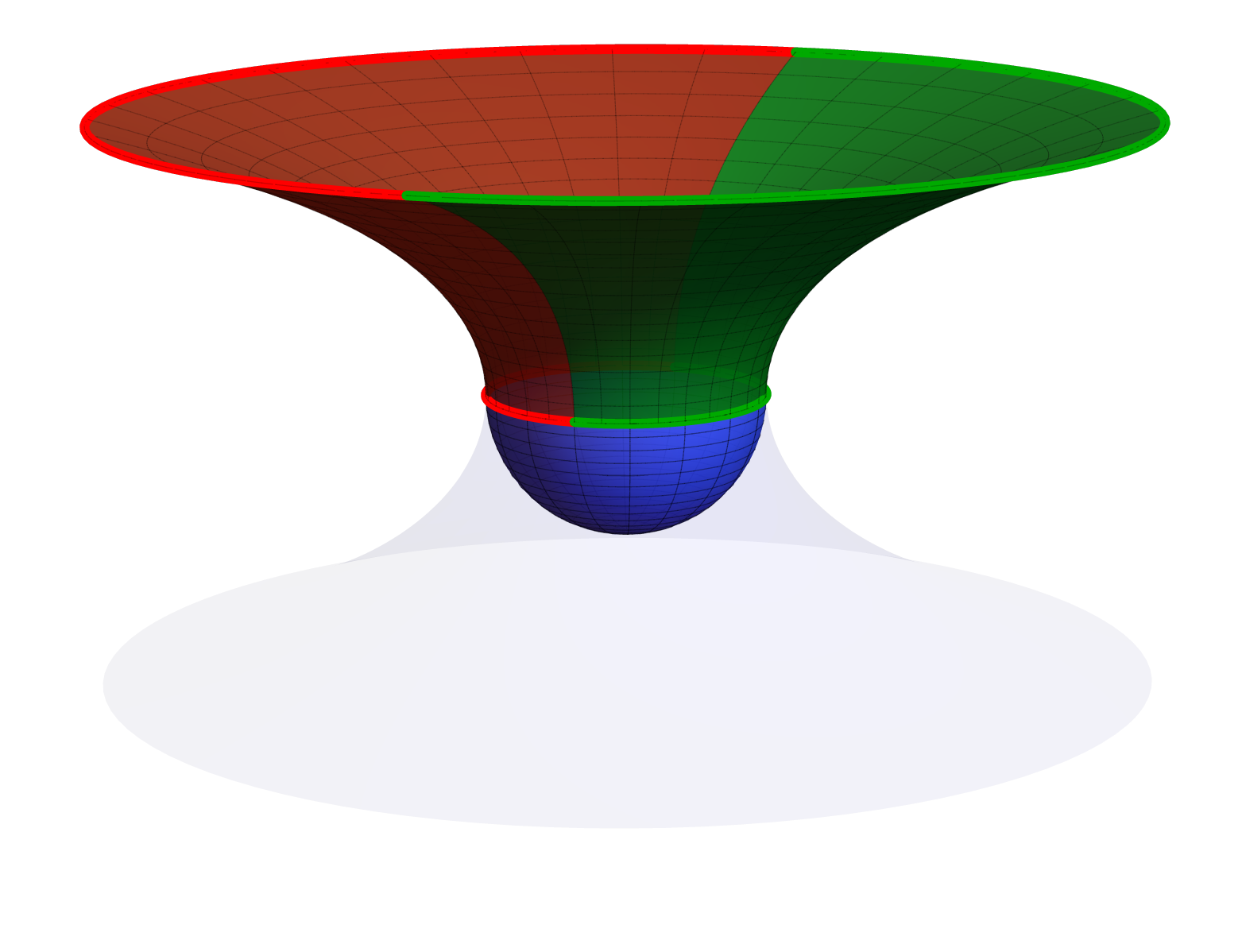} 
 	\includegraphics[width=3in]{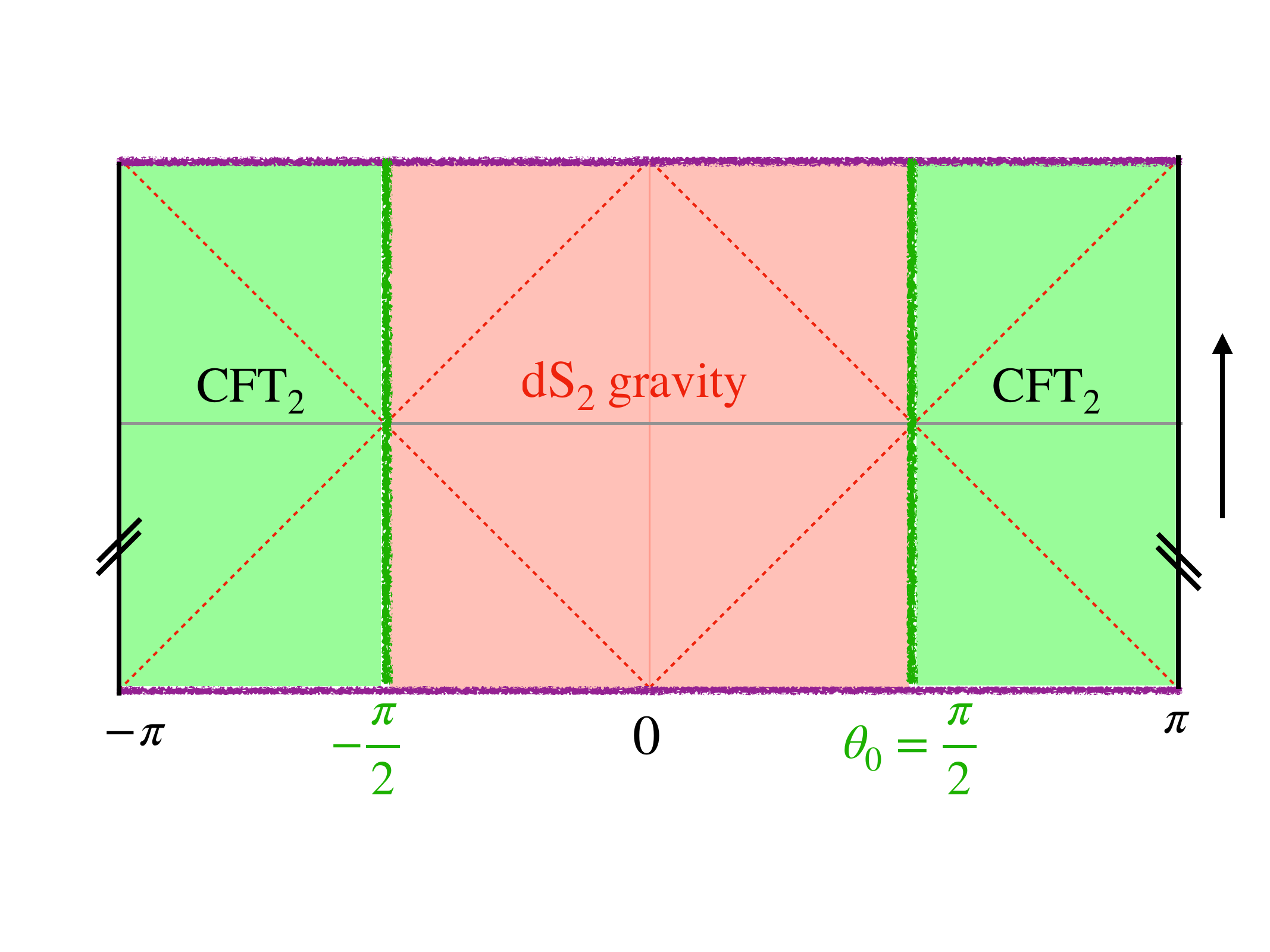}
	\caption{Left: The Hartle-Hawking state, which includes half of dS$_2$ space as the gravitational region and the other half as the bath region. The Euclidean part is represented by the blue hemisphere, while the Lorentzian part is indicated by the upper hyperboloid. Right: A time slice of the setup for gluing a half dS$_2$ gravity with a CFT$_2$ living on dS$_2$.}
	\label{fig:dSBath}
\end{figure}

In the following, we will assume that dS$_2$ gravity is confined to a half circle, \eg with angular coordinate
$\theta = [-\frac{\pi}{2}, \frac{\pi}{2}]$, where we impose the Neumann boundary condition. For the other half of the circle, the gravity is turned off by imposing the Dirichlet boundary condition.
For the sake of convenience in calculations, it is assumed that the dynamical matter on the full dS$_2$ background is still CFT$_2$. In other words, we glue a half dS$_2$ gravity with a CFT$_2$ bath system along a defect line at $\theta=\theta_0 = \pm\frac{\pi}{2}$.
See figure \ref{fig:dSBath} for the illustration of the system under consideration. 

It should be noted that there are various gravitational theories that can support a global dS$_2$ solution. One of the simplest theories which can have the dS$_2$ solution is the Liouville gravity (or the 2d induced gravity):
\begin{equation}\label{LVth}
S_{\mt{L}}=-\frac{c}{24\pi}\int d^2x \left[(\de_T\phi)^2-(\de_\theta\phi)^2+\frac{e^{2\phi}}{L^2} \right]\,,
\end{equation}
where $\phi$ is the Liouville field such that the 2d metric reads $ds^2=e^{2\phi}(-dT^2+d\theta^2)$. This action is induced from a 2d CFT with the central charge $c$ which is directly coupled with a dynamical metric, without adding any action of 2d gravity. As argued in \cite{Suzuki:2022xwv} (see also \cite{Chen:2020uac,Chen:2020hmv,Akal:2021foz,Neuenfeld:2024gta}), this theory is indeed realized as an effective theory on the EOW brane via the brane-world type setup, which we will employ in section \ref{sec:doubleholography}. It may also be helpful to write down a covariant form of the Liouville gravity action as the nonlocal Polyakov action \cite{Chen:2020uac}, namely
\ba
S_{\mt{L}}=\frac{c}{96\pi}\int dx^2 \s{-g}\left[-R\frac{1}{\Box}R-\frac{4}{L^2}\right].
\ea

Moreover, distinct types of JT gravity can be obtained through the implementation of different dimensional reductions from higher-dimensional spacetime. For instance, one can obtain a dS JT gravity with positive curvature coupled to a dynamical dilaton field, namely 
\begin{equation}\label{eq:JT01}
S_{\rm JT}=\frac{1}{16 \pi \GNN} \int d^2 x \sqrt{-g} \, \Phi\(R-\frac{2}{L^2}\)  + \text{boundary terms} \,, 
\end{equation}
by performing the spherical reduction from three-dimensional Einstein-Hilbert action, see \eg \cite{Sybesma:2020fxg,Kames-King:2021etp,Svesko:2022txo,Franken:2023ugu} for more details. The dynamical dilaton field, denoted by $\Phi$, is naturally interpreted as the radius in dS$_3$. It is important to note that the dilaton value in this theory must be constrained to be positive in order to ensure that the effective Newton constant remains positive. Consequently, its dS$_2$ solution encompasses the cosmological horizon and has two boundaries located at the south and north pole where the dilaton $\Phi$ vanishes. The corresponding Penrose diagram of this dS$_2$ JT gravity is thus identical to that of a half of global dS$_2$. Conversely, the dS version of AdS JT gravity is derived by taking the limit of a higher-dimensional extremal black hole \cite{Maldacena:2019cbz,Cotler:2019nbi}. Consequently, the induced two-dimensional gravitational theory incorporates an additional topological term, \ie 
\begin{equation}\label{eq:JT02}
S_{\rm JT'}=\frac{\Phi_0}{16 \pi \GNN} \int d^2 x \sqrt{-g} R+\frac{1}{16 \pi \GNN} \int d^2 x \sqrt{-g}  \, \Phi \(R-\frac{2}{L^2}\) + \text{boundary terms} \,, 
\end{equation}
where the constant field $\Phi_0$ is related to the ground state entropy of higher-dimensional extremal black hole and dynamical $\Phi$ controls the deviation from the extremality. In comparison to previous dS$_2$ gravity theories, the dS JT gravity defined in eq.~\eqref{eq:JT02} allows for both cosmological horizons and black holes due to the fact that the dilaton field $\Phi$ is permitted to take negative values. As a result, the Penrose diagram for this dS JT gravity is distinct from that of the global dS spacetime. In contrast, it is analogous to a Schwarzschild-de Sitter spacetime. 

In the remainder of this section, we will limit our focus to the most basic case where two-dimensional gravity theory is induced from a 2d CFT, where the integration of the CFT degrees of freedom lead to the Liouville action (\ref{LVth}). This setup is conveniently described by a classical gravity in higher dimensions via the double holography as will be studied in section \ref{sec:doubleholography}. In this case, there is no contribution to the entropy from the 2d gravity and the entropy all comes from the 2d CFT sector.
However, it is trivial to simply add the topological term as in the first term in (\ref{eq:JT02}), which is proportional to $\int d^2x\s{-g}R$, if we like. This simply produces an additional constant to the total gravitational entropy  (or Bekenstein-Hawking entropy), \ie 
\begin{equation}\label{eq:S0}
\frac{\Phi_0}{4 \GNN} \equiv S_{0} = \text{constant} \,.
\end{equation}
The remaining two distinct types of JT gravity will be discussed in appendix \ref{sec:JT}. As a summary, it can be stated that all conclusions presented in the main text also apply to the dS JT gravity with only cosmological horizon. However, it would be subtle for dS JT gravity defined in eq.~\eqref{eq:JT02} due to the appearance of the black hole singularity.

\subsection{Island formula}

We are interested in evaluating the entanglement entropy and its time evolution for a boundary interval $\mA$ situated within the CFT$_2$ bath. For the sake of simplicity, we will consider the symmetric interval defined by 
\begin{equation}\label{eq:intervalA}
\mA : \qquad  \{ t = t_A \,, \quad \theta \in [-\pi , -\theta_A ] \cup [\theta_A, \pi]  \} \,. 
\end{equation}
Drawing lessons from the breakthrough in understanding the fine-grained entropy of Hawking radiation and the black hole information paradox, it is pointed out that the correct formula for the fine-grained entropy of a non-gravitating region entangled with a gravitational system is given by the so-called island formula \cite{Penington:2019npb,Almheiri:2019psf,Almheiri:2019hni}, \ie 
\begin{equation}
    	S_{\mathrm{EE}} (\hat{\rho}_{\mathcal{A}})= \underset{\mI}{\operatorname{Min}} \left[ \underset{\mI}{\operatorname{Ext}}\left(\frac{\mathrm{Area}[ \partial \mI]}{4 \GNN} +  S_\mathrm{matter}\left(\mA \cup \mI         \right) \right)\right]\,,
\end{equation}
where the first area term denotes the gravitational entropy \footnote{It should be noted that the corresponding expression for the ``area" term is contingent upon the gravitational theory under consideration. In a more general sense, it is defined as the Wald-Dong entropy for a generic diffeomorphism-invariant theory.} and the second term, denoted as $S_{\mathrm{matter}}$, is often referred to as the bulk entropy and represents the quantum corrections, which are given by the von Neumann entropy of matter fields. The sum of the leading area term and the quantum correction $S_{\mathrm{matter}}$ is the so-called generalized entropy denoted by $S_{\rm gen}$. The position of island $\mI$ is determined by its boundary $\partial \mI$, \ie the quantum extremal surface (QES), which is obtained by extremizing the generalized entropy. 

Although the original island formula was developed in AdS gravity coupled to a bath, it is reasonable to expect that the same formula should also apply to dS or flat gravity. 
A substantial body of prior research has also investigated the island formula in dS gravity, with a particular focus on dS JT gravity, \eg \cite{Hartman:2020khs,Sybesma:2020fxg,Balasubramanian:2020xqf,Aguilar-Gutierrez:2021bns,Kames-King:2021etp,Svesko:2022txo,Aalsma:2022swk}.
In the following section, we will attempt to calculate the fine-grained entropy of the boundary interval $\mA$, by directly applying the island formula. Nevertheless, it will become evident that the subsequent naive island solution presents inherent issues from different perspectives.

\subsubsection{Naive antipodal island in dS$_2$ gravity}

In light of the assumption that gravitational entropy is merely a constant (eq.~\eqref{eq:S0}), the remaining objective for evaluating the island formula is to calculate the bulk entropy, or the von Neumann entropy of intervals in CFT$_2$.
Noting the fact the CFT$_2$ matter is living on global dS$_2$ space. To derive the entanglement entropy of a single interval $[z_1, z_2]$ in the global dS coordinates, we can use the flat-spacetime result with counting the appropriate Weyl factors for each endpoints of the interval. By employing the Weyl factor specified in eq.~\eqref{eq:dSflat}, it is straightforward to get 
\begin{equation}\label{eq:bulkentropy}
\begin{split}
    S_{\rm matter}\( [\theta_1, \theta_2] \) &= \frac{c}{6} \log \(  \frac{(z_1 -z_2 )(\bar{z}_1 - \bar{z}_2)}{\epsilon_1 \epsilon_2 \Omega_1 \Omega_2 }  \) \\
    &=  \frac{c}{6} \log \(  \frac{ 2\( \cos (T_1 - T_2) - \cos(\theta_1 -\theta_2) \)}{\epsilon_1 \epsilon_2\cos T_1 \cos T_2 }  \) \\
    &= \frac{c}{6} \log \( \frac{2(1+\sinh t_1 \sinh t_2- \cosh t_1 \cosh t_2 \cos(\theta_1 -\theta_2) ) }{\epsilon_1 \epsilon_2} \) \,,
    \end{split}
\end{equation} 
where we have introduced two UV cut-off $\epsilon_i$ associated with two individual endpoints. Note that we are always working on the Hartle-Hawking state. Here we have specified the von Neumann entropy in terms of the global Lorentzian time $t$, which is derived from the Euclidean signature by Wick rotation. In the next section we will see that the same answer can be obtained by evaluating the geodesic length in AdS$_3$.

In the non-island phase, in the absence of contributions from the island, the entanglement entropy of a single interval $ \mA$ is given by the following expression:
\begin{equation}\label{timecon}
 	S_{\mathrm{EE}}(\hat{\rho}_{\mathcal{A}}) \big|_{\rm{non-island}}= S_{\rm matter}\( \mA \) = \frac{c}{3} \log \( \frac{ 2\cosh t_A  \cdot  \sin\theta_A }{\epsilon}  \) \,,
\end{equation}
which presents a linear growth 
\begin{equation}
S_{\mathrm{EE}}(\hat{\rho}_{\mathcal{A}}) \big|_{\rm{non-island}} \simeq \frac{c}{3} t_A \,. 
\end{equation}
at late times. This can be understood as a reflection of the inflation of the dS$_2$ background.

\begin{figure}[t]
	\centering
	\includegraphics[width=5.5in]{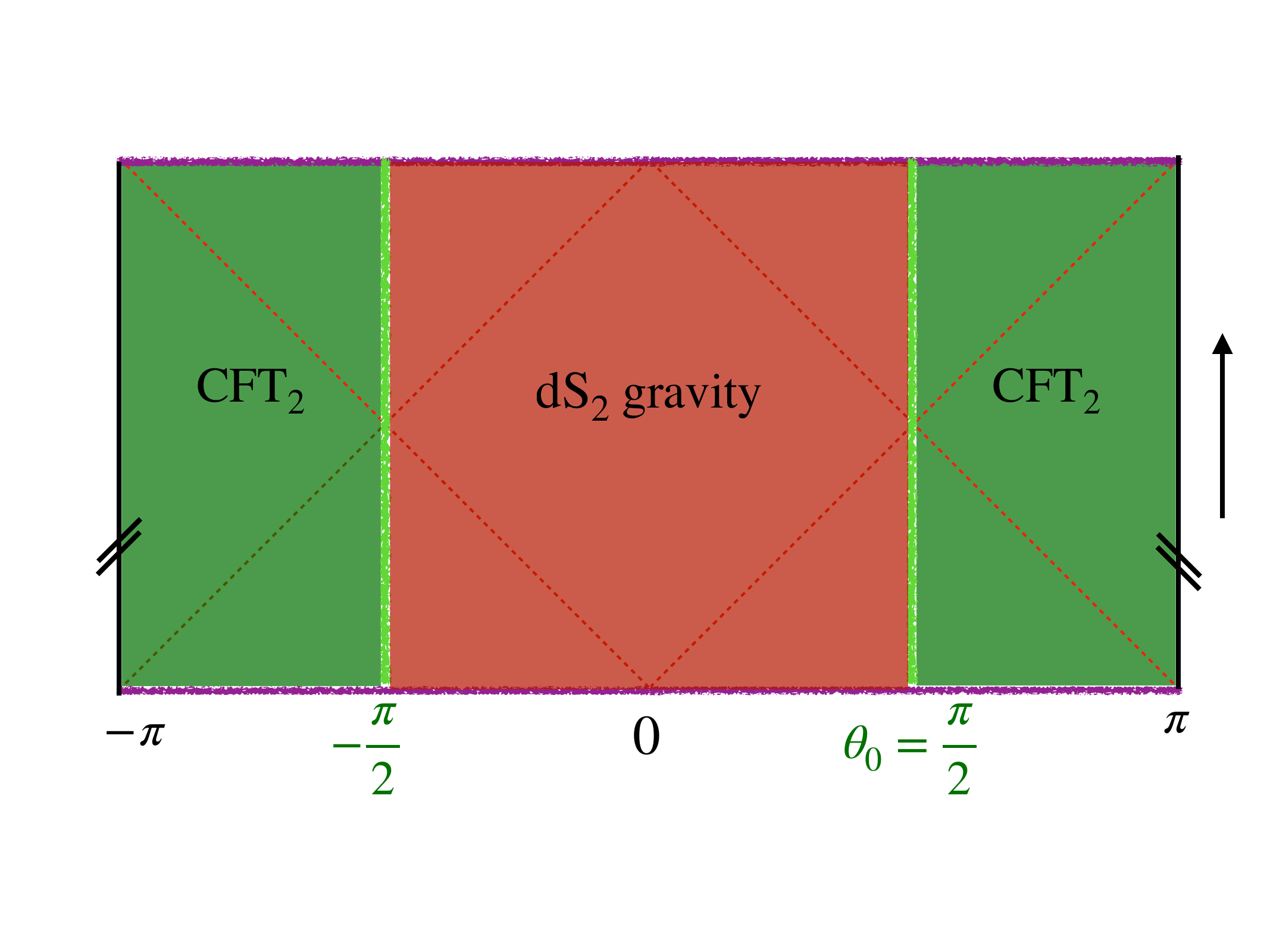}
	\caption{Left: A time slice of our setup with dS gravity coupled to the CFT$_2$ bath. The antipodal island is denoted by the fact that the endpoints of the island are the antipodal points of the bath interval $\mathcal{A}$. Right: The bath interval $\mathcal{A}$ and its corresponding antipodal island are shown in the Penrose diagram.}
	\label{fig:antipodal}
\end{figure}

Furthermore, let us consider the island phase with the assumption that the boundary of the island region is located at 
\begin{equation}
 \partial \mI : \qquad  \(  t=t_I, \theta = \pm \theta_I \)   \,.
\end{equation}
Since the global Hartle-Hawking state is pure, the bulk entropy in the island phase can be calculated by considering the complimentary region of $\mA \cup \mI$, which is the two-interval region $(\mA \cup \mI )^c$. This calculation for bulk entropy corresponds to evaluating two-point functions associated with endpoints at $\theta=\theta_A$ and $\theta=\theta_I$. The corresponding generalized entropy at the island phase thus can be obtained in a straightforward manner by applying the two-point function derived in eq.~\eqref{eq:bulkentropy}, \ie 
\begin{equation}
\begin{split}
 S_{\rm gen} &= S_0 +  S_{\rm matter}\( [\theta_I, \theta_A] \)  + S_{\rm matter}\( [-\theta_A, -\theta_I] \) \\
  &= S_0 + \frac{c}{3} \log \( \frac{2(1+\sinh t_I \sinh t_A- \cosh t_I \cosh t_A \cos(\theta_I -\theta_A) ) }{\epsilon \, \epsilon_I} \) \,,\\
 &= \tilde{S}_0 + \frac{c}{3} \log \( \frac{2(1+\sinh t_I \sinh t_A- \cosh t_I \cosh t_A \cos(\theta_I -\theta_A) ) }{\epsilon} \) \,,
\end{split}
\end{equation}
where the cutoff scale at the island region, $\epsilon_I$, has been absorbed into the Newton constant $S_0$, by introducing $\ti{S}_0=S_0+\frac{c}{3}\log\frac{1}{\ep}$. The extremization with respect to the spatial direction $\theta_I$ and the timelike direction $t_I$ is simply given by solving the following equations: 
\begin{equation}\label{solvbex}
    \begin{split}
  \partial_{\theta_I} S_{\rm gen}  =0 &\longrightarrow \quad    \cosh t_I \cosh t_A \sin(\theta_I -\theta_A) =0  \,, \\
   \partial_{t_I} S_{\rm gen}  =0 &\longrightarrow \quad   \cosh t_I \sinh t_A- \sinh t_I \cosh t_A \cos(\theta_I -\theta_A) =0 \,.
    \end{split}
\end{equation}
It is evident that our choice \eqref{eq:S0} facilitates the extremization over the gravitational region, as the derivative of the gravitational entropy vanishes. The solution for the extremization is unique and is given by the following equation: 
\begin{equation}\label{eq:QESantipodal}
 \theta_I = -\pi + \theta_A  \,, \quad t_I = -t_A \,,
\end{equation}
which will be called the antipodal island, as illustrated in figure \ref{fig:antipodal}. 

\begin{figure}[t]
    \centering
    \includegraphics[width=4.5in]{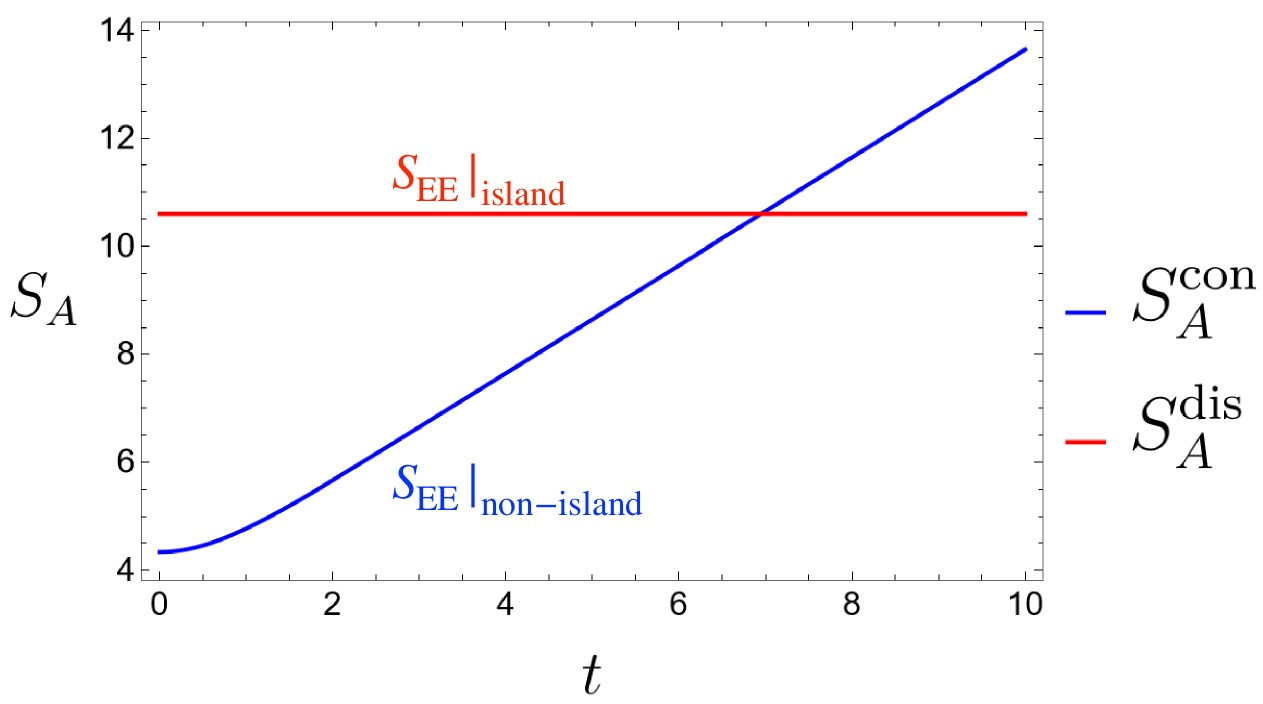}
    \caption{The phase transition between the entropy $S_{\mt{EE}}(\hat{\rho}_{\mt{A}})$ for a bath interval $\mathcal{A}$ in the non-island phase and that derived by the island formula. It gives the same results as the standard holographic entanglement entropy $S_{\mathcal{A}}$. We set $\theta_{A_1}- \theta_{A_2} = \frac{\pi}{4}$, $\epsilon = 0.01$ for this plot.}
    \label{fig:dS2EE}
\end{figure}

The distinguishing feature related to this type island is that it moves back in time as the bath interval involves in time. The trivial solution $\theta_I =\theta_A, t_I =t_A$ is naturally discarded because it is outside the gravitational region. Consequently, the island formula associated with this antipodal island gives rise to a constant, \ie
\begin{equation}\label{discgh}
	S_{\mathrm{EE}}(\hat{\rho}_{\mathcal{A}}) \big|_{\rm{island}} = S_0+\frac{2c}{3}\log\frac{2}{\ep}
= \tilde{S}_0  +\frac{c}{3} \log \( \frac{4}{\epsilon}  \)\,, 
\end{equation}
which would be always smaller than that from the non-island phase at late times. The phase transition from the non-island phase to the island phase is thus given by 
\begin{equation}
 t_\ast  =  \arccosh \(  \frac{2\exp ( \frac{3}{c}\tilde{S}_0 )}{\sin \theta_A} \)   \approx \frac{3}{c}\tilde{S}_0 - \log\( \frac{\sin \theta_A}{4} \) \,.
\end{equation}
The time evolution of the entanglement entropy, as calculated by the island rule, is depicted in figure \ref{fig:dS2EE}. Similar types of island in JT gravity has also been discussed in detail in \cite{Hartman:2020khs,Sybesma:2020fxg,Aguilar-Gutierrez:2021bns,Kames-King:2021etp,Svesko:2022txo}.

\subsubsection{Problems with antipodal islands}
As indicated in the title of this section, we will demonstrate that the naive antipodal island is not physical for various reasons. The attentive reader may have noticed that the previously identified antipodal island introduces ambiguity in defining the island region. As illustrated in figure \ref{fig:antipodal}, we define the island region by fixing its two endpoints, \ie 
\begin{equation}
\partial \mI : \qquad  \( - t_A \,, -\theta_A + \pi  \)   
\cup  \( - t_A \,, \theta_A - \pi  \)  \,. 
\end{equation}
However, it is important to note that the calculations for the matter entropy part are based on the calculations of the two-point functions associated with the two-interval subregion, which is defined as $(\mA \cup \mI )^c$. Substitution of the island solution derived by extremization, namely the antipodal island, reveals that the time slice consisting of $(\mA \cup \mI )^c \cup (\mA \cup \mI )$ becomes ill-defined. Noticing the fact\footnote{According to our convention, the endpoint located at $\theta_A$ in the bath region is constrained by $\theta_A \in [\pi/2 , \pi ]$.} 
\begin{equation}
 -\theta_A + \pi  \ge \theta_A - \pi   \,,
\end{equation}
It is evident that the corresponding time slice encompassing the bath interval and island region is counted three times, which is an obvious discrepancy. From this perspective, it can be concluded that the antipodal island cannot be considered a valid or physical solution, despite being the unique solution for extremalizing the generalized entropy. As will be seen in the following section, the entanglement wedge associated with the antipodal island is not well-defined when the so-called double holography is employed.

One might inquire as to the origin of the issue associated with antipodal islands. This inquiry may lead to the conclusion that the gravitational entropy is taken as a constant, which is a simple choice. Nevertheless, it is necessary to demonstrate that the antipodal island is invalid from a physical standpoint. This is because the quantum extremal surface may not always be the dominant saddle in calculations of fine-grained entropy. It is important to recall that the quantum extremal surface is a generalization of the RT surface, which is defined as the maximin surface \cite{Hubeny:2007xt,Wall:2012uf} in the bulk spacetime with the Lorentzian signature. It is reasonable to posit that the quantum extremal surface is also given by a maximin surface \cite{Akers:2019lzs} which extremizes the generalized entropy $S_{\rm gen}$. However, the antipodal island, which resides in the dS gravitational region, is instead maximal in the spatial direction and minimal in the timelike direction. To demonstrate this assertion, we may examine the second derivative of the generalized entropy, $S_{\rm gen}$, with respect to the location of islands. More explicitly, it is straightforward to get 
\begin{equation}
\begin{split}
 \frac{ \partial^2 S_{\rm gen}  }{\partial\theta_I \partial\theta_I } \bigg|_{t_I = -t_A, \theta_I = \theta_A - \pi  } &=   - \frac{1}{2  }\cosh^2 \theta_A  < 0  \,, \\
 \frac{ \partial^2 S_{\rm gen}  }{\partial t_I \partial t_I } \bigg|_{t_I = -t_A, \theta_I = \theta_A - \pi  } &=\frac{1}{2}   > 0  \,. \\
\end{split}
\end{equation}
The preceding two inequalities indicate that the antipodal island, as perceived from the perspective of the generalized entropy, is maximal in the spatial direction and minimal in the timelike direction, which is opposite to the property of RT surface for holographic entanglement entropy. In the Euclidean signature, the antipodal island thus corresponds to the maximum in the Euclidean time direction. Although the quantum extremal surface related to the antipodal island in dS gravity corresponds to a saddle point, it can be concluded that it cannot dominate the calculation for the fine-grained entropy due to the absence of the maximin surface. Regardless of the properties of the gravitational or non-gravitational region, it can be expected that this is a common problem: the quantum extremal surface corresponding to the dominated saddle does not exist. To address this issue, it is necessary to develop a novel approach for calculating entanglement entropy that does not rely on the island formula.

Furthermore, it is worth mentioning that a comparable configuration involving the coupling of dS space to a non-gravitational domain has been previously investigated in \cite{Shaghoulian:2021cef}, where a cutoff surface was introduced within the static patch of d-dimensional dS spacetime. The author has highlighted the analogous issues pertaining to the minimax surface in the dS bulk spacetime, observing that it contravenes the entanglement wedge nesting. In light of these considerations, the author puts forth the proposal of anchoring extremal surfaces to the cosmological horizon (bilayer proposal), thereby generalizing the (H)RT formula to dS space \cite{Shaghoulian:2022fop}. In the following section, we will instead seek to identify a solution by moving to a higher dimension. To this end, we employ the double holography in AdS$_3$, which reveals that the entanglement entropy of a subsystem in the non-gravitational region in the semi-classical limit should be given by the non-extremal island\footnote{The author in \cite{Shaghoulian:2021cef} has also suggested that the potential quantum extremal surface in dS space may be represented by the cutoff surface located at the boundary of the dS gravitational region, which is referred to as the degenerate surface.}.



\section{Non-extremal island from double holography}\label{sec:doubleholography}
 
In this section, we explore a doubly holographic setup involving a dS$_2$ braneworld coupled to a non-gravitational bath. This doubly holographic perspective provides additional evidence that the naive island formula cannot be directly applied in the de Sitter braneworld, as discussed in the previous section. Leveraging higher-dimensional holography, the entanglement entropy of a bath interval can be accurately computed using the holographic correlation function in the AdS/BCFT correspondence. We further show that the appropriate formula for evaluating the entanglement entropy associated with dS gravity is given by the non-extremal island situated at the edge of the dS gravitational region.

\subsection{Double holography with a dS braneworld}

\begin{figure}[t]
	\centering
	\includegraphics[width=6in]{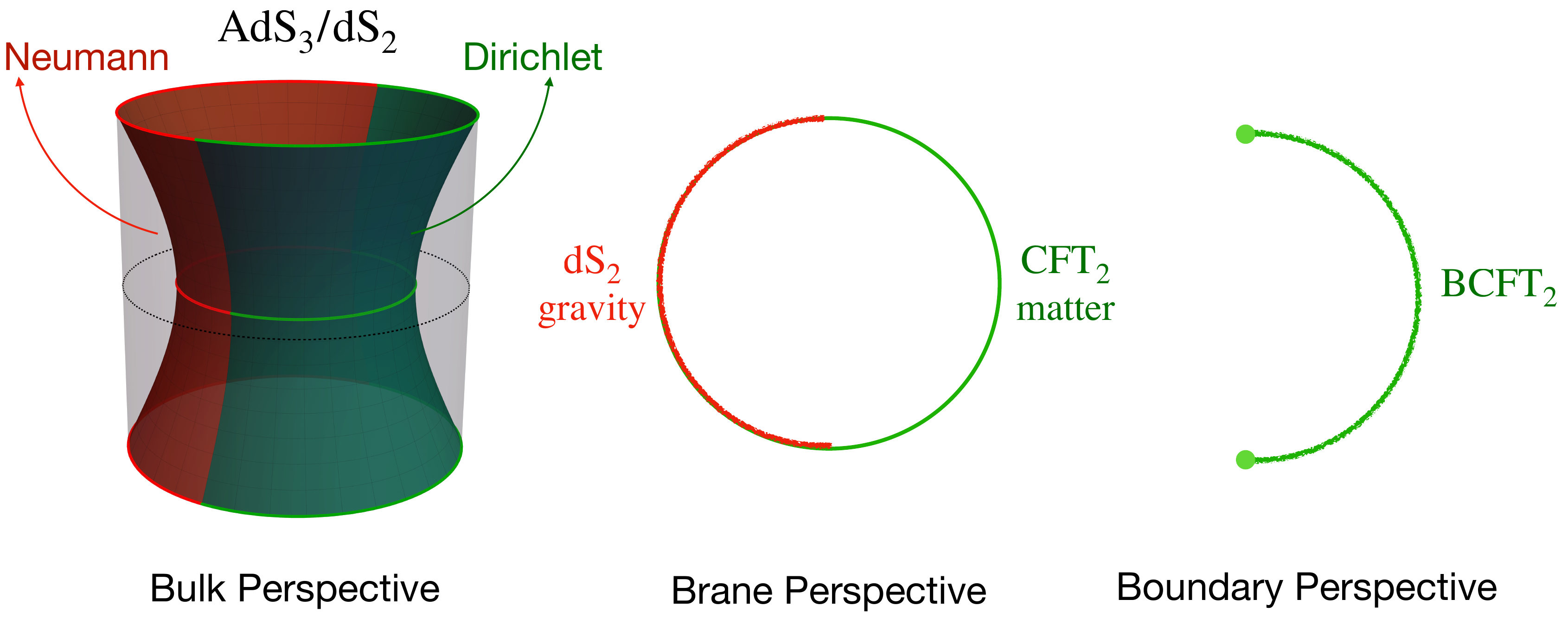}
	\caption{Three perspectives in double holography for dS gravity coupled to a non-gravitational bath residing in dS spacetime. The image in the middle shows the brane perspective, which is the same as shown in figure \ref{fig:dSBath} in section \ref{sec:island}.}
	\label{fig:double}
\end{figure}

We begin by constructing a doubly holographic model for the setup discussed in the previous section, specifically dS$_2$ gravity coupled to a CFT$_2$ bath. The essential concept entails the embedding of the dS$_2$ brane within an AdS$_3$ bulk spacetime, utilizing the AdS$_3$/dS$_2$ slicing. This can be expressed as follows:
\begin{equation}\label{eq:dS2_slice_AdS3}
    ds^2 =  L_{\mt{AdS}}^2\qty(d\eta^2 + \sinh^2{\eta}\qty(-dt^2 + \cosh^2{t}\;d\theta^2))\,, 
\end{equation}
where $L_{\mt{AdS}}$ denotes the AdS radius and each 
$\eta=\mathrm{const}>0$ surface is nothing but a dS$_2$ spacetime. The Euclidean AdS$_3$ spacetime with the Euclidean dS brane is obtained by performing the Wick rotation $t=-i\rmtE$:
\begin{equation}
    ds^2 = L_{\mt{AdS}}^2\qty(d\eta^2 + \sinh^2{\eta}\qty(d\rmtE^2 + \cos^2{\rmtE}\;d\theta^2)).
\end{equation}
The Euclidean time $\rmtE$ runs in the range $\rmtE\in[-\frac{\pi}{2},\frac{\pi}{2}]$. The asymptotic boundary of AdS$_3$ can be approached by taking the limit $\eta \to \infty$. In the following, we assume that the UV cut-off surface is located at $e^\eta= \coth \left(\frac{\epsilon}{2}\right) \approx \frac{2}{\epsilon} $ with identifying $\epsilon \, L_{\mt{AdS}}$ as the UV cutoff in the CFT$_2$. 

To align with the previous setup involving dS$_2$ gravity and a CFT$_2$ bath, we cut the global AdS$_3$ by a constant $\eta$ slice and assume that the boundary of the bulk spacetime is located at $\eta = \eta_\infty$, see figure \ref{fig:double} for an illustration. Since only part of the dS$_2$ brane is gravitational, we need to impose inhomogeneous boundary conditions on the dS$_2$ brane located at $\eta=\eta_\infty$:
\begin{enumerate}
 \item For $-\theta_D \leq \theta \leq \theta_D$, we identify this region as an End-of-World brane and thus impose the Neumann boundary condition, \ie 
 \begin{equation} \label{eq:Neumann}
    K[g]_{ij} - K[g]g_{ij} + \mathcal{T} g_{ij} = 0 \,.
\end{equation}
where $g_{ij}$ and $K[g]_{ij}$ denote the induced metric and extrinsic curvature of the brane, respectively. It is worth noting that a dS brane is only supported as a solution when its tension $\mathcal{T}$ satisfies $|\mathcal{T}|>1/L{\mt{AdS}}$. More explicitly, the tension of the brane determines its location in terms of $\mathcal{T}= \frac{\coth \eta_\infty}{L_{\mt{AdS}}}$.\footnote{Note that the sign of the tension, as well as the extrinsic curvature, will play an important role in the analysis in section \ref{sec:dif}. The trace of the extrinsic curvature of the dS brane, \ie, $K$, is positive with respect to the physical spacetime defined by the bulk region enclosed by the dS brane.} 
\item For $\theta_D \leq \theta \leq \pi$ and $-\pi \leq \theta \leq -\theta_D$, we assign the Dirichlet boundary condition as the normal boundary condition for the AdS/CFT correspondence. For simplicity, we can set $\theta_D = \frac{\pi}{2}$ in the following calculations. Since we expect that the field theory living on the dS brane is a holographic CFT$_2$, this region should be placed at the conformal boundary. This can be achieved by taking the large $\eta$ limit, \ie relating $\eta_{\infty}$ to the (dimensionless) UV cut-off $\epsilon$ as follows:
\begin{equation}
\eta_\infty  \approx  \log \( \frac{2}{\epsilon} \)\,.
\end{equation}
\end{enumerate}
 
This AdS$_3$ bulk spacetime with an EOW dS$_2$ brane serves as the holographic dual of the model discussed in the previous section. Furthermore, we argue that this is, in fact, a doubly holographic model. In other words, there are three equivalent dual descriptions from different perspectives at different dimensions:
\begin{enumerate}
    \item[(A)] \textbf{Bulk perspective:} AdS$_3$ bulk spacetime with an EOW dS$_2$ brane. We have the standard Einstein gravity living in AdS$_3$ spacetime with a dS$_2$ boundary, where the Dirichlet (and Neumann) boundary condition is imposed on one half of dS$_2$ (and the other half dS$_2$). The latter half dS$_2$ with the Neumann boundary condition can be regarded as an EOW brane, as shown in figure \ref{fig:double}.
    \item[(B)] \textbf{Brane perspective:} 2d induced gravity on a dS$_2$ braneworld is coupled to the 2d non-gravitational field theory living on dS$_2$ spacetime. We have a CFT on one half of dS$_2$ and induced gravity (or equivalently Liouville gravity) on the other half of dS$_2$, which are joined together along the timelike boundary. See the middle panel of figure \ref{fig:double}.
    \item[(C)] \textbf{Boundary perspective:} BCFT$_2$ living in a half dS$_2$ spacetime. There is a defect line located at $\theta=\theta_D$. This one-dimensional defect can be interpreted as the holographic dual of a half dS$_2$ gravitational spacetime \cite{Kawamoto:2023nki}.
\end{enumerate}
The equivalence between (A) and (B) is elucidated by braneworld holography \cite{Randall:1999ee, Randall:1999vf, Gubser:1999vj, Karch:2000ct} in the context of the dS brane. The induced gravity on the dS$_2$ brane, specifically the Liouville gravity \eqref{LVth}, can be derived by performing dimensional reduction along the radial direction \cite{Suzuki:2022xwv, Kawamoto:2023wzj, Akal:2020wfl, Neuenfeld:2024gta}. It is important to note that the reduced theory on dS$_2$ also includes a CFT matter sector, where divergences in the matter sector precisely cancel those appearing in the gravitational action. The duality between (B) and (C) can be supported by the half dS holography proposed in \cite{Kawamoto:2023nki}, which posits that the gravitational theory in a half dS$_{d+1}$ spacetime with a Dirichlet boundary is dual to a non-local quantum field theory \footnote{Although it is not immediately clear that the boundary dual theory is non-local, our findings regarding the non-extremal island (defined in later subsections) provide an alternative demonstration of this non-locality. Specifically, the island corresponding to a finite bath interval is observed to be spread across the entire gravitational region, rather than being confined by the quantum extremal surface.} residing on this $d$-dimensional timelike boundary. A qualitatively similar yet distinct holographic setup involving dS bulk duals is constructed through $T\overline{T}$ deformation \cite{Alishahiha:2004md, Dong:2018cuv, Gorbenko:2018oov}. We will not delve into this dS holography in this paper, as it is not critical for the calculations presented herein. Finally, the holographic duality between (A) and (C) follows from the AdS/BCFT correspondence \cite{Takayanagi:2011zk, Fujita:2011fp, Karch:2000gx}, where the dS brane setup was first considered in \cite{Akal:2020wfl}. Additionally, refer to the recent paper \cite{Wei:2024zez} for an interesting application of dS branes to cross-cap states. In this paper, we will focus on the holographic duality between (A) and (B).

The construction of this doubly holographic model is summarized in figure \ref{fig:double}. The advantage of this approach is that the bulk perspective, which employs a higher-dimensional AdS spacetime, offers an alternative method for calculating entanglement entropy. This aids in resolving the puzzles associated with the island formula in de Sitter gravity.

\subsection{Naive holographic entanglement entropy is problematic}\label{subsec:Naive Holographic Entanglement Entropy}

Similar to the previous section, we will focus on calculating the entanglement entropy of an interval $\mathcal{A}$ in the non-gravitational region. Without loss of generality, let us consider a subsystem $\mathcal{A}$ at a constant time slice $t=t_A$ on the dS$_2$ brane given by $\eta=\eta_\infty$ and choose a finite interval $\mathcal{A}=[\theta_{A_1},\theta_{A_2}]$ within the Dirichlet region\footnote{For a symmetric interval centered in the middle of the space, as shown in figure \ref{fig:antipodal}, one would choose $A=[\theta_A, \pi] \cup [-\pi, -\theta_A]$.}. Leveraging the holographic duality with the AdS$_3$ bulk spacetime, we can apply the holographic entanglement entropy (HEE) formula \cite{Ryu:2006bv,Ryu:2006ef,Hubeny:2007xt,Faulkner:2013ana,Engelhardt:2014gca} to compute the entanglement entropy of the subsystem $\mathcal{A}$ on the conformal boundary. We will explicitly demonstrate that this approach yields the same results as the island formula. However, we will also find that the HEE approach does not provide physically accurate results in this context.

\subsubsection{Holographic entanglement entropy}
Since there is an EOW brane in the bulk AdS$_3$ spacetime, there are two types of extremal surfaces according to the prescription of the AdS/BCFT correspondence \cite{Takayanagi:2011zk,Fujita:2011fp,Akal:2021foz,Akal:2022qei}. The first candidate is the connected extremal surface denoted by $\Gamma^{\con}_{\mathcal{A}}$, whose boundary is located only on the entangling surface $\partial \mathcal{A}$. The second candidate is the disconnected surface $\Gamma^{\dis}_{\mathcal{A}}$, which is anchored on the EOW brane. So $\Gamma_{A}^{\dis}$ denotes the geodesics which start from one of the entangling surfaces and end on the EOW brane. The endpoint of $\Gamma_{A}^{\dis}$ on the EOW will be determined by the extremization conditions. We denote the corresponding contributions to the entanglement entropy as $S_{\mathcal{A}}^{\mathrm{con}}$ and $S_{\mathcal{A}}^{\mathrm{dis}}$, respectively. Applying the holographic entanglement entropy (HEE) formula, the entanglement entropy of the bath interval $\mathcal{A}$ is given by
\begin{equation}\label{eq:BCFT_HEE}
S_{\mathcal{A}} = \min \{S_{\mathcal{A}} ^{\mathrm{con}}, S_{\mathcal{A}} ^{\mathrm{dis}}\} = \min \left\{ \frac{\text{Area}(\Gamma_{\mathcal{A}}^{\rm{con}})}{4\GN}, \frac{\text{Area}(\Gamma_{\mathcal{A}}^{\rm{dis}})}{4\GN}\right\} \,,
\end{equation}
where $\GN$ denotes the Newton's constant for the gravitational theory in the AdS$_3$ bulk spacetime and is distinct from the one for the induced gravity on the dS brane. To evaluate the area of extremal surfaces in AdS$_3$, we only need the dimensionless length of the spacelike geodesic $D_{ab}$, defined as:
\begin{equation}\label{eq:Dab}
\cosh D_{ab} = \cosh \eta_a \cosh \eta_b - \sinh \eta_a \sinh \eta_b \left(\cosh t_a \cosh t_b \cos \left(\theta_b - \theta_a\right) - \sinh t_a \sinh t_b \right) \,. 
\end{equation}
where the spacelike geodesic connects two points $(t_a, \eta_a, \theta_a)$ and $(t_b, \eta_b, \theta_b)$ in AdS$_3$ spacetime.

\textbf{Connected Extremal Surface $\Gamma_{\mathcal{A}}^{\rm{con}}$:} Considering the geodesics connecting the two endpoints of the interval $\mathcal{A}$, namely: $A_1 = (\eta_\infty, t_A, \theta_{A_1})$ and $A_2 = (\eta_\infty, t_A, \theta_{A_2})$, the corresponding entropy contribution is given by
\begin{equation}\label{eq:HEE_CON}
S_{\mathcal{A}}^\con = \frac{L_{\mt{AdS}} D_{12} }{4 G_N} = \frac{c}{3} \log \cosh{t_A} + \frac{c}{3} \log \qty(\frac{2}{\epsilon} \abs{\sin \qty(\frac{\theta_{A_1} - \theta_{A_2}}{2})}) \,. 
\end{equation}
Here, we use the Brown-Henneaux formula \cite{Brown:1986nw}:
\begin{equation}
c = \frac{3L_{\mathrm{AdS}}}{2 \GN}\,. 
\end{equation}
The linear growth along the global time $t_{\mA}$ reflects the exponential inflation of the dS background. This holographic formula from AdS$_3$ agrees with the result derived from the brane perspective in the non-island phase, as indicated by eq.~\eqref{timecon}, after identifying the size of the interval $\mA$.

\begin{figure}[t]
	\centering
    \includegraphics[width=6in]{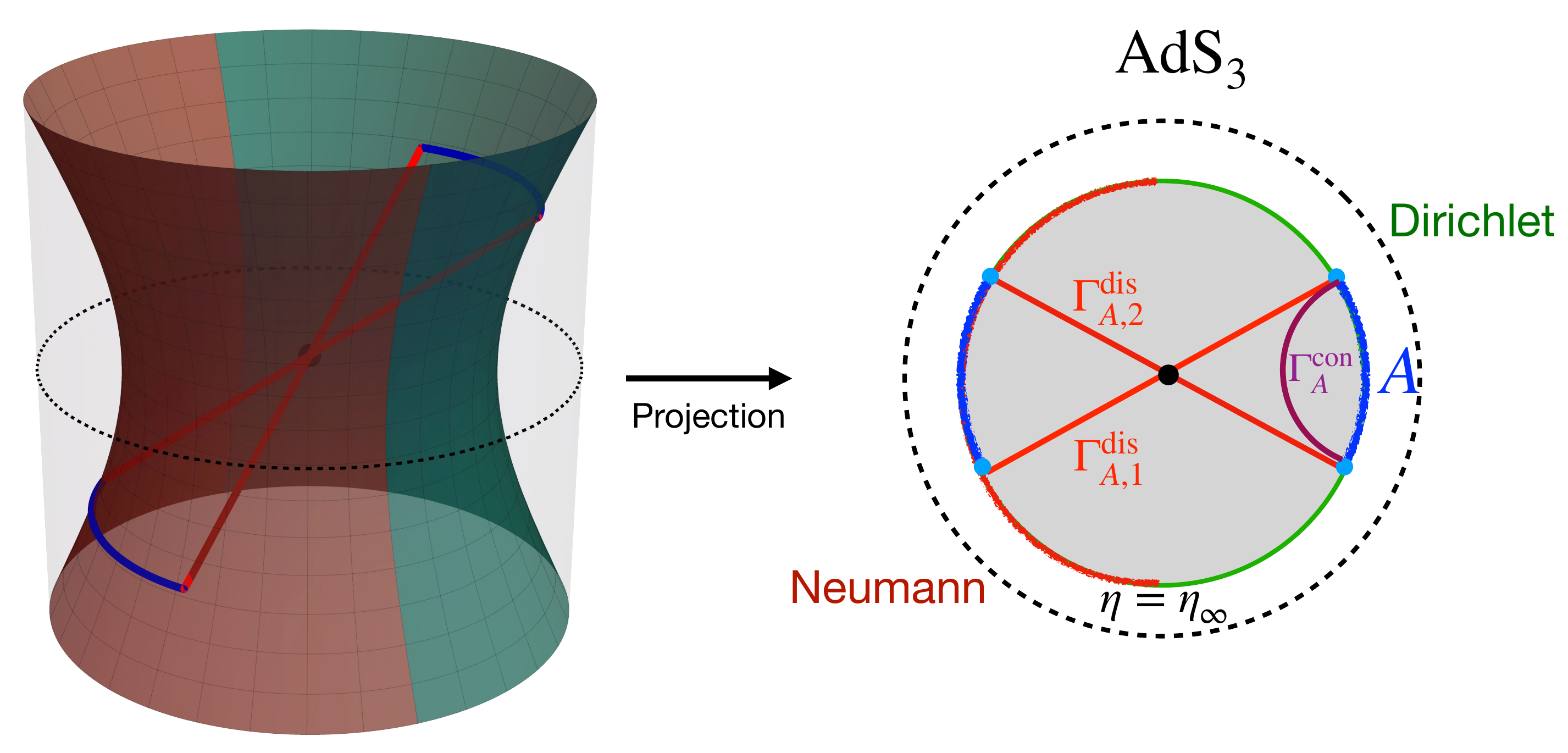}
    \caption{Left: A schematic diagram of the disconnected antipodal RT surface in AdS$_3$. The two antipodal RT surfaces (red lines) always intersect on the $t=0$ surface. Right: Sketch of a time slice in our setup. The red region denotes the EOW brane, while the green region represents the CFT bath with Dirichlet boundary conditions. The subsystem $\mathcal{A}$, for which we calculate the entanglement entropy, is indicated by the blue curve. The red lines in the bulk denote the RT surfaces for the disconnected phase, and the purple curve represents the RT surface for the connected phase.}
    \label{fig:dS_2EEdis}
\end{figure}

\textbf{Disconnected Extremal Surface $\Gamma_{\mathcal{A}}^{\rm dis}$:} 
In this phase, the extremal surfaces are disconnected, with each surface connecting the boundary of the interval $\mathcal{A}$ to the EOW brane. For a given extremal surface, we denote the endpoints on the AdS$_3$ boundary by $A_i= (\eta_\infty,t_{A_i}, \theta_{A_i})$  and the endpoints on the EOW brane by $I_i= (\eta_\infty, t_{I_i}, \theta_{I_i})$. Focusing on one geodesic, we simplify by omitting the subscript $i=1,2$. The location of $I_i$ can be fixed by solving the extremization conditions, namely
\begin{equation}\label{eq:extream_condition}
    \frac{\partial D_{AI}}{\partial t_I} = 0, \; \frac{\partial D_{AI}}{\partial \theta_I} = 0.
\end{equation}
Using the universal formula \eqref{eq:Dab} for the geodesic length, its derivatives are 
\begin{equation}
\begin{split}
    \sinh{D_{AI}} \frac{\partial D_{AI}}{\partial \theta_I} &= \sinh^2{\eta_\infty} \cosh{t_A} \cosh{t_I} \sin{(\theta_I - \theta_A)}  \,,\ \\
   \sinh{D_{AI}} \frac{\partial D_{AI}}{\partial t_I} &= \sinh^2{\eta_\infty}   \left(  \cosh{t_A} \sinh{t_I} \cos (\theta_A- \theta_I) -\sinh{t_A} \cosh{t_I}   \right) \,.
\end{split}
\end{equation}
Constrained to the EOW brane\footnote{In this computation, we do not need to specify the size of the Dirichlet region, \ie the value of $\theta_D$. However, if the Dirichlet region is larger than half the size of the entire space, there is no disconnected RT surface. We will assume that EOW brane on the spatial circle is bigger enough such that the disconnected extremal surface could exist.}, the extremal conditions yield a single solution: 
\begin{equation}\label{eq:extremal}
\theta_I = \theta_A - \pi \,, \qquad t_I = -t_A \,. 
\end{equation}
As illustrated in figure \ref{fig:dS_2EEdis}, the extremal disconnected RT surface always connects antipodal points on the spatial circle. This location on the EOW brane corresponds precisely to the quantum extremal surface, \ie the boundary of the antipodal island \eqref{eq:QESantipodal} derived from the island formula.

One can find that the holographic entanglement entropy associated with the extremal geodesic $\Gamma_{\mathcal{A}}^{\rm dis}$ reduces to a constant as follows 
\begin{equation}\label{eq:HEE_dis}
\begin{split}
        S_{\mathcal{A}}^{\dis} = \frac{L_{\mt{AdS}}}{4 \GN}\( D_{A_1I_1} + D_{A_2I_2} \)  = \frac{\eta_\infty}{\GN} &= \frac{2c}{3} \eta_\infty  \approx \frac{2c}{3} \log \( \frac{2}{\epsilon} \) \,,
\end{split}
\end{equation}  
After choosing $S_0=0$,\footnote{This is due to the fact that all contributions on the brane are assigned as originating from the CFT$_2$ (Liouville field) that resides on the EOW brane. An alternative approach would be to consider all contributions on the EOW as originating from the effective gravity on the dS$_2$ brane and rewrite $S_{\mathcal{A}}^{\dis}$ as 
\begin{equation*}
 S_{\mathcal{A}}^{\dis} =  2  \times \(  \frac{\eta_\infty}{4\GN}  +  \frac{c}{6} \log \( \frac{2}{\epsilon} \)  \)  \,,    
\end{equation*}
where the effective Newton's constant on dS$_2$ brane is given by $\frac{1}{\GN^{(2)}} = \frac{\eta_\infty}{\GN} $.} this reproduces our previous result \eqref{discgh} that is derived from the direct application of the island formula from the brane perspective. From the equivalent boundary perspective, we can also recast this result in terms of
\begin{equation}
        S_{\mathcal{A}}^{\dis} =2  \times \(   \frac{c}{6} \log \( \frac{2}{\epsilon} \)   +  S_{\rm bdy}  \) \,,
\end{equation}  
where the boundary entropy is defined as $S_{\rm bdy} = \frac{c}{6} \log \sqrt{ \frac{\mathcal{T}+L_{\mt{AdS}}}{\mathcal{T}-L_{\mt{AdS}}} } =\frac{c}{6} \eta_\infty$.

\subsubsection{Problems with the disconnected extremal surface}

The resulting HEE for an interval $\mathcal{A}$ is determined by the minimal value between $ S_{\mathcal{A}}^{\rm con}$ \eqref{eq:HEE_CON} and $ S_{\mathcal{A}}^{\rm dis}$ \eqref{eq:HEE_dis}. As illustrated by figure \ref{fig:dS2EE}, 
the connected extremal surface $\Gamma_{\mathcal{A}}^{\rm con}$  is favored at early times, leading to a HEE that exhibits linear growth with time. This growth can be described by eq.~\eqref{eq:HEE_CON}. After the critical time $t_\ast$, given by 
\begin{equation}\label{eq:tast01}
 t_\ast  \approx \log \left(\frac{4}{\epsilon} \right) -   \log \left( \abs{\sin \qty(\frac{\theta_{A_1} - \theta_{A_2}}{2})} \right) \,, 
\end{equation}
a phase transition occurs between the connected extremal surface $\Gamma_{\mathcal{A}}^{\rm con}$ and the disconnected extremal surface $\Gamma_{\mathcal{A}}^{\rm dis}$. After this transition, the HEE reaches a constant value derived in eq.~\eqref{eq:HEE_dis}.

Nevertheless, we argue that the holographic entanglement entropy calculation based on the conventional RT formula is not physical. Specifically, the disconnected extremal surface connecting two antipodal points on the dS space presents issues and appears unphysical. As discussed in the previous section, similar problems arise with the antipodal island. The application of double holography offers further evidence from the perspective of the bulk spacetime.

First and foremost, it is important to emphasize that the two disconnected extremal RT surfaces always intersect at the center of the bulk spacetime, specifically at $t=0, \theta=0$, as shown in figure \ref{fig:dS_2EEdis}. This intersection indicates that these extremal surfaces fail to meet the general conditions for RT surfaces and cannot be considered physical. Furthermore, this intersection makes the entanglement wedge \cite{Czech:2012bh,Wall:2012uf,Headrick:2014cta} ill-defined, as illustrated in the right panel of figure \ref{fig:dS_2EEdis}. On the other hand, the extremal geodesics anchored on the EOW brane represent the minimax surface along the Lorentzian dS brane. It is easy to get the second derivatives of the geodesic length at the antipodal solution. To wit, 
    \begin{equation}
        \begin{split}
         \left.\frac{\partial^2 D_{AI}}{\partial t_I^2} \right|_{\mathrm{antipodal}}&=-\left. \frac{\partial^2  D_{AI}}{\partial (\rmtE)_I^2} \right|_{\mathrm{antipodal}}=\frac{1}{2}\tanh{\eta_\infty}>0,\\  \left.\frac{\partial^2 D_{AI}}{\partial \theta_I^2} \right|_{\mathrm{antipodal}}&=-\frac{1}{2}\cosh^2{t_0}\tanh{\eta_\infty}<0 \,.
        \end{split}
    \end{equation}
These calculations show that a disconnected RT surface that is extremized along the dS brane is maximal with respect to variations along both the spatial and Euclidean time directions. Therefore, it is reasonable to infer that this type of extremal surface does not correspond to the dominant saddle in the Euclidean path integral for calculating the entanglement entropy. This behavior is particularly pathological when compared with the holographic entanglement entropy for an AdS brane. For a detailed comparison with the AdS brane scenario, please refer to appendix \ref{sec:appB}. 

Given these issues, we conclude that the standard holographic entanglement entropy formula and the island formula are not applicable to dS EOW branes or dS gravity coupled to a non-gravitational bath.

\subsubsection{Improved doubly holographic model}

Before proceeding, we would like to address an additional UV divergence that appears in the doubly holographic model with a dS EOW brane located at $\eta = \eta_\infty$. To join the EOW brane to the Dirichlet brane near the conformal boundary, we must choose $\eta_\infty \approx \log \left( \frac{2}{\epsilon} \right)$. Consequently, the contributions from the area of the disconnected RT surface will always contain twice the UV divergence compared to the connected ones. One might argue that these issues associated with the disconnected RT surface anchored to the dS brane are not problematic, as it would take an infinite time $t_\ast$ \eqref{eq:tast01} in the limit $\epsilon \to 0$ for the disconnected configuration to be favored.

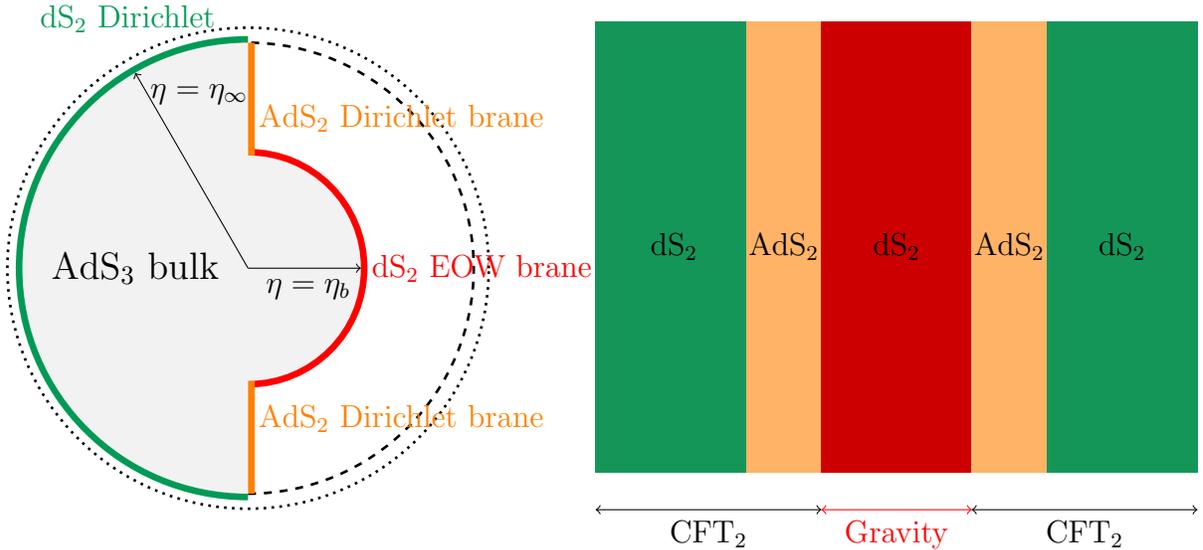
\begin{figure}[t]
    \centering
    \begin{minipage}[h]{0.495\linewidth}
       \centering
    \begin{tikzpicture}
        \draw[ForestGreen, line width=5pt] (0,3) arc(90:270:3);
        \draw[black, line width=1pt,dashed] (0,-3) arc(-90:90:3);
        \draw[black, line width=1pt,dotted] (3.2,0) arc(0:360:3.2);
        \draw[red, line width=5pt] (0,-1.5) arc(-90:90:1.5);
        \draw[orange, line width=5pt](0,1.5)--(0,3);
        \draw[orange, line width=5pt](0,-1.5)--(0,-3);
        \fill[lightgray!20!white](0,1.5)--(0,3) to[out=180,in= 90](-3,0)to[out=-90,in= -180](0,-3)--(0,-1.5)to[out=0,in=-90](1.5,0)to[out=90,in=0] (0,1.5);
        \draw[red](1.5,0)node[right]{dS$_2$ EOW brane};
        \draw[ForestGreen](-1.6,3)node[above]{dS$_2$ Dirichlet};
        \draw[orange](0,2)node[right]{AdS$_2$ Dirichlet brane};
        \draw[orange](0,-2)node[right]{AdS$_2$ Dirichlet brane};
        \draw[black](-1.5,0)node{\large{AdS$_3$ bulk}};
        \draw[black,->](0,0)--(1.5,0);
         \draw[black,->](0,0)--(-1.5,2.598);
        \draw[black](0.8,0)node[below]{$\eta=\eta_b$};
        \draw[black](-0.65,2.598)node[below]{$\eta=\eta_\infty$};
    \end{tikzpicture}
    \end{minipage}
    \begin{minipage}[h]{0.495\linewidth}
     \vspace{10mm}
        \centering
    \begin{tikzpicture}
       \fill[Green!80!black] (-2,-3)--(0,-3)--(0,3)--(-2,3)--cycle;
       \fill[orange!60!white](0,-3)--(1,-3)--(1,3)--(0,3)--cycle;
        \fill[red!80!black](1,-3)--(3,-3)--(3,3)--(1,3)--cycle;
        \fill[orange!60!white](3,-3)--(4,-3)--(4,3)--(3,3)--cycle;
        \fill[Green!80!black](4,-3)--(6,-3)--(6,3)--(4,3)--cycle;
\draw[black](-0.5,0)node[left]{dS$_2$};
\draw[black](0.5,0)node{AdS$_2$};
\draw[black](2,0)node{dS$_2$};
\draw[black](3.5,0)node{AdS$_2$};
\draw[black](5,0)node{dS$_2$};
\draw[<->](-2,-3.5)--(1,-3.5);
\draw[<->,red](1,-3.5)--(3,-3.5);
\draw[<->](3,-3.5)--(6,-3.5);
\draw[black](-0.5,-3.5)node[below]{CFT$_2$};
\draw[red](2,-3.5)node[below]{Gravity};
\draw[black](4.5,-3.5)node[below]{CFT$_2$};
    \end{tikzpicture}
    \end{minipage}
    \caption{Sketches of the doubly holographic setup with a dS brane at finite radius. The left panel shows the AdS$_3$ bulk spacetime with an EOW brane: Dirichlet boundary conditions are imposed on the large-radius dS$_2$ cut (green) and the AdS$_2$ cut (orange), while Neumann boundary conditions are imposed on the finite-radius dS brane (red). The dotted line indicates the conformal boundary of the global AdS spacetime. The right panel shows the picture from the brane perspective: there is no gravity in the green and orange regions, while induced gravity is present in the red region.}
    \label{fig:finite_cut_off_setup}
\end{figure}

Nevertheless, we posit that this phenomenon is merely a consequence of the intrinsic simplicity of the dS$_2$ brane solution within this doubly holographic model. To avoid this confusion, it may be beneficial to consider a more physical, but somewhat more complicated brane configuration. As shown in figure \ref{fig:finite_cut_off_setup}, we consider an EOW brane with a finite dS radius, choosing $\eta = \eta_b \ll \eta_{\infty}$ and imposing the Neumann boundary condition. In the CFT region at the conformal boundary with the Dirichlet boundary condition, we fix it as a dS$_2$ brane with $\eta=\eta_\infty \sim \log \left( \frac{2}{\epsilon} \right)$. To connect the two dS regions with different radii, it is necessary to include an AdS$_2$ region with the Dirichlet boundary condition. In the symmetric case with $\theta_D = \frac{\pi}{2}$, the intermediate AdS$_2$ region corresponds to a brane with vanishing tension. From the two-dimensional brane perspective, this doubly holographic model can be viewed as dS$_2$ gravity coupled to a non-gravitational bath with a background that includes both a half dS$_2$ and a finite AdS$_2$ spacetime. By taking the limit $\eta_b \to \eta_{\infty}$, we can recover the previous model where the AdS$_2$ region shrinks to nothing. In the next section, we will focus on the EOW dS$_2$ brane with a finite $\eta_b$ and demonstrate how the bulk holography in AdS$_3$ can help to solve the puzzles associated with the antipodal island and the disconnected RT surfaces. 


 
\subsection{Non-extremal Island on Euclidean dS brane}

Although we have argued that the standard holographic entanglement formula, which evaluates the area of extremal surfaces, cannot provide the correct answer, we can still calculate the entanglement entropy $S_{\mA}$ by taking the advantage of holography. Specifically, in the setting of a subsystem $\mathcal{A}$ in the non-gravitational half dS$_2$ region coupled to a gravity theory on another half dS$_2$, with or without an intermediate AdS$_2$ region, the entanglement entropy can be calculated using the replica method \cite{Calabrese:2004eu,Calabrese:2009qy}.
By taking the von Neumann limit with $n\to 1$, the entanglement entropy can be derived as the two-point function of twist operators at the endpoints, namely 
\begin{equation}\label{EErep}
S_{\mathcal{A}} \equiv \lim_{n\to 1}\frac{1}{1-n}\log \la \sigma_n(t_{A_1},\theta_{A_1})\bar{\sigma}_n(t_{A_2},\theta_{A_2})\lb \,,   
\end{equation} 
where the normalization of twist operators is fixed as $\la \sigma_1(t_{A_1},\theta_{A_1})\bar{\sigma}_1(t_{A_2},\theta_{A_2})\lb=1$ and their conformal dimension is $\Delta_n=\frac{c}{12}(n-\frac{1}{n})$. In the semi-classical limit of a holographic field theory, this is equivalent to taking a large central charge limit
$\Delta_n \sim c  \sim \frac{1}{\GN} \to \infty$, which allows for a bulk dual description \footnote{A meticulous reader may have noticed that the two limits, $c\to\infty$ and $n\to1$, are mutually exclusive from the point of view of the dimensionless parameter $\Delta_n$. This ambiguity could be circumvented by carefully dealing with the order of the limits. We first take $c \to \infty$ limit to compute the holographic correlation function $\frac{\log \ev{\sigma_n}}{\Delta_n}$ and then use $n \to 1$ to get the von Neumann entropy.}. For example, the two-point function of CFT$_2$ with a large charge can be derived by using the geodesic approximation \cite{Balasubramanian:1999zv,Louko:2000tp,Aparicio:2011zy,Balasubramanian:2012tu} in AdS$_3$, which reproduces the RT formula with using the geodesic length.  

For a holographic BCFT, the two-point functions at leading order in the large central charge limit $c\to\infty$ are dominated by two distinct channels: 
\begin{equation}\label{eq:twistors}
\ev{\sigma_n(t_{A_1},\theta_{A_1})\bar{\sigma}_n(t_{A_2},\theta_{A_2})}_{\mathrm{BCFT}}= \max 
  \begin{cases} \ev{\sigma_n(t_{A_1},\theta_{A_1})\bar{\sigma}_n(t_{A_2},\theta_{A_2})}  \\
\ev{\sigma_n(t_{A_1},\theta_{A_1})}\ev{\bar{\sigma}_n(t_{A_2},\theta_{A_2})}\\
  \end{cases} \,.
 \end{equation}
Here, the one-point function is non-vanishing due to the presence of a boundary in the background. These two channels correspond to the connected and disconnected Renyi entropies, analogous to the two types of RT surfaces in eq.~\eqref{eq:BCFT_HEE}. For the standard AdS$_3$/BCFT$_2$ correspondence with an AdS brane, the equivalence between eq.~\eqref{eq:twistors} and the holographic entanglement entropy formula in eq.~\eqref{eq:BCFT_HEE} can be demonstrated using the geodesic approximation \cite{Balasubramanian:1999zv,Louko:2000tp,Aparicio:2011zy,Balasubramanian:2012tu}. However, our findings suggest that the result from the disconnected RT surface anchored on a dS brane is problematic. In the following subsections, we will demonstrate that the factorized two-point function is not given by the disconnected extremal surface. Rather than employing the RT formula, we can calculate the two-point function directly through the holographic correspondence in AdS$_3$. The connected channel remains consistent with that derived from the connected RT surface. We will concentrate on the disconnected phase, meticulously evaluating the one-point function of the twistor operators in holographic field theory.  

\subsubsection{One-point functions in AdS/BCFT}
Let us first consider Euclidean AdS$_{d+1}$ bulk spacetime with a generic EOW brane. For example, the Euclidean AdS metric $G_{MN}$ in $(d+1)$ dimensions with the Poincar\'e coordinate reads
\begin{equation}
ds^2=G_{MN}\; dx^M dx^N =\frac{L_{\mt{AdS}}^2}{z^2} \left( dz^2  + \delta_{ij}
 dx^i dx^j  \right)\,,
\end{equation}
where the BCFT is living on the conformal boundary with the Dirichlet boundary condition and the EOW brane, satisfying the Neumann boundary condition, is assumed to be given by the hypersurface $z=F(x_b)$ with $x_b$ as the induced metric on the EOW brane. The expressions for the other coordinates for AdS${d+1}$ can be obtained by performing the coordinate transformations. Even though there is an assumed EOW brane in the bulk space, the bulk-to-bulk propagator $G{\mt{BB}}^\Delta(x,x')$ for the bulk scalar field still satisfies the same equation of motion as that for the AdS space, namely
\begin{equation}\label{eq:EOMGBB}
\left(\square_G +m^2\right) G_{\mt{BB}}^\Delta(x ; x')=\frac{\delta^{(d+1)}(x- x')}{\sqrt{G}}\,,
\end{equation}
where the mass of the bulk field is related to the conformal dimension of the dual operator on the boundary via the standard AdS/CFT dictionary $m^2 = \Delta (\Delta -d)$. For $(d+1)$-dimensional AdS space, the solution is known as \cite{Witten:1998qj,Freedman:1998tz,Ammon:2015wua} 
\begin{equation}\label{eq:bulk-to-bulk_Aropagator}
G_{\mt{BB}}^\Delta (x ; x')= G_{\mt{BB}}^\Delta (\xi)=  \frac{C_{\Delta}}{2^{\Delta}(2 \Delta-d)} \xi^{\Delta} \cdot{ }_2 F_1\left(\frac{\Delta}{2}, \frac{\Delta+1}{2} ; \Delta-\frac{d}{2}+1 ; \xi^2\right) \,,
\end{equation}
with the normalization constant given by 
\begin{equation}
C_{\Delta}=\frac{\Gamma(\Delta)}{\pi^{d / 2} \Gamma\left(\Delta-\frac{d}{2}\right)} \,. 
\end{equation}
Here we have introduced the so-called chordal distance $\xi$ that is associated with the (dimensionless) geodesic distance $D(x,x')$ as the following 
\begin{equation}\label{eq:geodesic_length}
D(x ; x') =  \, \ln \left(\frac{1+\sqrt{1-\xi^2}}{\xi}\right) \,, \qquad  \xi = \frac{1}{\cosh  \( D(x ; x')  \)}.
\end{equation}
For the case with one point on the conformal boundary, \eg $z=0$ in the Poincar\'e coordinates, it is more useful to define the bulk-to-boundary propagator $K_{\mt{Bb}}^\Delta (x,x')$ by taking the boundary limit of the bulk-to-bulk propagator. For example, in the Poincar\'e coordinate, we have 
\begin{equation}\label{eq:bulk-to-boundary_propagator}
K_{\mt{Bb}}^\Delta(\vec{x}; z',\vec{x}') =\lim _{z \rightarrow 0} \frac{2 \Delta-d}{(z)^{\Delta}} G_{\mt{BB}}^\Delta (z,\vec{x}; z', \vec{x}')  \,,
\end{equation}
where the normalization factor is chosen to be the same as most literature. The bulk-to-boundary propagator in other coordinates is obtained from the conformal transformations.

The biggest advantage we take from the holographic bulk spacetime is that the one-point function of BCFT on the conformal boundary can be derived using the bulk-to-boundary propagator as follows \cite{Suzuki:2022xwv,Izumi:2022opi,Kastikainen:2021ybu}:
\begin{equation}\label{eq:Onepoint}
  \ev{\mO(x)}_{\mt{BCFT}} = \alpha \cdot \, \int_{\mt{EOW}} d^d x_b \sqrt{g} K_{\mt{Bb}}^\Delta (x ; x_b) \,.
\end{equation}
where $\alpha$ represents the normalization factor which is ignored for simplicity, and the coordinates $x$ and $x_b$ denote the two points on the Dirichlet boundary and the EOW brane, respectively. Due to this holographic dictionary, we can then evaluate the one-point function of the twistor operator and thus obtain the corresponding entanglement entropy in the disconnected phase by applying the replica trick \eqref{EErep}.

The integral along the EOW brane indicates that the gravitational region on the brane is correlated with the boundary. It seems that this contradicts the island story because the correlation is smeared over the whole EOW brane. However, it is interesting to note that this integral along the brane could be dominated by a particular saddle point, which can reproduce the island formula. This saddle point approximation is essentially the known geodesic approximation for the correlation functions in the AdS/CFT duality.
 
The first key point is that we only need correlation functions for the operator with a large conformal dimension $\Delta \to \infty$ to evaluate the entanglement entropy \eqref{eq:twistors} using the twistor operators with $\Delta_n \to \infty$. By taking $\Delta \to \infty$, we can approximate the bulk-to-bulk propagator and the bulk-to-boundary propagator by the geodesic distance \cite{Balasubramanian:1999zv}, \viz 
\begin{equation}
\begin{split}
G_{\mt{BB}}^\Delta(x ; x_b)    
    & \approx  \frac{1}{2\Delta-d}e^{-\Delta D(x ; x_b)}e^{1-\frac{d}{2}}\qty(\frac{\Delta}{2\pi}\frac{1+\sqrt{1-\xi^2}}{\sqrt{1-\xi^2}})^{\frac{d}{2}} \,, 
          \\
 K_{\mt{Bb}}^\Delta(\vec{x};z_b,\Vec{x}_b)&=\lim _{z \rightarrow 0} e^{-\Delta\qty(D(x ; x_b)+\log{z})}e^{1-\frac{d}{2}}\qty(\frac{\Delta}{\pi})^{\frac{d}{2}} \,. 
\end{split} 
\end{equation}
In this limit, the bulk-to-bulk and bulk-to-boundary propagators become dominated by the geodesic distances, which simplifies the calculation of the one-point function and, consequently, the entanglement entropy. For the derivation of these expressions, see appendix \ref{sec:appB}. 

The one-point function of the primary operator on the conformal boundary of BCFT is thus approximated by
\begin{equation}\label{eq:one_pt}
    \ev{\mO(x)}_{\mt{BCFT}}  \approx \int_{\mt{EOW}} d^d x_b \sqrt{g} e^{-\Delta\bar{D}(x ; x_b)} \,, 
\end{equation}
where $\bar{D}(x ; x_b)$ denotes the renormalized geodesic distance, excluding the UV divergent part. To calculate the entanglement entropy, we apply this large dimension limit to the twist operator $\sigma_n$ with conformal dimension $\Delta_n= \frac{c}{12}\left(n - \frac{1}{n}\right)$. Considering the holographic field theory with $c\to\infty$, the non-vanishing one-point function is evaluated as
\begin{equation}\label{eq:one_pt_twist}
    \ev{\sigma_n(x_A)}  \approx \const\times \int_{\mt{EOW}} d^2 x_b \sqrt{g} e^{-\Delta_n{D}(x_A ; x_b)} \,, 
\end{equation}
where we use the original geodesic length $D(x_A, x_b)$ to match the UV divergence for the holographic entanglement entropy.

The second key approach is the saddle point approximation for evaluating the integral along the brane. Assuming there exists a saddle point satisfying
\begin{equation}
 \partial_{x_b} D(x_A ; x_b)  \big|_{x_b= x_b^\ast} =0 \,,
\end{equation}
along {\it all} directions on the EOW brane, the saddle point approximation\footnote{A simple example of the saddle point approximation is the integral of the form
\begin{equation*}
I \equiv \int_{-\infty}^{\infty} e^{-f(x)}\, d x \approx e^{-f\left(x_0\right)} \int_{-\infty}^{\infty}  e^{-\frac{1}{2}\left(x-x_0\right)^2 f^{\prime \prime}\left(x_0\right)} \,d x= e^{-f\left(x_0\right)} \sqrt{\frac{2 \pi}{f^{\prime \prime}\left(x_0\right)}} \,. 
\end{equation*}
This approximation only holds for locally minimal points where
\begin{equation*}
 f^{\prime}\left(x_0\right) =0 \,, \qquad f^{\prime \prime}\left(x_0\right) > 0 \,.
\end{equation*}}
applied to the integral defined in eq.~\eqref{eq:one_pt_twist} yields
\footnote{The volume measure of the brane, \ie $\sqrt{g(x_b)}$ would also influence the position of the saddle point. However, its effect can be ignored after taking the limit $\Delta_n \to \infty$.} 
\begin{equation}
    \ev{\sigma_n(x_A)}  \approx  \sqrt{\frac{g(x_b^*)}{ \det (D''(x_A ; x_b^\ast))}}\; e^{-\Delta_n{D}(x_A; x_b^\ast)} \,.
\end{equation}
Finally, we find that this approximation of the one-point function reproduces the holographic entanglement entropy for the disconnected extremal surface, \ie 
\begin{equation}\label{formulaHEER}
\begin{split}
S_{\mathcal{A}} &\approx  \lim_{n\to 1}\frac{1}{1-n}\( \lim_{\Delta_n \to \infty} \( \log \ev{\sigma_n(x_{A_1})} + \log \ev{\sigma_n(x_{A_2})}  \) \)     \\ 
&=  \frac{c}{6} \, \(    D(x_{A_1}; x_{b_2}^\ast)  +D(x_{A_2}; x_{b_2}^\ast)  \)  \,, \\
\end{split}
\end{equation}
where we take the leading contribution in the limit $\Delta_n \sim c \to \infty$. The saddle point associated with the integral on the EOW brane corresponds to the extremal geodesic orthogonal to the EOW brane. This saddle point condition is equivalent to the extremization condition defined in eq.~\eqref{eq:extream_condition}. For a detailed illustration of this saddle point approximation in the AdS/BCFT correspondence, see appendix \ref{sec:appB}. 

However, a caveat associated with the saddle point approximation for the integral on the EOW brane is that it only applies to locally minimal points satisfying
\begin{equation}
 \partial_{x_b} \partial_{x_b} {D}(x_A ; x_b) \big|_{x_b = x_b^\ast}= D''(x_A ; x_b) > 0 \,,
\end{equation}
or all directions along the EOW brane \footnote{More precisely, this condition requires that the matrix $\partial_{x_b'} \partial_{x_b} {D}(x_A, x_b) \big|_{x_b = x_b^\ast}$ is positive definite.}. Saddle points that do not satisfy this condition cannot dominate the integral along the EOW brane. In our doubly holographic models with a dS EOW brane, we will demonstrate that the saddle point approximation fails because there are no locally minimal points on the dS EOW brane.


\subsubsection{One-point function with the dS brane}\label{subsec:one_point_dS}
In the following, we focus on computing the entanglement entropy with a dS gravity by using the one-point function of the twistor operator, which avoids the issues of the standard island formula and the disconnected RT surface. We begin with the Euclidean AdS$_3$/dS$_2$ slicing coordinates:
\begin{equation}
 ds^2= L_{\mt{AdS}}^2\(     d\eta^2 + \sinh^2{\eta}
 \(dt_{\mt{E}}^2+\cos^2{t_{\mt{E}}}d\theta^2\)  \) \,,
\end{equation}
where the corresponding chordal distance is given by
\begin{equation}
\xi = \frac{2  \sech({\eta_1}) \sech(\eta_2)}{1-\tanh (\eta_1) \tanh (\eta_2) \(\cos (t_{\mt{E}1}) \cos (t_{\mt{E}2}) \cos (\theta_1-\theta_2)+\sin (t_{\mt{E}1}) \sin (t_{\mt{E}2})\) } \,.
\end{equation}
We assume that the dS EOW brane with the Neumann boundary condition is located at $\eta = \eta_b$, as depicted in figure \ref{fig:finite_cut_off_setup}. The induced metric on the (Euclidean) EOW brane is given by the sphere metric:
\begin{equation}
ds^2 \big|_{\mt{EOW}}= L_{\mt{AdS}}^2\sinh^2{\eta_b}
 \(dt_{\mt{E}}^2+\cos^2{t_{\mt{E}}}d\theta^2\) \,.
\end{equation}
with the dS radius given by $ L \equiv L_{\mt{AdS}}\sinh{\eta_b}$.

Substituting the chordal distance to the bulk-to-boundary propagator $K_{\mt{Bb}}^\Delta$ defined in eq.~\eqref{eq:bulk-to-boundary_propagator} with $d=2$, one can derive the non-vanishing one-point function \eqref{eq:Onepoint} on the conformal boundary at $\eta_\infty \to \infty$ by performing the integral along the Euclidean dS EOW brane (\ie a half sphere), \viz
\begin{equation}\label{eq:dSintegral}
    \ev{\mO (t_{\mt{E}}, \theta)}_{\mt{dS}} = \int_{-\frac{\pi}{2}}^{\frac{\pi}{2}} dt_{\mt{Eb}}  \int_{-\frac{\pi}{2}}^{\frac{\pi}{2}} d\theta_b \, 
    \frac{\cos{t_{\mt{Eb}}}}{\qty(\cosh{\eta_b}-\sinh{\eta_b}(\cos{t_{\mt{Eb}}}\cos{t_{\mt{E}}}\cos(\theta-\theta_b)+\sin{t_{\mt{E}b}}\sin{t_{\mt{E}}}))^\Delta} \,,
\end{equation}
the normalization factor has been ignored and the edge of the EOW brane has been chosen at $|\theta_D |= \frac{\pi}{2}$ for clarity.
In particular, the one-point function simplifies to 
\begin{equation}\label{numintonea}
\begin{split}
  \ev{\mO (t_{\mt{E}}, \theta)}_{\mt{dS}} \approx \int_{-\frac{\pi}{2}}^{\frac{\pi}{2}} dt_{\mt{Eb}}  \int_{-\frac{\pi}{2}}^{\frac{\pi}{2}} d\theta_b \,  \frac{\cos t_{\mt{Eb} }}{\qty(1- \cos t_{\mt{E}} \cos t_{\mt{Eb}} \cos (\theta-\theta_b) - \sin t_{\mt{E}} \sin t_{\mt{Eb}} )^\Delta}\ \,,
\end{split}
\end{equation}
for the simpler model where the dS EOW brane is placed at $\eta_b =\eta_\infty \sim \infty$. In the following, we focus on deriving the leading contribution of the one-point function in eq.~\eqref{eq:dSintegral} in the limit $\Delta \to \infty$ for the purpose of computing the entanglement entropy. \\

\textbf{Saddle point is a local maximum:} 
In the limit of large $\Delta$, we apply the standard geodesic approximation and use the saddle point approximation to evaluate the integral for the one-point function $\ev{\sigma_n}_{\mt{dS}}$ of the twistor operator. The relevant integral is approximately given by 
\begin{equation}\label{eq:one_pt_dS}
 \ev{ \mO (t_{\mt{E}}, \theta)}_{\mt{dS}}  \approx \int_{-\frac{\pi}{2}}^{\frac{\pi}{2}} dt_{\mt{Eb}} \int_{-\frac{\pi}{2}}^{\frac{\pi}{2}} d\theta_b \; \cos{(t_{\mathrm{Eb}})}\; e^{-\Delta D(t_{\mt{E}}, \theta; t_{\mt{Eb}}, \theta_b)} \,, 
\end{equation}
where $ D(t_{\mt{E}}, \theta; t_{\mt{Eb}}, \theta_b)$ denotes the geodesic distance between the endpoints on the boundary and the EOW brane, which is the same as in eq.~\eqref{eq:Dab} following the Wick rotation. Ignoring the volume measure on the EOW brane\footnote{
In fact, the integration contour can be deformed to the complex plane, allowing the identification of complex saddles. However, the complex saddles are located at $\rmtE=\frac{\pi}{2}+\pi n, \; n\in \mathbb{Z}$. This results in a vanishing prefactor in eq.~\eqref{eq:one_pt_dS}. Additionally, we observe a complex geodesic length, which lacks a straightforward physical interpretation. In this paper, we concentrate on directly evaluating the integral on the EOW brane.}, we can find only one saddle point for the integral, \ie 
\begin{equation}\label{eq:antipodalxb}
  x_b^\ast  =  \(  \theta_b = \theta -\pi \,, t_{\mt{Eb}} = - t_{\mt{E}}   \) \,,  
\end{equation}
whose Lorentzian continuation corresponds to the endpoint of the disconnected RT surface derived in eq.~\eqref{eq:extremal}. This saddle point indeed represents the antipodal island. However, it is straightforward to see that this antipodal saddle point $x_b^\ast$  is a local maximum for the geodesic distance function because of 
\begin{equation}
    \frac{\partial^2  D\(t_{\mt{E}}, \theta; t_{\mt{Eb}}, \theta_b\) }{\partial t_{\mathrm{Eb}}^2} \bigg|_{x_b = x_b^\ast} <0, \quad \frac{\partial^2  D\(t_{\mt{E}}, \theta; t_{\mt{Eb}}, \theta_b\)}{\partial \theta_{b}^2} \bigg|_{x_b = x_b^\ast} <0 \,. 
\end{equation}
This indicates that the saddle point approximation using the antipodal saddle point at $x_b^\ast$ cannot be applied to the integral for calculating the one-point function of the twistor operator. In other words, the island formula cannot be applied using the quantum extremal surface anchored on the dS gravitational space. This nullity of the saddle point approximation highlights the issues associated with the island formula and the disconnected extremal surface in dS gravity.

\begin{figure}[t]
\centering
  \includegraphics[width=3.5in]{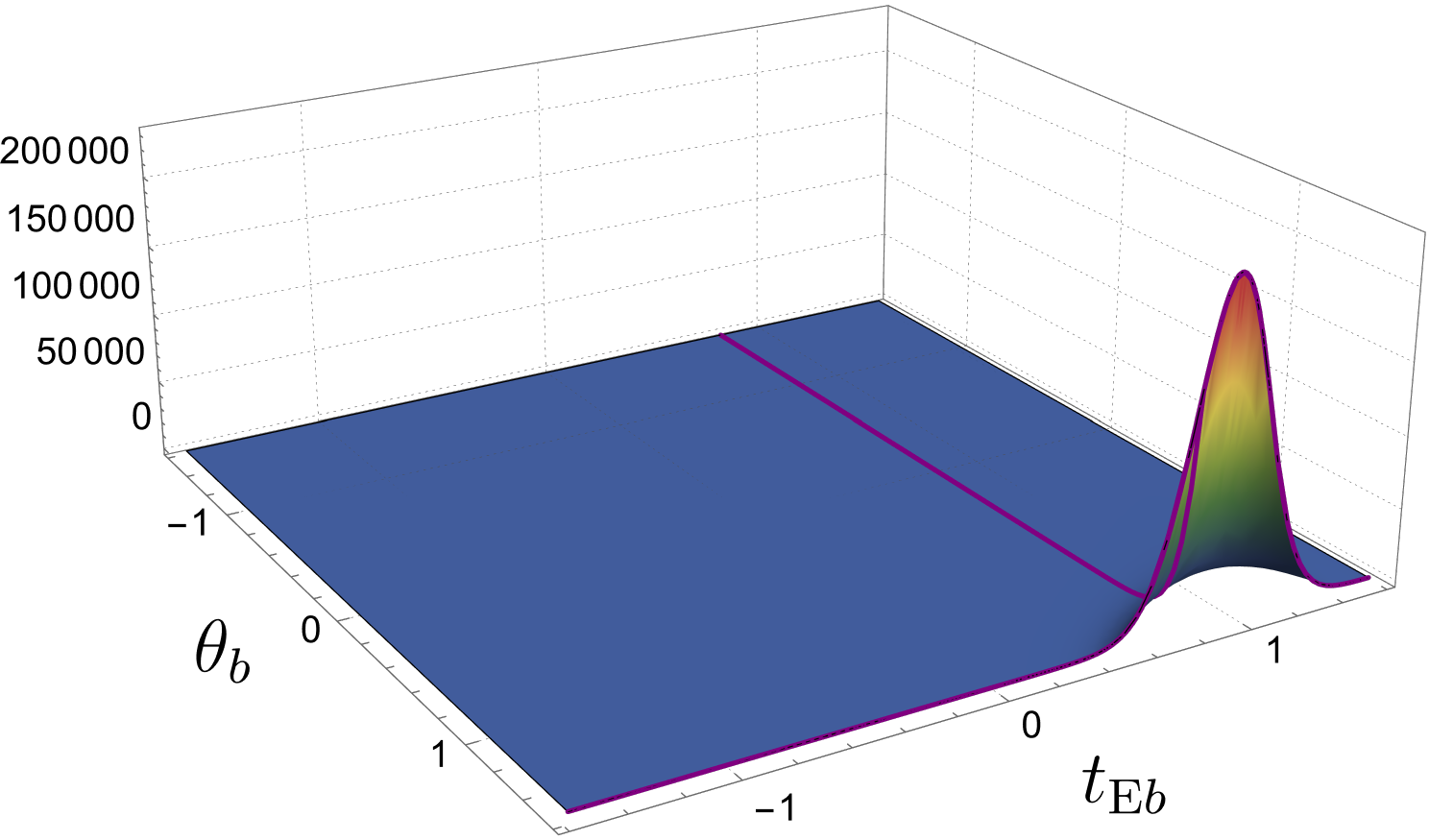}\\
  \medskip
    \includegraphics[width=2.9in]{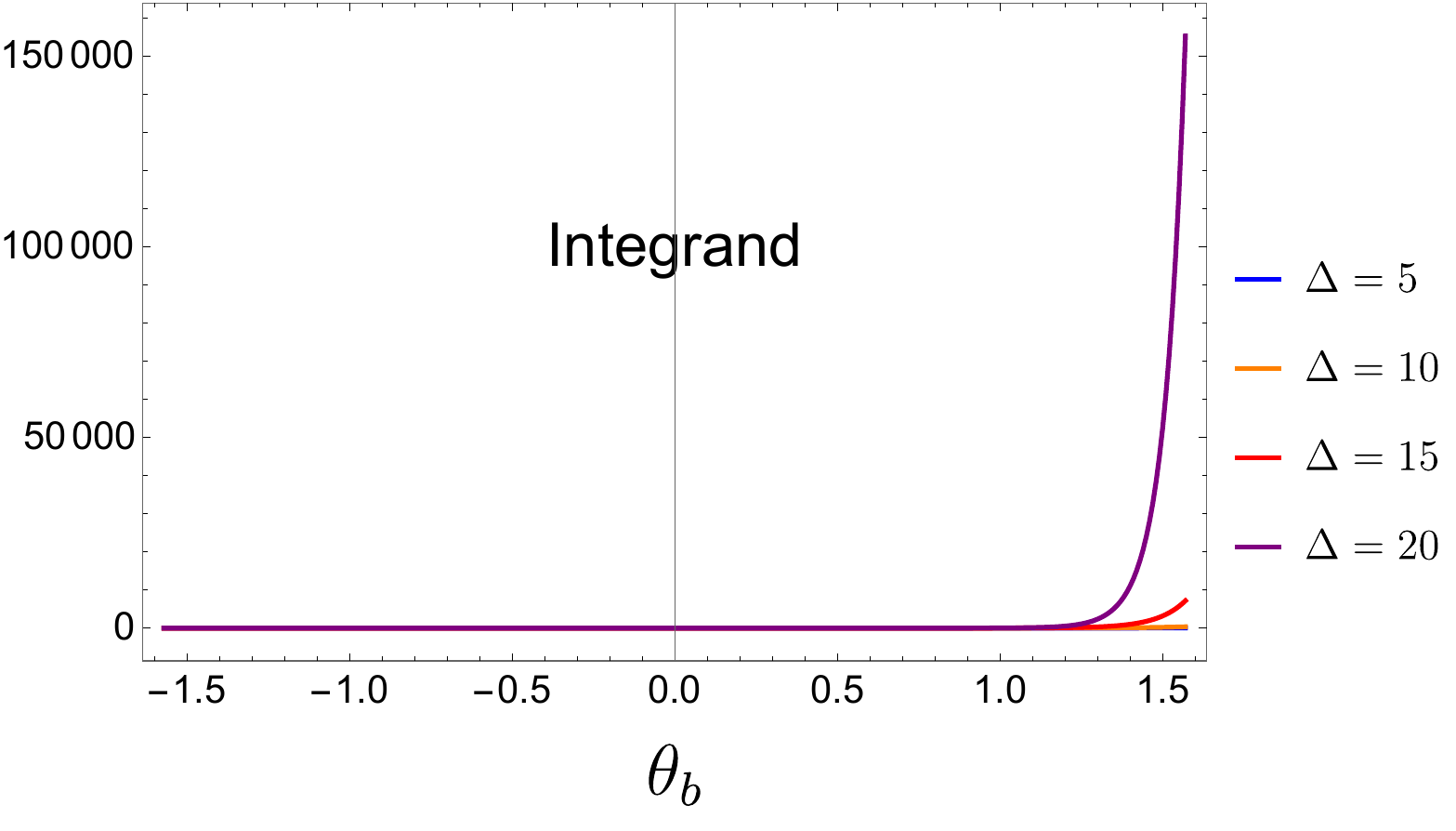}
    \includegraphics[width=2.9in]{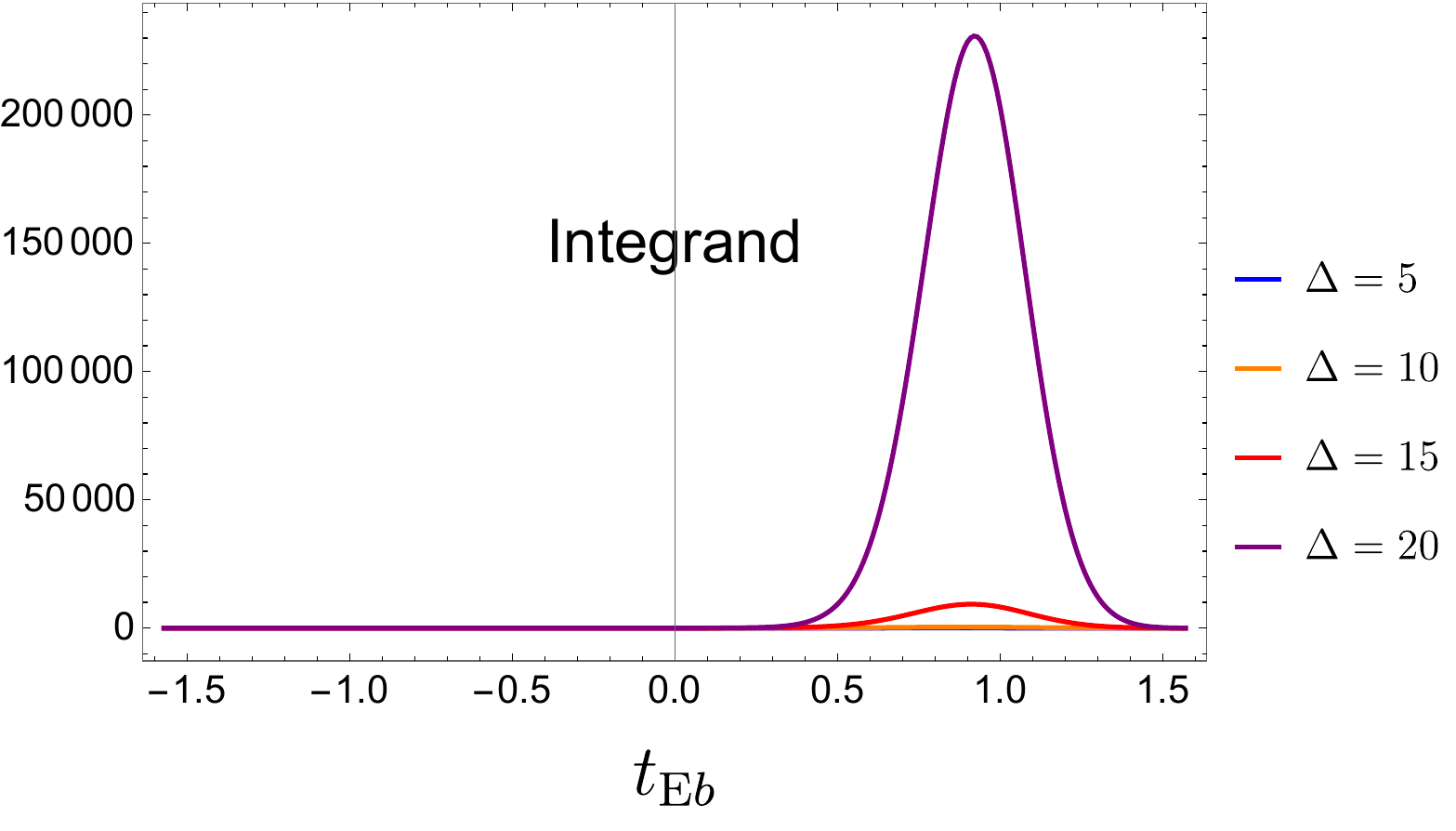}
     \caption{The integrand $\sqrt{g(x_b)} K_{\mt{Bb}}^\Delta (x ; x_b) $ for evaluating the one-point function $ \ev{\mO (t_{\mt{E}}, \theta)}_{\mt{dS}}$ defined in eq.~\eqref{eq:dSintegral} is a function of the endpoint $(t_{\mt{Eb}}, \theta_b)$ on the EOW brane. We take the boundary point at $\theta=\frac{3\pi}{4}, t_{\mt{E}}= \frac{\pi}{4}$ with choosing $\eta_b = 1$. For two bottom plots with various values of $\Delta$, one more parameter is chosen as $t_{\mt{Eb}}= \frac{\pi}{4}$, $\theta_b = \frac{\pi}{2}$, respectively.}\label{fig:3Dplot}
\end{figure}

\textbf{Dominant contribution from the edge:} 
Instead of using the geodesic and saddle point approximations, we can derive an analytic expression for the $\theta_b$ integral in terms of Appell hypergeometric functions. However, a key observation is that the integral along the dS EOW brane is always dominated by contributions from the edge, \ie $\theta_b = \frac{\pi}{2}$ for $\theta \in \left( \frac{\pi}{2}, \pi \right)$ and $\theta_b = -\frac{\pi}{2}$ for $\theta \in \left( -\pi, -\frac{\pi}{2} \right)$. This is illustrated in figure \ref{fig:3Dplot}, where the integrand $\sqrt{g(x_b)} K_{\mt{Bb}}^\Delta (x ; x_b)$ is plotted as a function of the brane point. Moreover, it is important to note that this edge contribution becomes increasingly dominant as the operator dimension $\Delta$ grows. Consequently, in the limit $\Delta \to \infty$, the spatial integral over $\theta_b$ can be approximated by its value at the edge of the EOW brane. This can be expressed as:
\begin{equation}\label{eq:tEbintegral}
\begin{split}
    \lim_{\Delta \to \infty}   \ev{ \mO (t_{\mt{E}}, \theta)}_{\mt{dS}} &= \lim_{\Delta \to \infty}  \( \int_{-\frac{\pi}{2}}^{\frac{\pi}{2}} dt_{\mt{Eb}}  \int_{-\frac{\pi}{2}}^{\frac{\pi}{2}} d\theta_b  \sqrt{g(t_{\mt{Eb}}, \theta_b)} K_{\mt{Bb}}^\Delta (t_{\mt{E}}, \theta ; t_{\mt{Eb}}, \theta_b)  \)  \\
   & \approx \lim_{\Delta \to \infty}  \int_{-\frac{\pi}{2}}^{\frac{\pi}{2}} dt_{\mt{Eb}} \( \sqrt{g(t_{\mt{Eb}}, \theta_b)} K_{\mt{Bb}}^\Delta (t_{\mt{E}}, \theta ; t_{\mt{Eb}}, \theta_b)  \) \bigg|_{ |\theta_b| =\frac{\pi}{2}} \\ 
   &=  \lim_{\Delta \to \infty}  \int_{-\frac{\pi}{2}}^{\frac{\pi}{2}} dt_{\mt{Eb}} \, 
    \frac{\cos{t_{\mt{Eb}}}}{\left(\cosh{\eta_b}-\sinh{\eta_b}(\cos{t_{\mt{Eb}}}\cos{t_{\mt{E}}}|\sin\theta|+\sin{t_{\mt{E}b}}\sin{t_{\mt{E}}})\right)^\Delta} \,,
\end{split}
\end{equation}
where the sign depends on the range of $\theta$. By focusing on the contributions from the edge, we simplify the calculation and capture the dominant behavior of the integral in the large $\Delta$ limit. We have also conducted a more detailed examination of the errors resulting from the utilization of this non-saddle point approximation. Our findings show that the sub-leading terms for computing the entanglement entropy are negligible in comparison to the leading term given by the edge \footnote{We have checked that the error between the precise integral $\ev{ \mO (t_{\mt{E}}, \theta)}_{\mt{dS}}$ and the edge contribution always vanishes in the limit $\Delta \to \infty$, \ie 
    \begin{equation*}
    \lim_{\Delta \to \infty} \left(  \frac{ \ev{ \mO (t_{\mt{E}}, \theta)}_{\mt{dS}} - \langle O\rangle_{|\theta_b |=\frac{\pi}{2}}}{\ev{ \mO (t_{\mt{E}}, \theta)}_{\mt{dS}} }   \right) = 0  \,,
    \end{equation*}
    by performing numerical integrals.}. Thus, we achieve a reliable and precise evaluation of the one-point function with $\Delta \to \infty$, despite not relying on the traditional saddle point approximation. This edge-dominant method is particularly useful in the context of dS gravity, where the standard saddle point approach encounters significant challenges.

Once the approximation for the spatial integral has been taken, it becomes more straightforward to track the integral along the Euclidean time direction on the EOW brane, given that the integrand behaves in a manner akin to a Gaussian function, as shown in figure \ref{fig:3Dplot}. In other words, we can then apply the saddle point approximation to the $t_{\mt{Eb}}$-integral. The time derivative of the integrand in eq.~\eqref{eq:tEbintegral} is derived as
\begin{equation}
\frac{-\Delta \sinh \eta_b \( \sin t_{\mt{E}}\cos t_{\mt{Eb}} - \sin t_{\mt{Eb}}  \cos  t_{\mt{E}} |\sin \theta |  \)}{ \left(\cosh{\eta_b}-\sinh{\eta_b}(\cos{t_{\mt{Eb}}}\cos{t_{\mt{E}}}|\sin\theta|+\sin{t_{\mt{E}b}}\sin{t_{\mt{E}}})\right)  } + \tan t_{\mt{Eb}}  + \mathcal{O} \(\frac{1}{\Delta}\) \,,   
\end{equation}
where we have performed a series expansion with for large $\Delta$  and keep the first two leading orders. At leading order, one can get one saddle point located at
\begin{equation}
\tan t_{\mt{Eb}}^\ast = \frac{\tan t_{\mt{E}}}{|\sin\theta|} \, \,, 
\end{equation}
which corresponds the peak appearing in the $t_{\mt{Eb}}$-integral. One may worry that the volume measure of the EOW brane may influence this saddle point approximation due to the fact that Euclidean brane is a compact sphere. It is indeed true that there are two extra saddle points at 
\begin{equation}
  t_{\mt{Eb}} \approx \frac{\pi}{2}  + \frac{1}{\Delta} \frac{\cosh \eta_b -\sin t_{\mt{Eb}}}{|\sin \theta | \cos t_{\mt{Eb}}} +\mathcal{O} \(\frac{1}{\Delta^2}\) \quad \text{and} \quad  t_{\mt{Eb}} \approx  - \frac{\pi}{2}  + \frac{1}{\Delta} \frac{\cosh \eta_b + \sin t_{\mt{Eb}}}{|\sin \theta | \cos t_{\mt{Eb}}} +\mathcal{O} \(\frac{1}{\Delta^2}\) \,, 
\end{equation}
which are close to the edge of the time direction. However, these saddle points can only appear and dominate the integral in the regime with $\cos t_{\mt{E}} < 0$ which is not included by the Hartle-Hawking state with a compact Euclidean time $t_{\mt{E}} \in (-\frac{\pi}{2}, \frac{\pi}{2})$. 

\begin{figure}[t]
	\centering
	\includegraphics[width=2.7in]{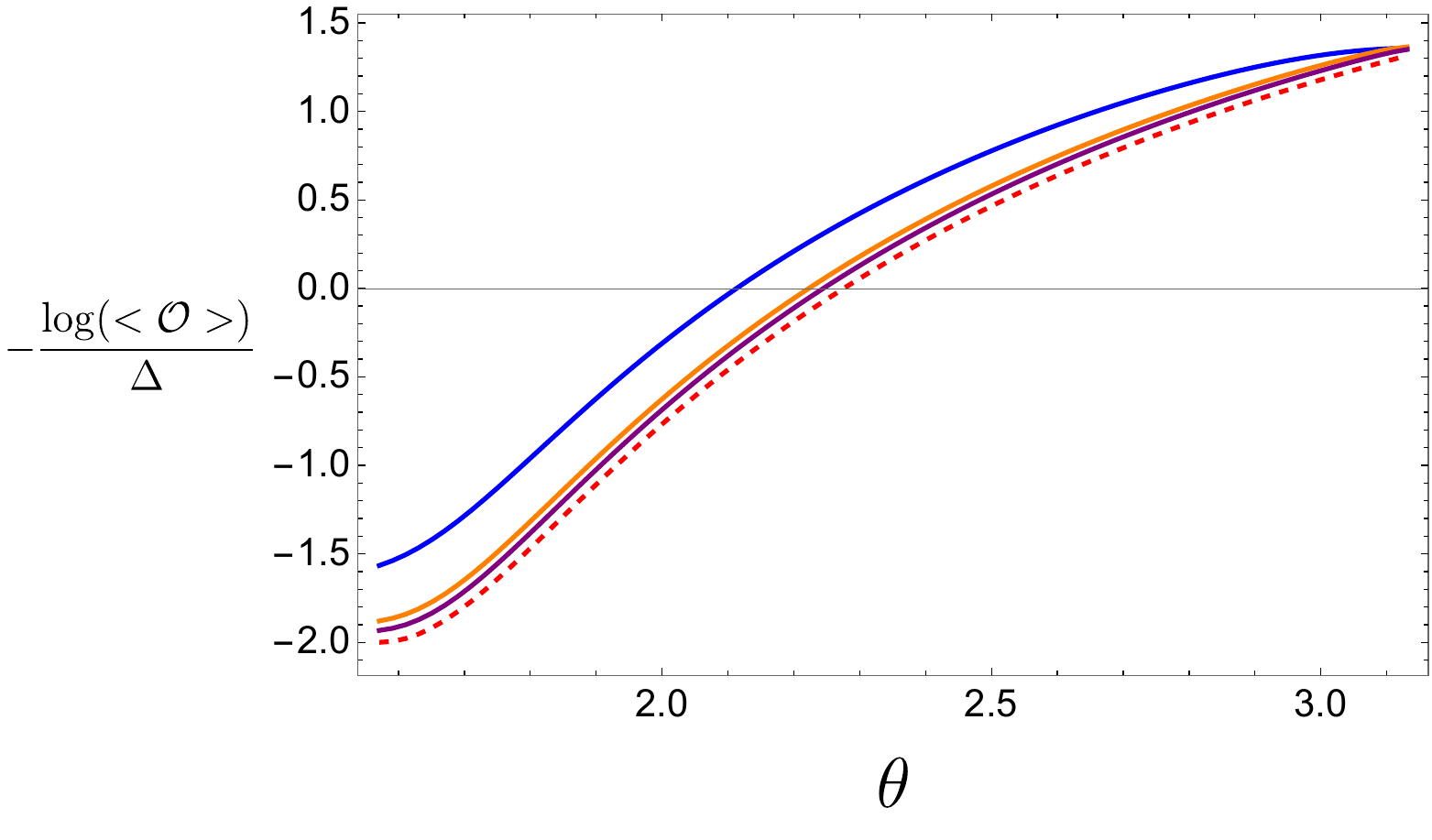}
	\includegraphics[width=3.3in]{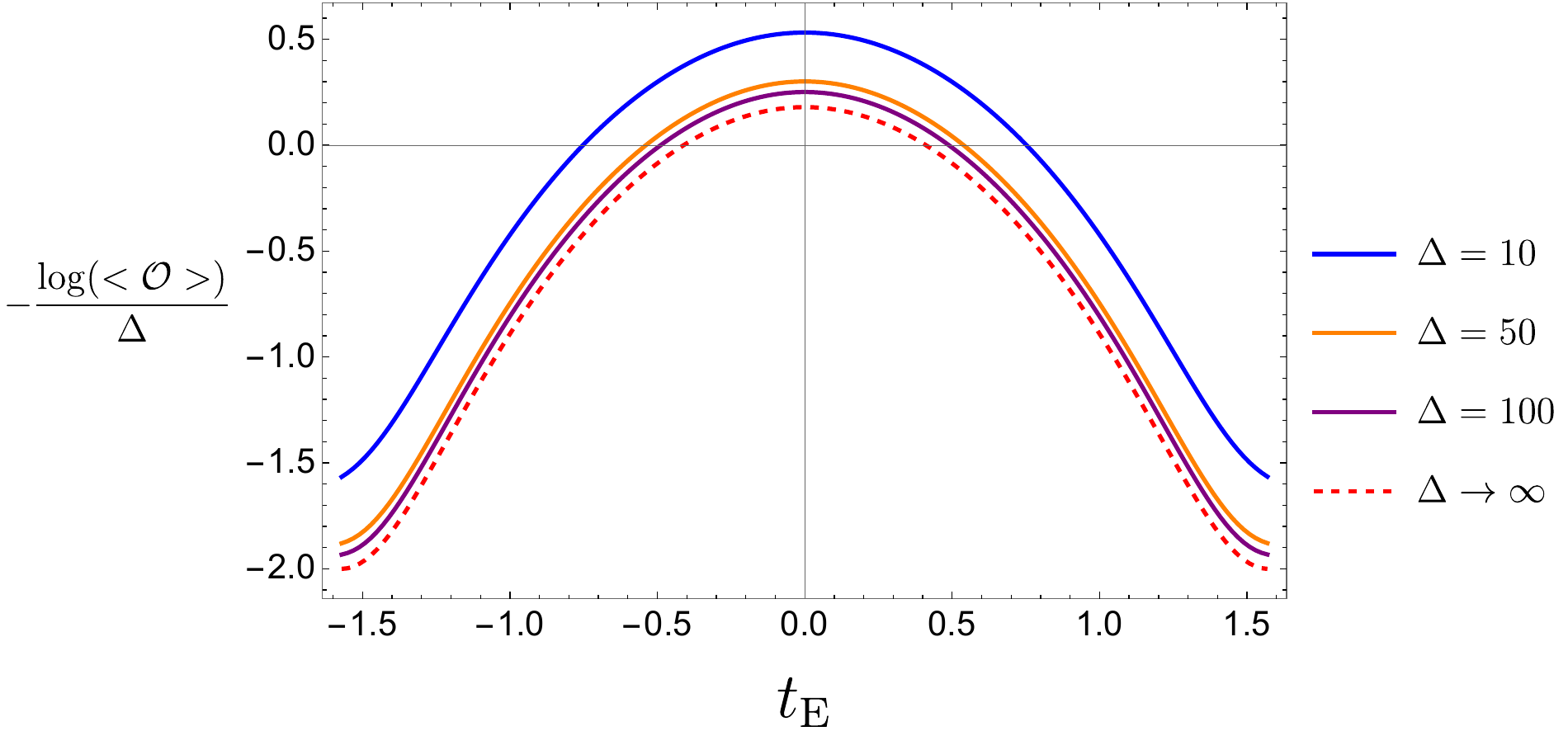}
	\caption{Numerical integrals of the one-point function $\frac{- \log \ev{ \mO (t_{\mt{E}}, \theta)}_{\mt{dS}}}{\Delta}$ with different values of the dimension $\Delta$. As $\Delta$ increases, the one-point function approaches the analytical expression derived in eq.~\eqref{eq:onepointapp} for $\Delta \to \infty$. For these two numerical plots, we choose $\eta_b =2$ and set $t_{\mt{E}}=0$ for the left plot and $\theta=\frac{3\pi}{4}$ for the right plot.}\label{fig:onepoint}
\end{figure} 

As a summary, we conclude that the $t_{\mt{Eb}}$-integral for the one-point function \eqref{eq:tEbintegral} at the large dimension limit is dominated by the saddle point located at $t_{\mt{Eb}}= t_{\mt{Eb}}^\ast$. We are interested in computing the entanglement entropy by using the one-point function of the twistor operator as defined in eq.~\eqref{EErep}. More explicitly, we can get 
\begin{equation}\label{eq:onepointapp}
\begin{split}
 \lim_{\Delta  \to \infty} \(  \frac{- \log \ev{ \mO (t_{\mt{E}}, \theta)}_{\mt{dS}}   }{\Delta} \)   &\approx \log \left(\cosh{\eta_b}-\sinh{\eta_b}(\cos{t^\ast_{\mt{Eb}}}\cos{t_{\mt{E}}}|\sin\theta|+\sin{t^\ast_{\mt{E}b}}\sin{t_{\mt{E}}})\right)  \\
 &= \log \( \cosh \eta_b - \sinh{\eta_b} \sqrt{ \cos^2 t_{\mt{E}} \sin^2\theta  + \sin^2 t_{\mt{E}}  }      \) \,, 
\end{split}
\end{equation}
where the sub-leading terms all vanish in the limit $\Delta \to \infty$.
By comparing with the numerical integrals, we can find that the one-point function approaches the analytical expression \eqref{eq:onepointapp} by taking large dimension limit, as shown in figure \ref{fig:onepoint}.

\begin{figure}[t]
\centering
    \includegraphics[width=2.9in]{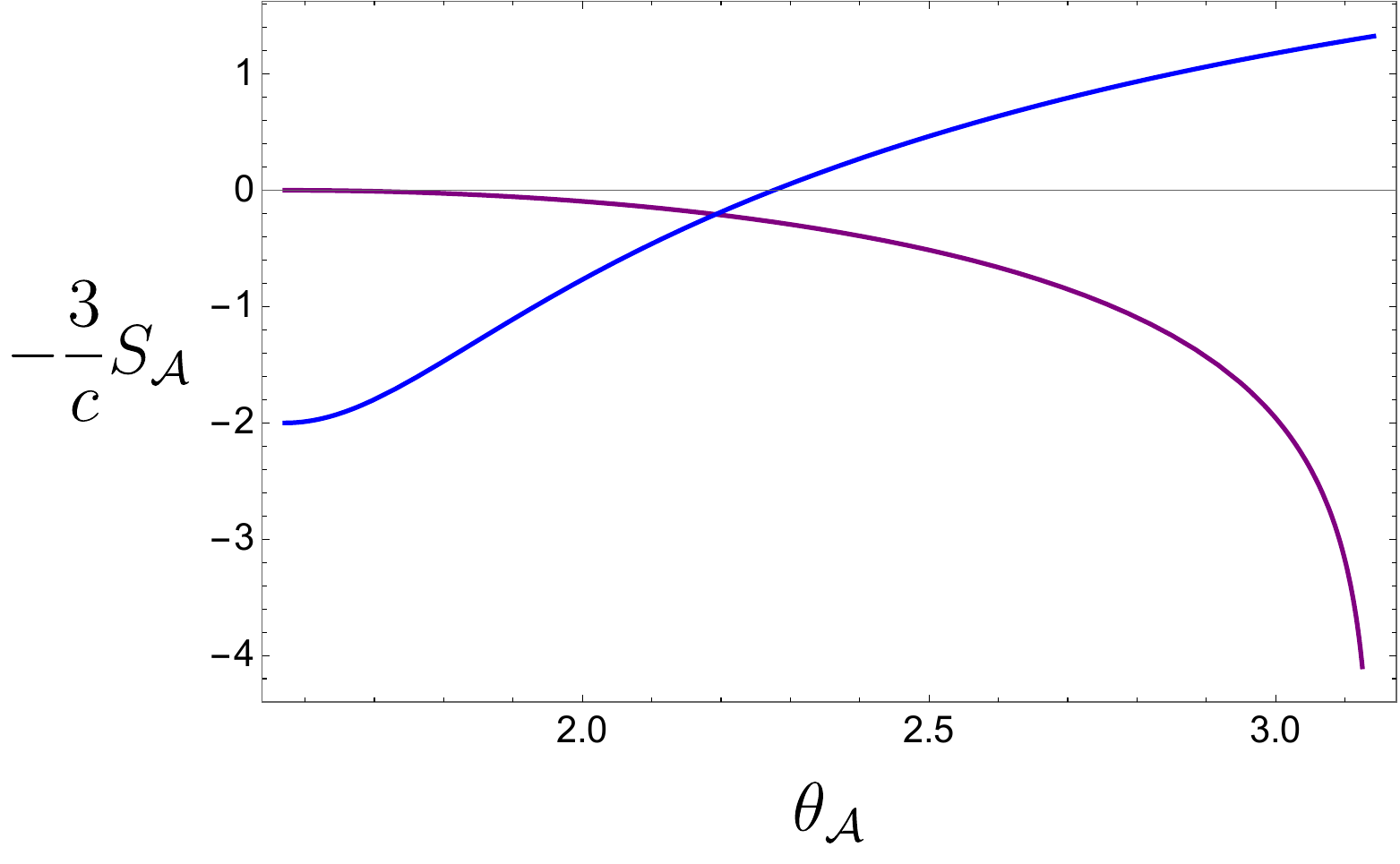}
    \quad 
    \includegraphics[width=2.95in]{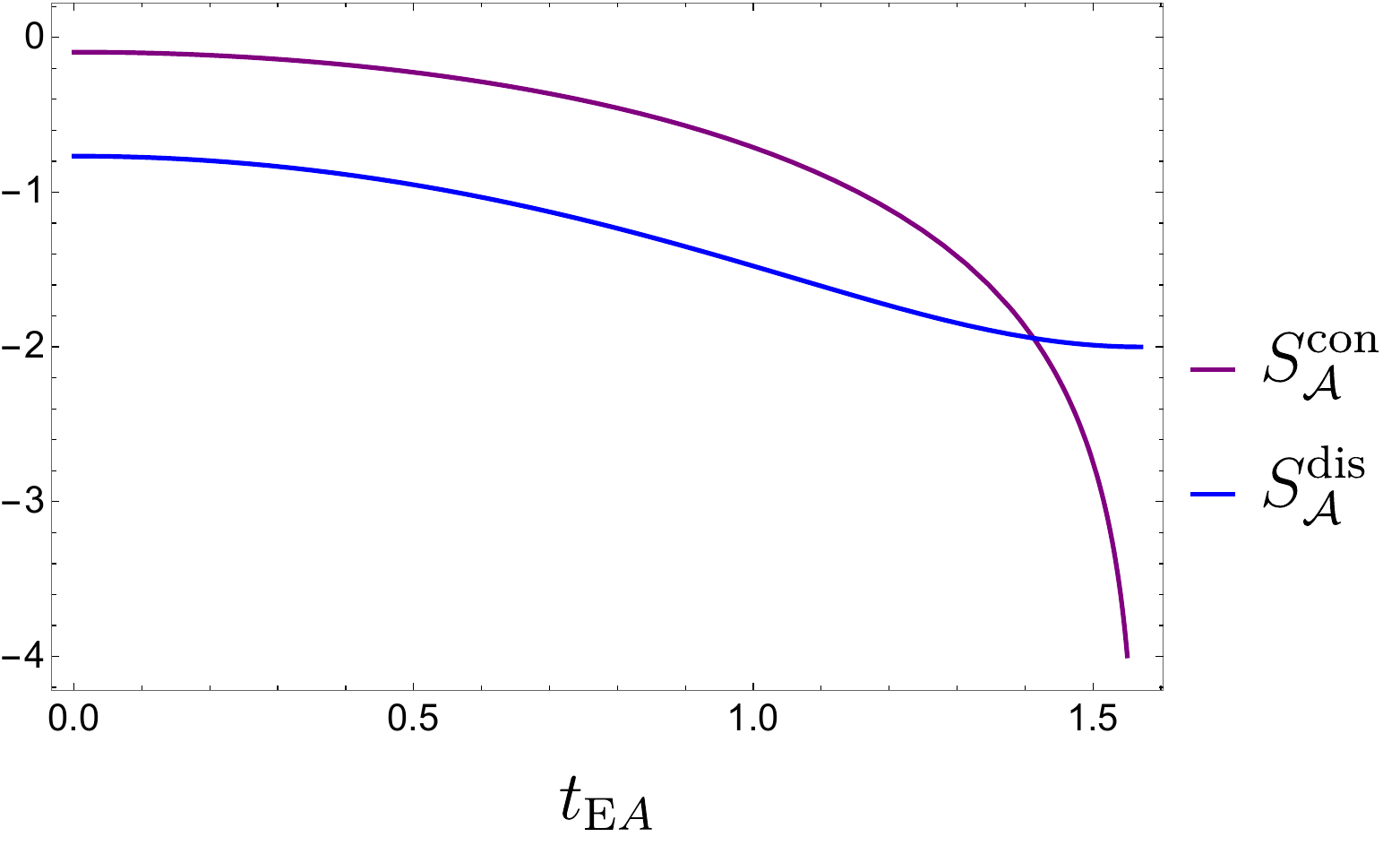}
     \caption{The entanglement entropy $S_{\mA}$ derived in eq.~\eqref{eq:SAtwophases} is given by the minimum value between the connected phase $S_{\mathcal{A}}^{\mathrm{con}}$ and the disconnected phase $\tilde{S}_{\mathcal{A}}^{\mathrm{dis}}$. In both plots we subtract the UV divergence $\frac{c}{3}\log{\frac{2}{\ep}}$ from the entanglement entropy and take $\eta_b=2$. For the left/right plot we set $t_{\mt{E}A}=0$ and $\theta_A = 2$, respectively.}\label{fig:phasetransiiton}
\end{figure}

Equipped with the result of the one-point function $\mO \left(t_{\mt{E}}, \theta\right)$ in the large dimension limit, we can finally apply the replica trick \eqref{EErep} to calculate the entanglement entropy of a boundary interval. Taking the symmetric interval $\mA$ \eqref{eq:intervalA} on the Euclidean boundary, \ie  
\begin{equation}\label{eq:EintervalA}
\mA : \qquad  \{ t_{\mt{E}} = t_{\mt{E}A} \,, \quad \theta \in [-\pi , -\theta_A ] \cup [\theta_A, \pi]  \} \,, 
\end{equation}
the entanglement entropy $S_{\mA}$ derived from the replica trick and double holography is given by the following minimization 
\begin{equation}\label{eq:SAtwophases}
S_{\mA} =    \min 
  \begin{cases}  
  S_{\mA}^{\rm con} = \frac{c}{3} \log \( \frac{ 2\cos t_{\mt{E}A} \cdot \sin\theta_A }{\epsilon}  \) \\
  \,\\
\tilde{S}_{\mA}^{\rm dis} =\frac{c}{3}\log \( \frac{2}{\epsilon} \(   \cosh \eta_b - \sinh{\eta_b} \sqrt{ \cos^2 t_{\mt{E}A} \sin^2\theta_A  + \sin^2 t_{\mt{E}A}  } \)     \)  \\
  \end{cases} \,,
\end{equation}
where we have restored the universal divergent part for the disconnected phase $S_{\mA}^{\rm dis}$. When the subsystem $\mA$ is large (\ie $\theta_A$ is close to $\frac{\pi}{2}$), the disconnected phase is favored in general. At late times $t_{\mt{E}A} \sim \frac{\pi}{2}$, the connected phase with $S_{\mA}^{\rm con}$ would be always dominant since it approaches negative infinity. These phase transition are illustrated by the example shown in figure \ref{fig:phasetransiiton}.

\subsection{Holographic dual and non-extremal island on the dS brane}

In the preceding subsection, we computed the entanglement entropy $ S_{\mathcal{A}}$ utilizing the holographic correlation functions of the twistor operators. Although we have previously demonstrated that neither the standard holographic entanglement entropy formula, which involves extremal surfaces, nor the conventional island formula is applicable to a subsystem coupled with dS gravitational spacetime, our results obtained from the holographic correlation functions suggest that a novel holographic formula for evaluating entanglement entropy may be identified.

Despite the limitations of the saddle point approximation, we continue to derive the holographic entanglement entropy at leading order with $\frac{1}{\GN} \sim c \to \infty $. Consequently, both the one-point function and two-point function of the twistor operator at large dimensions are consistently described by the geodesic approximation, as discussed around eq.~\eqref{eq:one_pt}. For instance, the one-point function $\ev{\sigma_n}_{\mt{dS}}$ for $\Delta_n \to \infty $ can be expressed as
\begin{equation}\label{eq:one_pt_dS02}
 \lim_{\Delta_n  \to \infty} \ev{ \sigma_n (t_{\mt{E}}, \theta)}_{\mt{dS}}  \approx \int_{-\frac{\pi}{2}}^{\frac{\pi}{2}} dt_{\mt{Eb}} \int_{-\frac{\pi}{2}}^{\frac{\pi}{2}} d\theta_b \; \cos{(t_{\mathrm{Eb}})}\; e^{-\Delta_n D(t_{\mt{E}}, \theta; t_{\mt{Eb}}, \theta_b)} \,, 
\end{equation}
where the geodesic distance $D(t_{\mt{E}}, \theta; t_{\mt{Eb}}, \theta_b)$ from a boundary point to the Euclidean EOW brane is given by 
\begin{equation}
D(t_{\mt{E}}, \theta; t_{\mt{Eb}}, \theta_b) = \eta_\infty + \log{\qty(\cosh
\eta_b-\sinh{\eta_b}\qty(\cos{\rmtE}\cos{\rmtE_b}\cos(\theta - \theta_b )+\sin{\rmtE}\sin{\rmtE_b}))} \,. 
\end{equation}
By applying the insights gained from evaluating the two integrals associated with the one-point function, we find that the dominant contribution is equivalent to the length of the geodesic connecting the boundary point to the brane point at the edge with $ |\theta_b | =\frac{\pi}{2}$, \ie 
\begin{equation}
\tilde{S}_{\mA}^{\rm dis} = \frac{L_{\mt{AdS}} }{4\GN}  \(   D(t_{\mt{E}A}, \theta_{A}; t_{\mt{Eb}}, \theta_b) \bigg|_{\theta_b = \frac{\pi}{2}, t_{\mt{E}b}=  t^\ast_{\mt{E}b}} + D(t_{\mt{E}A}, -\theta_{A}; t_{\mt{Eb}}, \theta_b)  \bigg|_{\theta_b =-\frac{\pi}{2}, t_{\mt{E}b}=  t^\ast_{\mt{E}b}} \)\,, 
\end{equation}
where the time slice $t_{\mt{Eb}}^\ast$ on the brane is determined by
\begin{equation}\label{eq:tast}
\tan t_{\mt{Eb}}^\ast = \frac{\tan t_{\mt{E}A}}{|\sin\theta_A|} \,.
\end{equation}
It is important to note that this type of disconnected geodesic, denoted as $\tilde{\Gamma}^{\rm dis}_{\mA}$, differs from the extremal geodesic ${\Gamma}^{\rm dis}_{\mA}$ anchored on the EOW brane. The disconnected geodesic is non-extremal with respect to the spatial direction but is locally minimal with respect to variations along the Euclidean timelike direction on the EOW brane. It is straightforward to verify that its second time derivative is always positive:
\begin{equation}
    \frac{\partial^2 D(t_{\mt{E}}, \theta; t_{\mt{Eb}}, \theta_b)}{\partial t_{\mt{E}b}^2} \bigg|_{\theta_b = \frac{\pi}{2}, t_{\mt{E}b} = t^\ast_{\mt{E}b} } = \frac{1}{\frac{\coth{\eta_b}}{\sqrt{\sin^2{t_{\mt{E}}}+\sin^2{\theta}\cos^2{t_{\mt{E}}}}}-1} >  \frac{1}{\coth{\eta_b}-1}>0 \,. 
\end{equation}

\begin{figure}[t]
    \centering
    \begin{minipage}[h]{0.495\linewidth}
       \centering
    \begin{tikzpicture}
        \draw[ForestGreen, line width=5pt] (0,3) arc(90:270:3);
        \draw[black, line width=1pt,dashed] (0,-3) arc(-90:90:3);
        \draw[black, line width=1pt,dotted] (3.2,0) arc(0:360:3.2);
        \draw[red, line width=5pt] (0,-1.5) arc(-90:90:1.5);
        \draw[orange, line width=5pt](0,1.5)--(0,3);
        \draw[orange, line width=5pt](0,-1.5)--(0,-3);
        \fill[lightgray!20!white](0,1.5)--(0,3) to[out=180,in= 90](-3,0)to[out=-90,in= -180](0,-3)--(0,-1.5)to[out=0,in=-90](1.5,0)to[out=90,in=0] (0,1.5);
\fill[blue!20!white,opacity=0.5](-2.59808,1.5)to[out=-30,in= 30](-2.59808,-1.5)--cycle;
\fill[blue!20!white,opacity=0.5](-2.59808,1.5) arc(150:210:3)--cycle;
    \draw[blue, line width=3pt] (-2.59808,1.5) arc(150:210:3);
         \draw[blue, line width=2pt,double] (-2.59808,1.5)to[out=-30,in= 30](-2.59808,-1.5);
         \draw[blue](-1.5,0)node[right]{\Large{$\Gamma_{\mathcal{A}}^{\mathrm{con}}$}};
          \draw[blue](-3.2,0)node[left]{\Large{$\mathcal{A}$}};
          \draw[blue](-2.5,0)node{{EW}};
          \draw[black,line width=5pt](0,3.2)node[above]{\Large{Small $\mathcal{A}$}};
    \end{tikzpicture}
    \end{minipage}
    \begin{minipage}[h]{0.495\linewidth}
        \centering
    \begin{tikzpicture}
      \draw[ForestGreen, line width=5pt] (0,3) arc(90:270:3);
        \draw[black, line width=1pt,dashed] (0,-3) arc(-90:90:3);
        \draw[black, line width=1pt,dotted] (3.2,0) arc(0:360:3.2);
        \draw[cyan, line width=6pt] (0,-1.5) arc(-90:90:1.5);
        \draw[orange, line width=5pt](0,1.5)--(0,3);
        \draw[orange, line width=5pt](0,-1.5)--(0,-3);
        \fill[lightgray!20!white](0,1.5)--(0,3) to[out=180,in= 90](-3,0)to[out=-90,in= -180](0,-3)--(0,-1.5)to[out=0,in=-90](1.5,0)to[out=90,in=0] (0,1.5);
        \draw[blue, line width=3pt] (-1.5,2.59808) arc(120:240:3);
          \draw[blue, line width=2pt,double] (-1.5,-2.59808)to[out=60,in= 180](0,-1.5);
          \draw[blue, line width=2pt,double] (-1.5,2.59808)to[out=-60,in= -180](0,+1.5);
         \draw[blue](-1.3,1.5)node[below]{\Large{$\tilde{\Gamma}_{\mathcal{A}}^{\mathrm{dis}}$}};
          \draw[blue](-1.3,-1.5)node[above]{\Large{$\tilde{\Gamma}_{\mathcal{A}}^{\mathrm{dis}}$}};
          \draw[blue](-3.2,0)node[left]{\Large{$\mathcal{A}$}};
           \draw[cyan](0.2,1.8)node[right]{Non-extremal Island};
           \draw[blue](0,0)node[left]{EW};
          \fill[blue!20!white,opacity=0.4](-1.5,2.59808) arc(120:240:3)--cycle;
           \fill[blue!20!white,opacity=.4](-1.49,2.59808)to[out=-60,in= 180](0,1.5)to(0,-1.5)
           to[out=180,in= 60](-1.49,-2.59808)--cycle;
           \fill[blue!20!white,opacity=.45](0,-1.5) arc(-90:90:1.5);
          \draw[black,line width=5pt](0,3.2)node[above]{\Large{Large $\mathcal{A}$}};      
    \end{tikzpicture}
    \end{minipage}
    \caption{Left: The non-island phase for a small interval $\mathcal{A}$. Right: The island phase for a large interval $\mA$ coupled to the dS gravity. Note that the {\it non-extremal island}, indicated by the blue curve, includes the entire dS gravity region.}
        \label{fig:dSphase}
\end{figure}
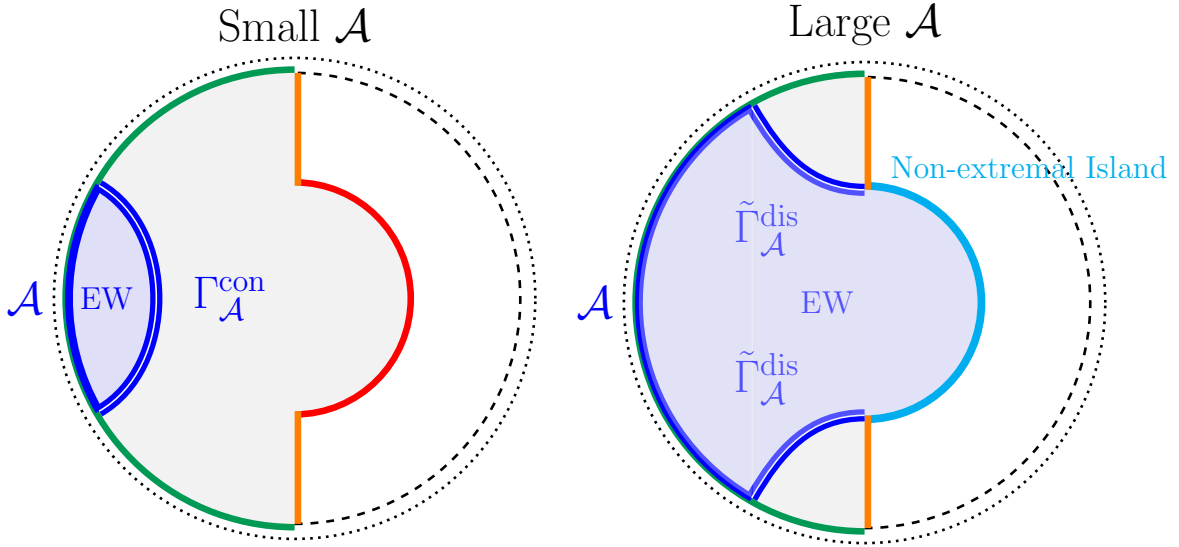

In summary, as illustrated in figure \ref{fig:dSphase}, the key result from the calculation of the one-point function$\ev{ \sigma_n (t_{\mt{E}}, \theta)}_{\mt{dS}}$ is the derivation of a new holographic entanglement formula for an interval connected to the dS brane. Specifically, we have
\begin{equation}\label{eq:HEE}
S_{\mathcal{A}} = \min \qty{S_{\mathcal{A}} ^{\mathrm{con}}, \tilde{S}_{\mathcal{A}}^{\mathrm{dis}}} = \min \left\{ \frac{\text{Area}(\Gamma_{\mathcal{A}}^{\rm{con}})}{4\GN}, \frac{\text{Area}(\tilde{\Gamma}_{\mathcal{A}}^{\rm{dis}})}{4\GN}\right\} \,, 
\end{equation}
where the disconnected surface $\tilde{\Gamma}_{\mathcal{A}}^{\rm{dis}}$, which connects the endpoints of the boundary interval and the edge of the dS EOW brane, is extremal only with respect to the time direction on the dS brane.

Finally, we turn our attention to the island formula for a non-gravitational bath coupled to de Sitter gravity. It remains intriguing to interpret the disconnected phase as the island phase from the perspective of the brane. However, in this case, the boundary of the island is no longer determined by the quantum extremal surface. Our findings thus indicate the presence of a {\it non-extremal island} whose boundary is fixed at the edge of the dS gravitational region. In the connected phase, or non-island phase, where the size $\mathcal{A}$ is small, the entanglement wedge does not extend to the EOW brane. Consequently, an observer situated within $\mathcal{A}$ cannot probe the dS gravitational region. As the size of the bath interval$\mathcal{A}$ increases, a phase transition occurs from the non-island phase to the island phase. Thus, in the island phase, the entanglement wedge of the non-gravitational interval will always encompass the entire dS gravitational region. This marks a clear distinction between dS gravity and AdS gravity, where a portion of the AdS gravitational region can still be entangled with a non-gravitational bath.

\subsection{Lorentzian dS brane}

\begin{figure}[t]
    \centering
     \includegraphics[width=2.8in]{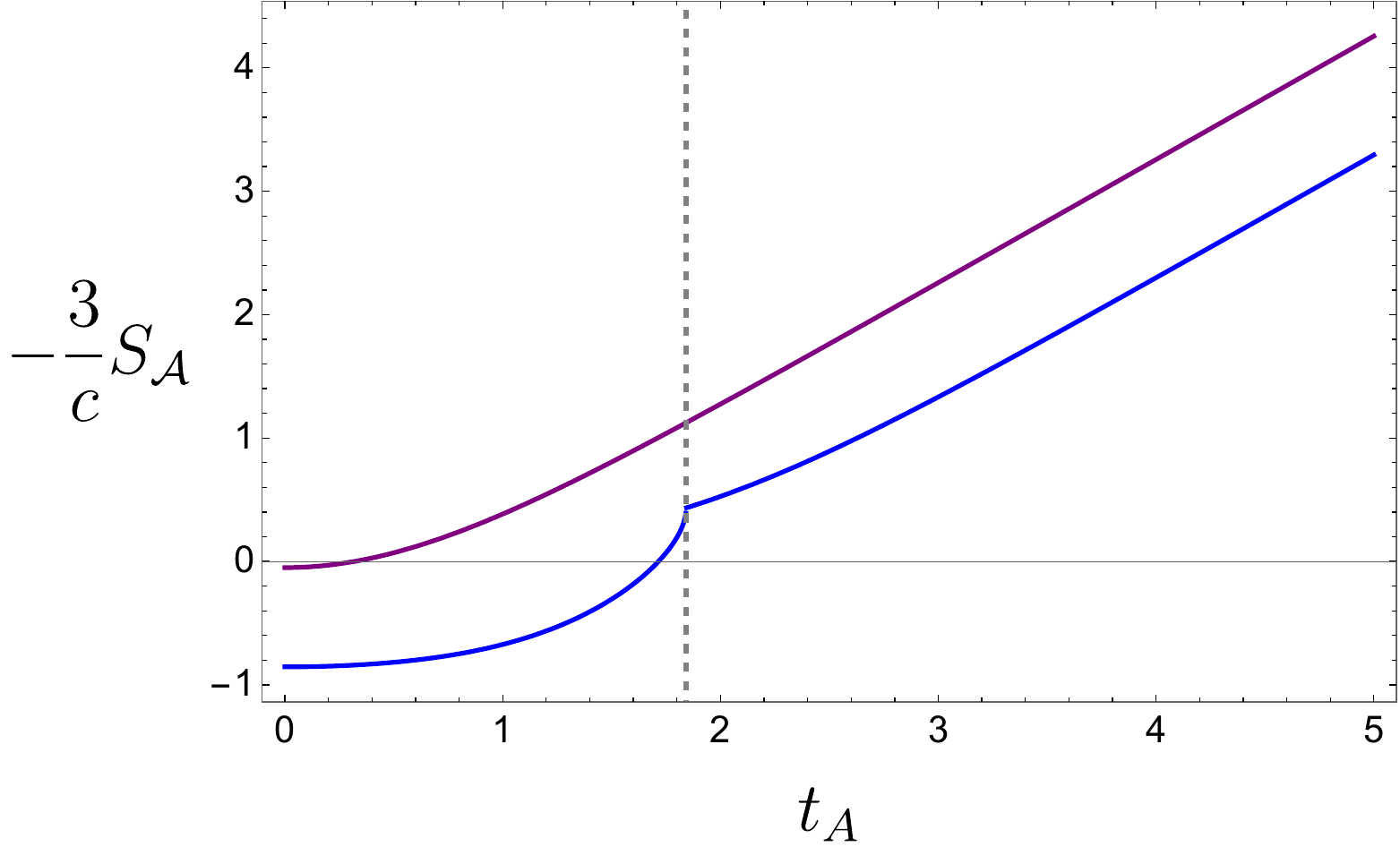}
     \qquad
      \includegraphics[width=2.9in]{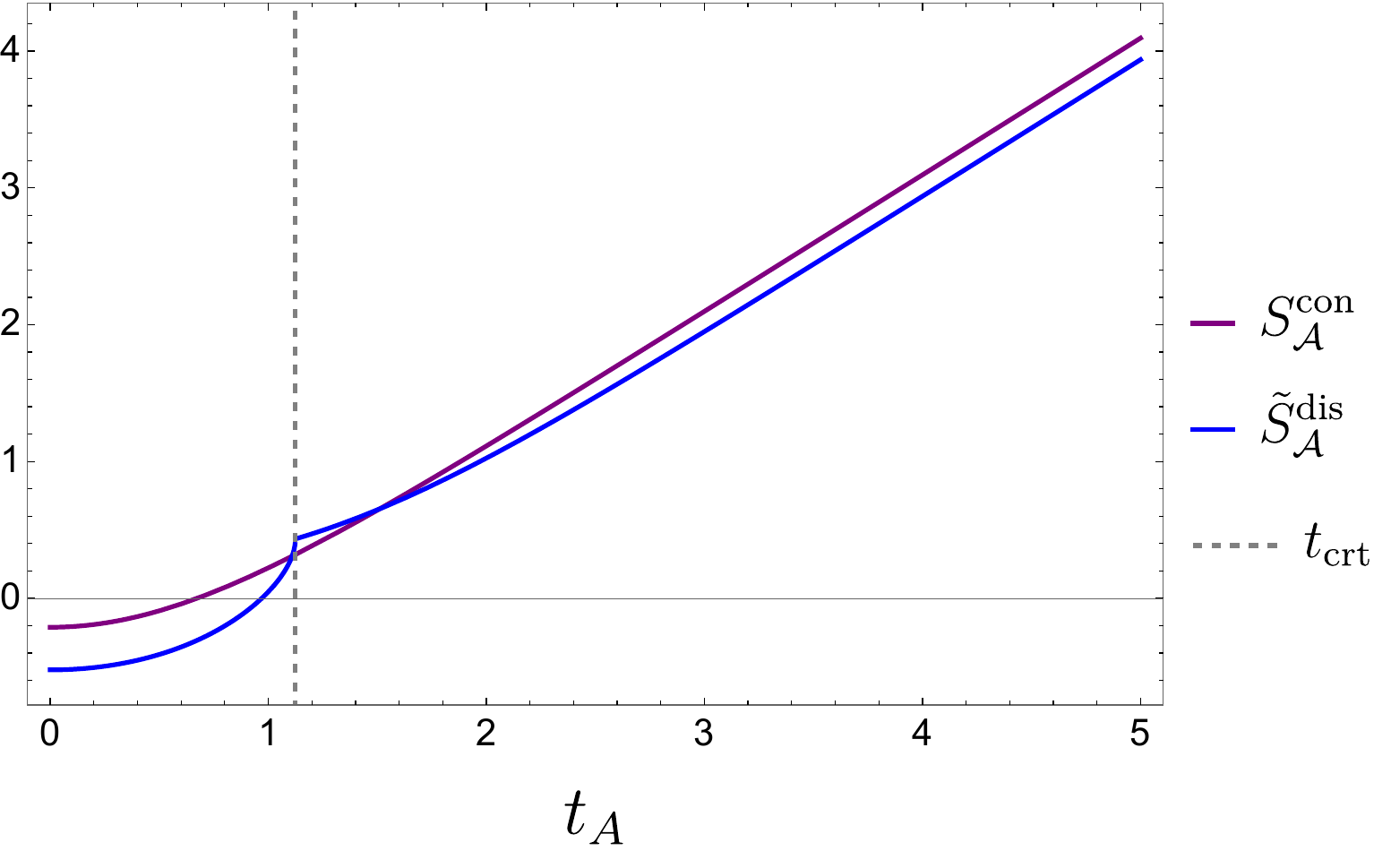}
    \caption{The Lorentzian time evolution of the connected and disconnected holographic entanglement entropy. We subtract the UV divergence $\frac{c}{3} \log{\frac{2}{\epsilon}}$. In the left and right panels, we set the subsystem to be $\mathcal{A} = [\theta = 0.6 \pi, -0.6 \pi]$ and $\mathcal{A} = [\theta = 0.7 \pi, -0.7 \pi]$, respectively, with $\eta_b = 2$. After the critical time $t_A=t_{\mathrm{cr}}$, denoted by the dashed line, we plot the real part of the disconnected holographic entanglement entropy.}    \label{fig:HEE_LORENTZIAN_THETA}
\end{figure}

The analysis in the previous section focused on the Euclidean holographic correlation function with a Euclidean EOW brane, from which we derived the holographic entanglement entropy $S_{\mA}$. Finally, we turn to the Lorentzian time evolution of holographic entanglement entropy. By taking the Euclidean result $S_{\mA}$ derived in eq.~\eqref{eq:SAtwophases} and performing the Wick rotation $t_{\mt{E}A} = i t_{A}$, we obtain  
\begin{equation}\label{eq:SAtwophases02}
S_{\mA} =    \min 
  \begin{cases}  
  S_{\mA}^{\rm con} = \frac{c}{3} \log \( \frac{ 2\cosh t_{A} \cdot \sin\theta_A }{\epsilon}  \) \\
  \,\\
\tilde{S}_{\mA}^{\rm dis} =\frac{c}{3}\log \( \frac{2}{\epsilon} \(   \cosh \eta_b - \sinh{\eta_b} \sqrt{ \cosh^2 t_{A} \sin^2\theta_A  -  \sinh^2 t_{A}  } \)     \)  \\
  \end{cases} \,. 
\end{equation}
However, it is obvious that this disconnected phase $\tilde{S}_{\mA}^{\rm dis}$ becomes complex valued after the critical time $t_{\mathrm{cr}}$ given by
\begin{equation}
    \tanh{t_\mathrm{cr}}= | \sin{\theta_A} |.
\end{equation}
To understand the origin of the critical time, we note that the saddle point on the time direction along the EOW brane \eqref{eq:tast}, \ie the endpoint of the geodesic $\tilde{\Gamma}_{\mA}^{\rm dis}$ reads
\begin{equation}
\tanh t_b^\ast = \frac{\tanh t_A}{|\sin\theta_A|}\,,
\end{equation}
after the Wick rotation. We found that the endpoints of the disconnected geodesics $\tilde{\Gamma}_{\mA}^{\rm dis}$ on the dS EOW branes reach future infinity $t_b \to \infty$ at the critical time $t_A=t_\mathrm{cr}$. The fact that the geodesic length becomes complex-valued for $t_A > t_\mathrm{cr}$ indicates that the geodesic extends into the complexified coordinate direction. In this situation, as argued in the holographic dual of a half de Sitter space \cite{Kawamoto:2023nki}, we can regard the real part of $\tilde{S}_{\mathcal{A}}^{\dis}$ as the entanglement entropy, assuming the Schwinger-Keldysh formulation of the time-dependent density matrix. With an appropriate future projection to a final state, we may also interpret the complex entropy $\tilde{S}_{\mathcal{A}}^{\dis}$ as holographic pseudo entropy \cite{Nakata:2020luh} rather than holographic entanglement entropy, which is analogous to time-like entanglement entropy \cite{Doi:2022iyj, Doi:2023zaf,Narayan:2022afv}. In the former interpretation, we plotted the Lorentzian time evolution of both the connected and disconnected contributions in figure \ref{fig:HEE_LORENTZIAN_THETA}. For a very large subsystem $\mathcal{A} $, where its endpoints are close to the edges at $\theta = \pm \frac{\pi}{2}$, the disconnected phase is always dominant. It exhibits a monotonic growth until the critical time. After the critical time, it grows linearly as $\sim \frac{c}{6} t_A$, as shown in figure \ref{fig:HEE_LORENTZIAN_THETA}. As the size of $\mA$ decreases, the connected phase can compete with the disconnected phase, leading to phase transitions as depicted in the right panel of figure \ref{fig:HEE_LORENTZIAN_THETA}. It is noteworthy that, regardless of the dominant phase, there is a linear growth in $\mathcal{A}$ at late times, \ie 
\begin{equation}
    S_{\mA} = \min \qty{  S_{\mA}^{\rm con} ,  \text{Re}\(  \tilde{S}_{\mA}^{\rm dis} \)  } \, \sim \, \frac{c}{3} t_{A} \,, \quad \text{for} \quad t_A \gg 1 \,. 
\end{equation}
This can be attributed to the inflation of dS space in global time, which affects the entanglement entropy through the Weyl factor.  
Given the success of the holographic one-point function in computing the holographic entanglement entropy within the Euclidean signature, there is reason to believe that the Lorentzian one-point function of the twistor operator may prove to be an effective means of determining the entanglement entropy at late times. To illustrate, the holographic one-point function $ \ev{\mO (t, \theta)}_{\mt{dS}}$ with the Lorentzian signature can be expressed as follows 
\begin{equation}\label{eq:dSintegral02}
    \ev{\mO (t, \theta)}_{\mt{dS}} = \int_{-\frac{\pi}{2}}^{\frac{\pi}{2}} dt_{\mt{Eb}}  \int_{-\frac{\pi}{2}}^{\frac{\pi}{2}} d\theta_b \, 
    \frac{\cos{t_{\mt{Eb}}}}{\qty(\cosh{\eta_b}-\sinh{\eta_b}(\cos{t_{\mt{Eb}}}\cosh t \cos(\theta-\theta_b)+ i \sin{t_{\mt{E}b}}\sinh t  ))^\Delta} \,.
\end{equation}
which is always real by definition\footnote{The imaginary part is precisely canceled due to the fact that the $t_{\mt{E}b}$-integral is symmetric between $\pm t_{\mt{E}b}$}. However, the above integral is not well-defined due to the pole at $t=0$ slice. For example, we can find explicit expression for some particular $\Delta$ while there is a branch cut 
Nevertheless, the above integral is not well-defined due to the presence of a pole at the $t=0$ slice. For instance, the explicit expression for some specific value of $\Delta$ can be derived but with the presence of a branch cut at $\cosh t = \frac{\coth \eta_b}{|\sin \theta |}$. It is anticipated that the complex deformation of the integral contour may circumvent this issue and yield the same result as the Wick rotation. Nevertheless, we discovered that the result of the integral for $\ev{\mO (t, \theta)}_{\mt{dS}} $ is highly sensitive to the choice of the complex contour, and thus, We leave this issue for future investigation.  

\section{Non-extremal island in dS or AdS braneworld}\label{sec:dif}
The island formula in the AdS setup where the gravitational theory is coupled to the non-gravitational bath is captured by the quantum extremal surface, which extremizes the generalized entropy denoted by $S_{\rm gen}$. In this setup, the generalized entropy is minimized when the position of the island is varied along the spatial direction and maximized  when the position is varied along the temporal direction. However, as previously demonstrated, the quantum extremal surface in dS gravity coupled to the bath is instead minimized along the timelike direction, but maximized along the spatial direction. This compelling evidence suggests that this type of quantum extremal surface in dS gravity is not a physically relevant quantity for calculating the correct entanglement entropy. One possible explanation for the discrepancy between quantum extremal surfaces in AdS gravity and dS gravity is the sign of the Ricci curvature of the gravitational spacetime where the island resides. However, this section will demonstrate that the stability of the extremal surfaces is contingent upon the extrinsic curvature of the gravitational brane-world from the perspective of the double holography. In a more generic context, the stability of the quantum extremal surface is determined by the second variation of the generalized entropy along the gravitational region. 
By using the doubly holographic setup, we derive the generalized entropy in the brane perspective as the area of the extremal surface anchored on the two-dimensional (AdS or dS) brane-world living in the AdS$_3$ bulk spacetime. The double holography allows us to ascertain whether a quantum extremal surface is a maximin surface by examining the variation of the geodesic length along the EOW brane.

In section \ref{sec:geovar}, we will examine the first and second variation of geodesic length in an arbitrary Euclidean spacetime. By employing Synge’s formula for the second variation of geodesics, we derive a necessary and sufficient condition for the stability of the extremal surface. The stable extremal surface is referred to as the maximin surface in the Lorentzian signature and minimal surface in the Euclidean signature. In our doubly holographic setup, we will demonstrate that the sufficient condition is related to the extrinsic curvature of the EOW brane. In section \ref{sec:insES}, we will examine explicit examples in the AdS$_3$/BCFT$_2$ setup with time-symmetric profiles of EOW branes. It is demonstrated that both the dS brane and the AdS brane are susceptible to unstable extremal surfaces.

\subsection{Geodesic variation and stability condition}\label{sec:geovar}

As motivated before, we first focus on exploring the variations of geodesic length in a generic Euclidean spacetime in this subsection. The background results in this subsection can be also found in textbooks about Riemannian geometry, \eg  \cite{frankel2011geometry,chow2023hamilton,Hawking:1973uf}. Even we do not show the results explicitly, similar results can be also derived for the variation of area of codimension$-n$ surfaces in higher dimensional space. We believe that it can be applied to explore the stability of generalized entropy for a higher-dimensional system. We will leave this for future exploration and focus on the geodesic variations for simplicity.

\begin{figure}[t]
	\centering
	\includegraphics[width=4in]{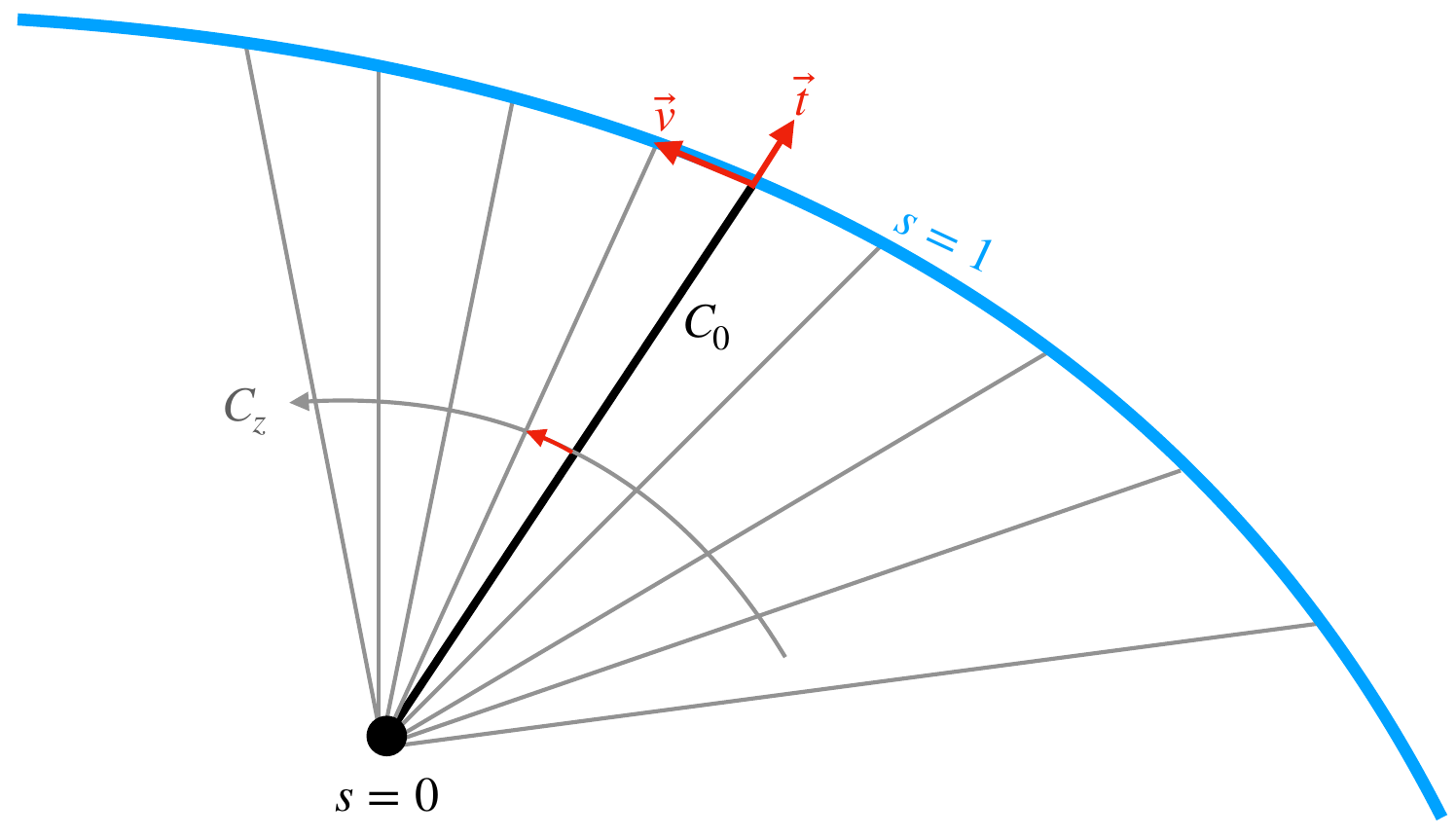}
	\caption{The geodesic variations of $C_0$ form a one-parameter family of neighboring lines, $C_z$, parametrized by $z$. In particular, we are interested in the variation along a brane, which is obtained by anchoring all perturbed geodesics to the brane.}
	\label{fig:variations}
\end{figure}

Supposing $C_0(s)$ is a curve in the Riemannian manifold with a given metric $g(x^\mu)$, we parametrize the curve by a parameter $s$. To explore the variation from the original curve $C_0(s)$, we define a one-parameter family of neighboring lines $C_z(s)$ parameterized by the variation parameter $z$, see figure \ref{fig:variations} for the illustration. The arc length $L(C_z)$ for a single curve is defined by 
\begin{equation}
L(C_z)= \int_0^1 \sqrt{g\(\frac{\partial \mathbf{x}}{\partial s},\frac{\partial \mathbf{x}}{\partial s}\)} ds \,, 
\end{equation}
where $g(\vecv,\vecv')$ denotes the inner product of two vectors associated with the Riemannian metric $g$ and we choose the geodesic parameter $s$ as the affine parameter. By defining the tangent vector and variation vector as 
\begin{equation}
\vecT =\frac{\partial \mathbf{x}}{\partial s} , \qquad  \vecV=\frac{\partial \mathbf{x}}{\partial z} \,,
\end{equation}
respectively, we derive the first variation of the arc length which is known as 
\begin{equation}
\frac{d}{dz} L(C_z) \bigg|_{z=0}= \frac{1}{L_0} \( g\(\mathbf{t},\mathbf{v} \)\bigg |_0^1 - \int_0^1 g\(\nabla_{\vect} \vect, \vecv\) ds \)\,,
\end{equation}
where the geodesic length reduces to $L_0=L(C_0)= ||\vect||$, $\mathbf{t}=\vecT |_{z=0}$ is the tangent vector along the curve $C_0$, and $\mathbf{v}=\vecV |_{z=0}$ is the variation vector for the original curve.

Since we are interested in the extremal surface, \ie geodesics , the tangent vector of the original curve should satisfy the geodesic equation, \ie $\nabla_{\vect} \vect  = 0$. As a result, 
the first variation formula for geodesics reduces to 
\begin{equation*}\label{variation_first}
\frac{d}{dz} L(C_z) \bigg|_{z=0}= \frac{1}{L_0}  g\(\mathbf{t},\mathbf{v} \)\bigg |_0^1 \,,
\end{equation*} 
where only the boundary terms at the endpoint contribute. Furthermore, the second variation of the arc length of the geodesic is given by Synge's formula, namely\footnote{Our definition of Riemannian tensor is given by 
\begin{equation*}
 R(\vecX,\vecY) \vecZ=\nabla_\vecX\nabla_\vecY \vecZ - \nabla_\vecX\nabla_\vecY \vecZ - \nabla_{\[\vecX,\vecY\]}\vecZ \,.
\end{equation*}
It is also useful to define the component of any vector $\vecX$ perpendicular to tangent vector $\vect$ as 
\begin{equation*}
\vecX^\bot = \vecX - \frac{g\(\vecX,\vect\)}{||\vect||^2} \vect\,.
\end{equation*}
} 
\begin{equation}\label{Synge_for}
\frac{\partial^2 L(C_z)}{\partial z \partial z} \bigg |_{z=0}=\frac{1}{L_0} g\(\vect, \nabla_\vecv \vecv\)\bigg |^1_0 +\frac{1}{L_0} \int^1_0 \[ || (\nabla_\vect \vecv )\wedge \vect||^2 -g\(R(\vecv, \vect)\vect, \vecv\)\] ds,
\end{equation}
where $|| (\nabla_\vect \vecv )\wedge \vect||^2=g\left((\nabla_\vect \vecv )^\bot, (\nabla_\vect \vecv )^\bot\right)$ is the area of the parallelogram spanned by the vectors $\vect,\nabla_\vect \vecv$ \footnote{For the second order variation of geodesic, one can also introduce another independent parameter $w$ with new variation vector $\vecw$ and derive the variation of arc length in terms of 
\begin{equation}
\frac{\partial^2 L(C_z)}{\partial z \partial w} \bigg |_{z=w=0}=  \frac{1}{L_0} g\(\vect, \nabla_\vecw \vecv\)\bigg |^1_0 +\frac{1}{L_0} \int^1_0 \[ g\((\nabla_\vect \vecv )^\bot, (\nabla_\vect \vecw )^\bot\) +g\(R(\vecw, \vect)\vecv, \vect\)\] ds  \,.
\end{equation}
In a more general sense, these variation formulas can be generalized to the case with an arbitrary functional, leading to similar geometric conclusions. For more details, see \eg \cite{Bernamonti:2019zyy,Bernamonti:2020bcf,chow2023hamilton}
}. Note that the variation proportional to the tangent vector is trivial since it does not change the curve. In the following, we only need to consider the orthogonal variation with satisfying $g(\vect,\vecv)=0$ as well as $ g\left(\nabla_{\vect}\vecv,\vect\right)=0$.  Consequently, the orthogonal variation for a geodesic reduces to 
\begin{equation}
L''(C_0)=  \frac{1}{L_0} g\(\vect, \nabla_\vecv \vecv\)\bigg |^1_0 +\frac{1}{L_0} \int^1_0 \[ || (\nabla_\vect \vecv )||^2 -g\(R(\vecv, \vect)\vect, \vecv\)\] ds\,.
\end{equation}
By using this second variation formula, one find that the geodesic $C_0$ with fixed endpoints is stable and locally minimizes the arc length in a negatively curved Riemannian manifold.  If we consider the perturbation by varying the geodesic from one to another one, which means that we only focus on a one-parameter family of neighboring geodesic $C_{z}$. Only special variation vector $\vecV$ can vary one geodesic to another one. The restriction is nothing but the geodesic deviation equation, \ie
\begin{equation}\label{geodesic_deviation}
\frac{D^2 \vecV}{\partial s^2} + R\(\vecV,\vecT\) \vecT=0\,,
\end{equation}
which is a group of second order linear ordinary differential equations for $\vecV$. In differential geometry, the geodesic deviation equation is also known as Jacobi equation, whose solution is called the Jacobi field. The variation of arc length of geodesic $C_0$ generated by Jacobi field $\vecv$ is thus given by
\begin{equation}\label{variation_second}
\begin{split}
L''(C_0)&=  \frac{1}{L_0}\( g\(\vect, \nabla_\vecv \vecv\) +g\(\vecv, \nabla_\vect \vecv\) \)\bigg |^1_0\,,\\
\end{split}
\end{equation}
where the contributions only come from the initial and end points like the first variation for geodesic.

By using necessary ingredients reviewed before, we move to analyzing the stability of the extremal surfaces anchored on a brane living in (Euclidean) AdS bulk spacetime, as illustrated in figure \ref{fig:variations}. First of all, we are only interested in the cases with one endpoint fixed at $s=0$ but non-vanishing variation at another endpoint $s=1$ on the brane. The variation of the geodesic is realized by varying the endpoint on the brane, which indicates that the variation vector $\vecV$ is nothing but the tangent vector along the brane. 
Since we assume that the original geodesic is extremal with respect to the brane profile, the first variation of the geodesic vanishes, implying $g\left(\mathbf{t},\mathbf{v} \right)=0$ on the brane. It gives rise to a fact that the tangent vector $\vect$ of the geodesic is related to the unit normal vector $\vecn$ of the brane, namely
\begin{equation}
 \vecn  =  \frac{1}{L_0} \vect \big |_{\rm{brane}}  \,, \qquad \text{with} \quad  || \vecn|| =1 \,. 
\end{equation}
We thus show that the boundary term in the second variation formula \eqref{Synge_for} is related to the second fundamental form $K_{\mu\nu}$ of the brane because of 
\begin{equation}
\frac{1}{L_0} g\(\vect, \nabla_\vecv \vecv\) \big |_{\rm{brane}} = g\(\vecn, \nabla_\vecv \vecv\) = - g\(\nabla_\vecv \vecn,  \vecv\) =-K(\vecv, \vecv)  \,.
\end{equation}
Also noting that the bulk Riemannian tensor term in the integral is related to the sectional curvature\footnote{The sectional curvature in our convention is defined by 
\begin{equation}
K_{\rm{set}} (\vecX, \vecY )  \equiv   \frac{ \langle R(\vecX, \vecY)\vecY, \vecX \rangle}{\langle \vecX, \vecX \rangle\langle \vecY, \vecY \rangle-\langle \vecX, \vecY \rangle^2}  =\frac{R(\vecX, \vecY, \vecX, \vecY)}{\langle \vecX, \vecX \rangle\langle \vecY, \vecY \rangle-\langle \vecX, \vecY \rangle^2}  \,.
\end{equation}} 
\begin{equation}
    g\(R(\vecv, \vect)\vect, \vecv\)=K_{\rm{set}} (\vecv, \vect )||\vecv\wedge\vect||^2,
\end{equation}
Combining all previous expressions, the variation generated by an arbitrary orthogonal variation vector along the brane is rewritten as 
\begin{equation}
\begin{split}
L''(C_0) &=  -K(\vecv, \vecv)+\frac{1}{L_0} \int^1_0 \[ || (\nabla_\vect \vecv )||^2 -K_{\rm{set}} (\vecv, \vect )||\vecv\wedge\vect||^2\] ds\,. \\
\end{split}
\end{equation}
As the first result, we conclude that the {\it necessary and sufficient condition} for the extremal geodesic anchored on the brane to be stable is given by requiring 
\begin{equation}\label{eq:finalcondition}
\begin{split}
L''(C_0)&=  \( -K(\vecv, \vecv) +g\(\vecv, \nabla_\vecn \vecv\) \)\big |_{\rm brane}  > 0\,.\\
\end{split}
\end{equation}
for any tangent vector $\vecv$ on the brane. 

Furthermore, it is remarkable that AdS$_{d+1}$ or Euclidean AdS space has constant {\it negative sectional curvature}, a characteristic feature of AdS spaces. The value of its sectional curvature does not depend on the chosen plane or point in the manifold, indicating uniformity in curvature throughout the spacetime. Together with the fact that $|| (\nabla_\vect \vecv )||^2$ and $||\vecv\wedge\vect||^2$ are positive, we find that the integral term  appearing in the second variation is always positive. As a result, the second variation is bounded below by the extrinsic curvature of the brane. If the brane has a {\it negative extrinsic curvature everywhere}, \ie\footnote{Note that this is only a sufficient but not necessary condition since the integral term is always non-negative.} 
\begin{equation}\label{eq:sufficient}
L''(C_0)  \ge   -K(\vecv, \vecv)\big |_{\rm brane}  > 0  \,,
\end{equation}
the geodesic orthogonal to the brane is stable, that is, locally minimizes the arc length. In other words, the only term that could spoil the stability of the geodesics anchored on the brane is the positive extrinsic curvature.

In the last part of this subsection, we shall now apply the aforementioned results to the EOW brane in AdS space, which satisfies the Neumann boundary condition \eqref{eq:Neumann}, as discussed in the preceding sections. As a consequence of the Neumann boundary condition, the extrinsic curvature of the corresponding d-dimensional brane is determined by its tension, \ie 
\begin{equation}
  K = \frac{d}{d-1} \mathcal{T} \,, \qquad  K_{ij} = \frac{\mathcal{T}}{d-1} g_{ij} \,.
\end{equation}
Then the sufficient condition  \eqref{eq:sufficient} is consistently satisfied when the brane with negative tension is considered. In the case of positive tension, the extremal geodesic anchored on the dS brane is inherently unstable, exhibiting a locally maximal length along the dS brane. In contrast, the extremal surface anchored on the AdS brane remains stable. This distinction can be attributed to the fact that the extrinsic curvature of the AdS brane is always relatively minor. More explicitly, we get
$K = \frac{d}{d-1} \mathcal{T} \le \frac{d}{L_{\mt{AdS}}}$ for a AdS brane since it only exists with a lower tension $ |T| <  \frac{d-1}{L_{\mt{AdS}}}$ \cite{Kawamoto:2023wzj}. Then the sufficient and necessary condition derived in eq.~\eqref{eq:finalcondition} will always be satisfied for the AdS brane, regardless of the sign of its extrinsic curvature. In contrast, the dS brane, which satisfies the Neumann boundary condition, is supported by a large tension, with $|T| > \frac{d-1}{L_{\mt{AdS}}}$. In consequence, the dS brane with positive tension will invariably satisfy 
\begin{equation}
 K(\vecv,\vecv)|_{\rm{dS\,\,brane}} \propto  K |_{\rm{dS\,\,brane}}  >  \frac{d}{L_{\mt{AdS}}} \,,
\end{equation}
resulting in the violation of the condition given by eq.~\eqref{eq:finalcondition}.

\begin{figure}[t]
	\centering
	\includegraphics[width=4.5in]{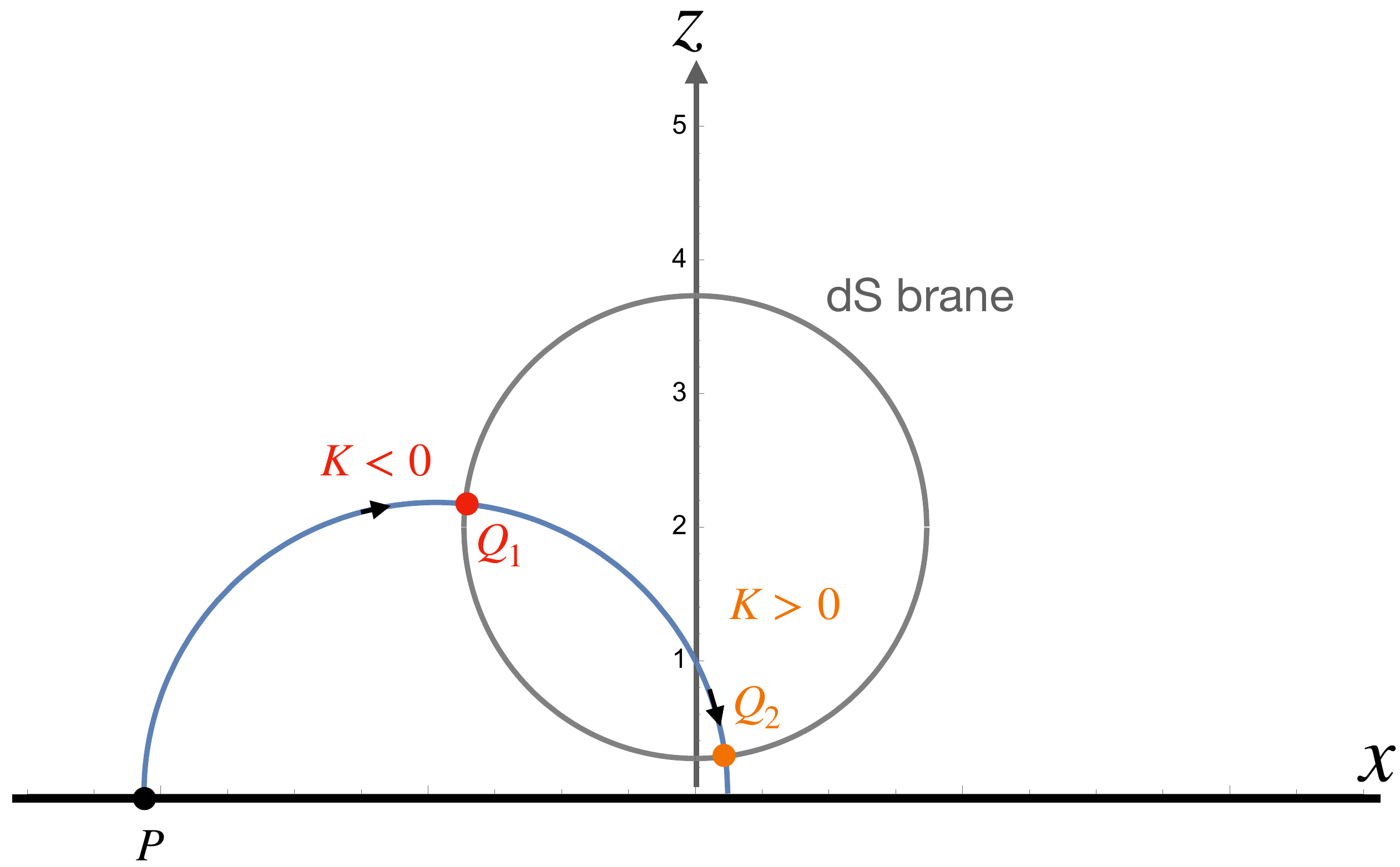}
	\caption{The extremal geodesic (denoted by the blue curve) starting from the boundary at a fixed point $P$ in Poincar\'e AdS spacetime intersects the dS brane, which is shown by the gray circle. If the geodesic intersects the dS brane from the inside (\ie at the yellow point $Q_2$ in this figure), the extrinsic curvature is positive, indicating that the extremal geodesic is unstable at this intersection.}
	\label{fig:dSgeodesic}
\end{figure}

In conclusion, the non-stable geodesic on the dS brane is attributed to the positive sign and the magnitude of its extrinsic curvature. As illustrated in figure \ref{fig:dSgeodesic}, the extremal geodesic is stable outside the dS bubble, where $K < 0$, and is non-stable inside the dS bubble, where $K > 0$.

\subsection{Instability of extremal surfaces in AdS/BCFT}\label{sec:insES}

\begin{figure}[t]
	\centering
	\includegraphics[width=4in]{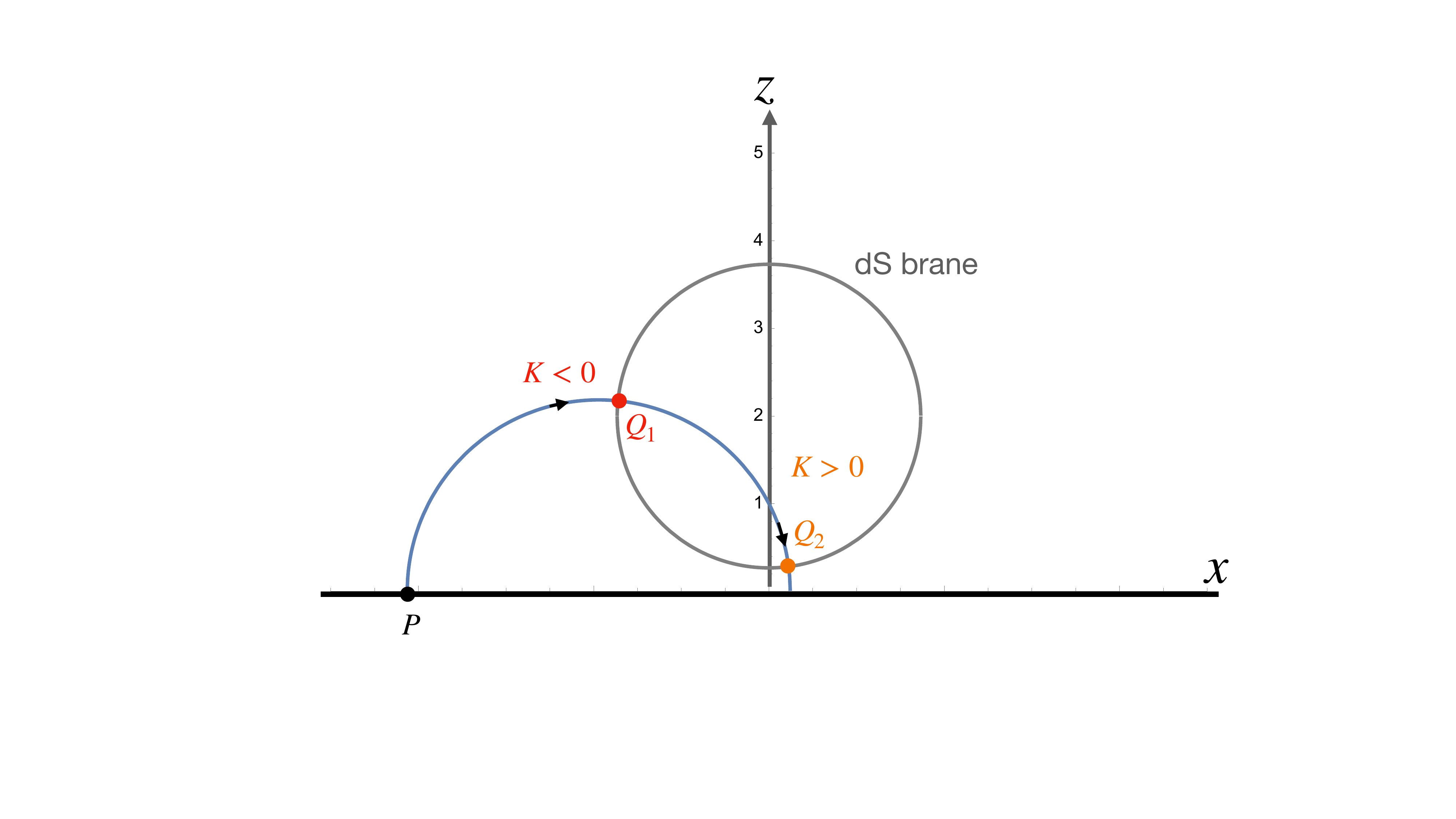}
	\caption{The setup for computing the holographic entanglement entropy in AdS/BCFT.}
	\label{fig:HEEBCFT}
\end{figure}

To elucidate the conclusion from the geodesic variation with more concrete examples, let us examine the AdS$_3$/BCFT$_2$ setup, where the spacetime is described by the  Poincar\'e  AdS$_3$ metric, as follows:
\begin{equation}
ds^2=\frac{-dt^2+dx^2+dz^2}{z^2} \,. 
\end{equation}
We assume that the physical space is bounded by a static two dimensional EOW brane. Assuming the worldsheet of the EOW brane is symmetric in time at $t=0$, we characterize its profile at this instant as $z = Z(x)$. In the subsequent analysis, we concentrate on the holographic entanglement entropy at $t=0$, which allows us to confine the three-dimensional space to the two-dimensional slice delineated by the domain $0 < z \leq Z(x)$. This region is illustrated in blue in figure \ref{fig:HEEBCFT}.

The dual BCFT resides on the half line $x \geq 0$, extended in the trivial time direction. We are interested in the holographic entanglement entropy $S_{\mathcal{A}}$ for the interval $\mA= [0, \xi]$ at $t = 0$. The intersection of the geodesic $\Gamma$ emanating from $x = \xi$ with the EOW brane is characterized by the coordinates $(x, z) = (a, Z(a))$. According to the standard holographic prescription for entanglement entropy in the AdS/BCFT correspondence, the value of $a$ is determined by minimizing the geodesic length $D$ with respect to the parameter $a$ \cite{Takayanagi:2011zk, Fujita:2011fp}.

The geodesic distance $D$ between $(\xi, \epsilon)$ and $(a, Z(a))$ is defined by
\begin{equation}
   \cosh D=\frac{(Z(a)-\ep)^2+(\xi-a)^2}{2Z(a)\ep}\equiv \mL \,,  
\end{equation}
where $z = \epsilon$ is the standard UV regulator. By requiring that the derivative of $D$, or equivalently $\mathcal{L}$, with respect to $a$ vanishes, we find
\begin{equation}
\xi-a=-\frac{Z(a)}{\dot{Z}(a)}+\s{\frac{Z(a)^2}{\dot{Z}(a)^2}+Z(a)^2-\ep^2} \,, 
\end{equation}
where $\dot{Z} = \frac{dZ(a)}{da}$. This determines the value of $a$ as a function of $\xi$. In the limit $\epsilon \to 0$, which corresponds to the BCFT living at the AdS boundary, we proceed with the analysis.

Now, we inquire whether the saddle point is stable, that is, whether it represents a local minimum. To determine this, we need to compute the second-order derivative, 
\begin{equation}
\frac{d^2\mL}{da^2}=\frac{2Z^2}{\dot{Z}^2}\left(\dot{Z}^2+\dot{Z}^4-Z\ddot{Z}+Z\ddot{Z}\s{1+\dot{Z}^2}\right) \,. 
\end{equation}
From this, it is evident that for the saddle point to become unstable, \ie, for $\frac{d^2\mL}{da^2} < 0$, it is necessary to have $\ddot{Z} < 0$. This condition is indeed generally possible. The null energy condition stipulates that
\begin{equation}
\ddot{Z}+(1+\dot{Z}^2)\de^2_tZ\leq 0 \,, 
\end{equation}
where $\de_t$ denotes the derivative with respect to the time variable, assuming the time symmetry condition $Z(x,t) = Z(x,-t)$. Specifically, for a static EOW brane, denoted by $z = Z(x)$, the time derivative is absent, and this condition simplifies to $\ddot{Z} \leq 0$.

\begin{figure}[t]
	\centering
	\includegraphics[width=2.8in]{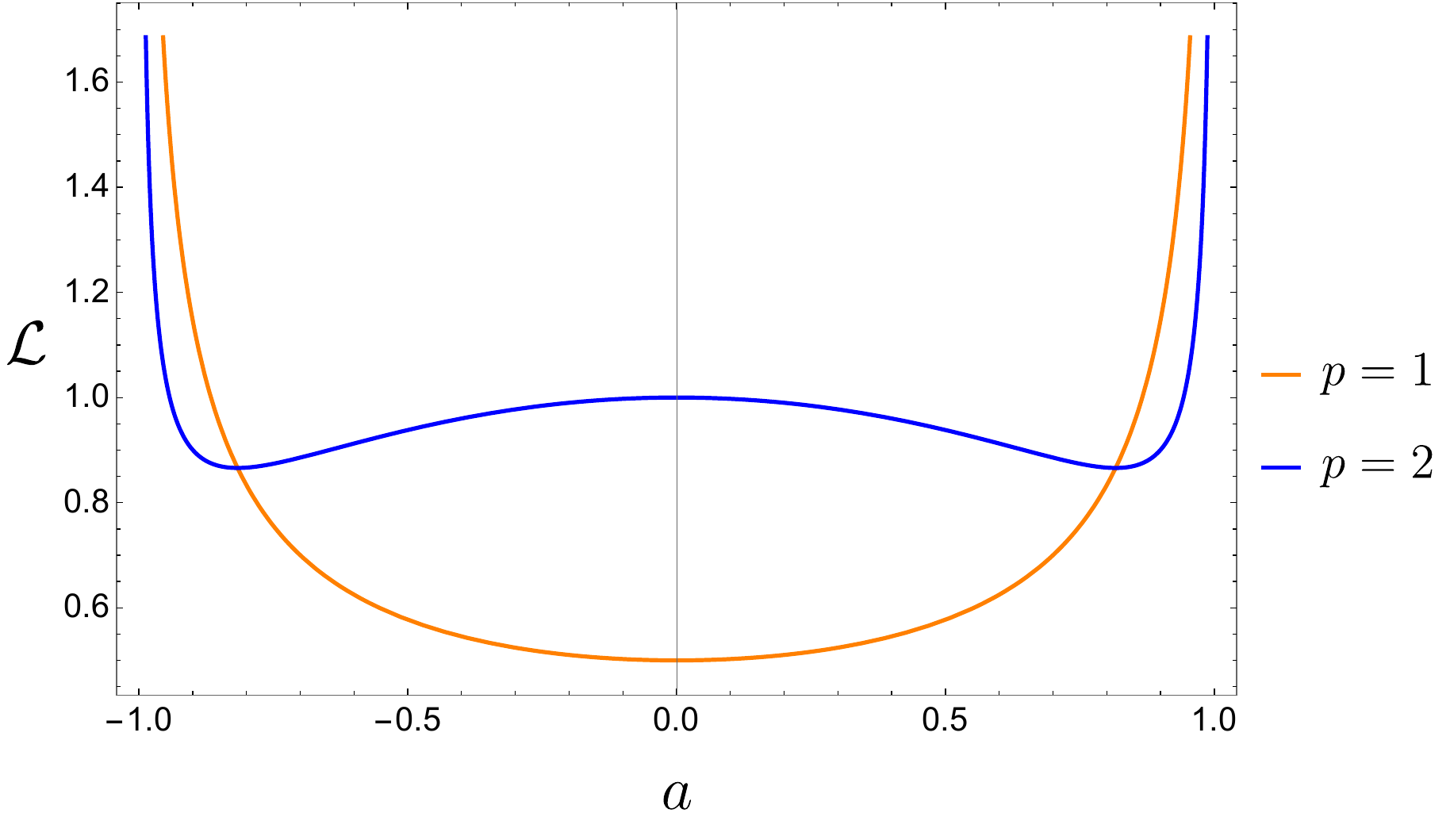}\quad
	\includegraphics[width=3.1in]{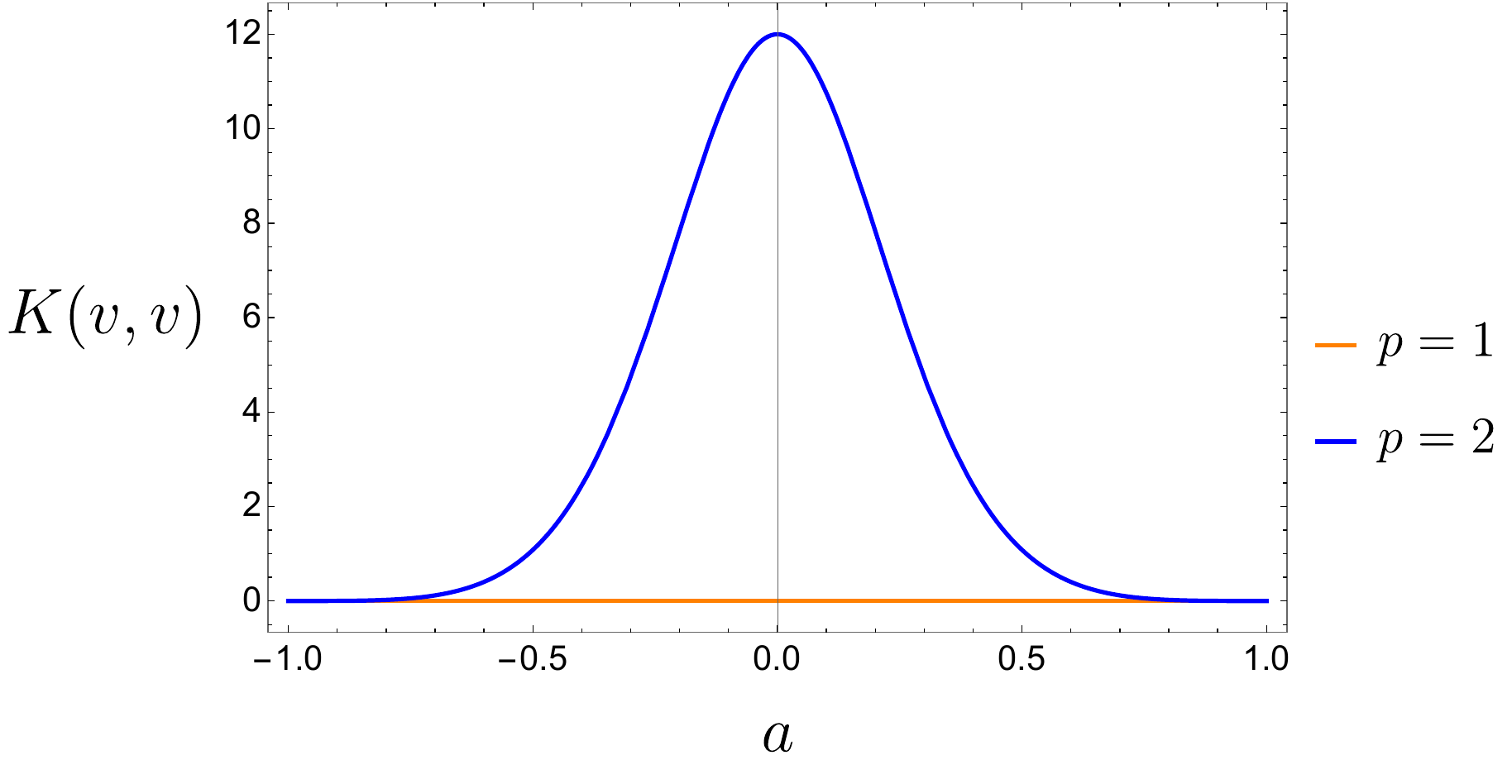}
	\caption{Left: The geodesic function $\mathcal{L}(a)$ with setting $\xi=0, \epsilon=1$ is illustrated as a function of the intersection point $a$ on the EOW brane, which is parameterized by eq.~\eqref{eowpro}. Right: The extrinsic curvature component, defined as $K(\vecv, \vecv)$, is plotted as a function of $a$.}\label{fig:HEEBCFT2}
\end{figure}

For instance, consider the profile of an EOW brane described by
\begin{equation}\label{eowpro}
z = Z(x) = p \sqrt{1 - x^2} \,,
\end{equation}
where $p=1$ corresponds to the conventional constant tension brane. It is evident that $\frac{d^2 \mathcal{L}}{da^2} \propto 2p^2 - p^4$ at $a = \xi = 0$. For $p=2$, the geodesic distance between $(0,0)$ and $(a, Z(a))$ is depicted in figure \ref{fig:HEEBCFT2}. It is observed that the saddle point at $a = 0$ is unstable, whereas there are two stable saddle points. In the AdS/BCFT framework, where the brane extends within the region $-1 \leq x \leq 1$ and intersects the AdS boundary at $z=0$, the stable saddle points yield the correct holographic entanglement entropy, and the corresponding geodesic is extremal and locally minimal. However, if the EOW brane is confined to a smaller range, for example, $-0.5 < x < 0.5$, there exists no extremal point that minimizes the length. This situation is analogous to our non-extremal island for the CFT on a half dS$_2$ coupled to gravity on a half dS$_2$. It is noteworthy that this phenomenon is not exclusive to positive curvature on the EOW brane. To illustrate this, let us consider the EOW brane as static. The scalar curvature on the EOW brane is then given by
\begin{equation}\label{curda}
R = -\frac{2(\dot{Z}^2 + \dot{Z}^4 - Z \ddot{Z})}{(1 + \dot{Z}^2)^2} \,.
\end{equation}
Applying the null energy condition $\ddot{Z} < 0$, we deduce that this brane invariably exhibits a negative curvature $R \leq 0$.

On the other hand, the extrinsic curvature for the static EOW brane described by the profile $z=Z(x)$ is given by
\begin{equation}
    K_{tt}=\frac{1}{Z^2 \sqrt{\dot{Z}^2+1}},\ \ K_{xx}=-\frac{Z \ddot{Z}+\dot{Z}^2+1}{Z^2 \sqrt{\dot{Z}^2+1}}.
\end{equation}
Additionally, the contraction with respect to the variation vector $v^\mu = P^\mu_x$ ($P_{\nu}^\mu$ being the projection tensor on the brane) along the $x$ direction is given by
\begin{equation}
   K(\vecv,\vecv)= -\frac{Z^2(Z \ddot{Z}+ \dot{Z}^2+1)}{\left(\dot{Z}^2+1\right)^{5/2}} \,.
\end{equation}
Here, we have chosen the outward-pointing unit normal vector of the brane. Imposing the null energy condition $\ddot{Z} < 0$, the quantity $K(\vecv, \vecv)$ can be either positive or negative, and its magnitude can be arbitrarily large in certain regions of $x$. Thus, in general, there exist both stable and unstable saddle points, as analyzed in section \ref{sec:geovar}. We can then consider the example of the profile in eq.~\eqref{eowpro} and plot the function $K(\vecv, \vecv)$ as a function of the intersection point $a$ on the EOW brane, as shown in figure \ref{fig:HEEBCFT2}. We find that $K(\vecv, \vecv)$ is always positive in this region, and the minimal stable points appear in regions where $K(\vecv, \vecv)$ is small. However, the maximal point appears when $K(\vecv, \vecv)$ is large, rendering the second variation negative at that point.

Next, we consider a setup where the EOW brane is a dS brane or its deformation. Using the AdS/dS slicing of AdS$_3$,
\begin{equation}
   ds^2=  d\eta^2 + \sinh^2{\eta}\qty(-dt^2+\cosh^2{t}d\theta^2)  \,,
\end{equation}
the dS brane is located at the hypersurface $\eta =\eta_b$. We can generalize this setup to consider a brane determined by a more generic profile $\eta = f(\theta)$. We examine variations along the $\theta$ direction, where the variation vector is $v^\mu = P^\mu_\theta$, with $P^\mu_\nu$ being the projection tensor on the brane. The contraction of the extrinsic curvature with the variation vectors, \ie $K(\vecv,\vecv)$ is given by 
\begin{equation}
\frac{\cosh ^2t \sinh f \sqrt{f'^2\sech^2t 
   \csch^2f+1} \left(\cosh f \left(2 f'^2+\cosh ^2t \sinh ^2f\right)-f'' \sinh f\right)}{\left(f'^2+\cosh ^2t \sinh ^2f\right)^3} \,.
\end{equation}
While the denominator of $K(\vecv, \vecv)$ is always positive, the numerator can be either positive or negative and is time-dependent. For the dS brane where $f(\theta) = \eta_b$, we have $K(\vecv, \vecv) = \coth \eta_b , 
\csch^2\eta_b \sech^2 t$, which retains the sign of $\eta_b$. A positive $\eta_b$ may indicate instability of the extremal surface. To explore the deformed dS brane, we consider modifying $f(\theta)$, such as
\begin{equation}\label{eq:dSdeformation}
f(\theta) = \eta_b - q  \cos \theta \,, 
\end{equation}
with $\theta \in [\frac{\pi}{2}, \frac{3\pi}{2}]$. For positive $q$, the brane is deformed towards the boundary, and $K(\vecv, \vecv)$ remains positive, potentially leading to instability. Conversely, for negative $q$, the brane contracts inwards, which can result in $K(\vecv, \vecv)$ being negative in certain regions, thus ensuring that the geodesic remains minimal under variations in the spatial direction.

\begin{figure}[t]
   \centering
\includegraphics[width=2.8in]{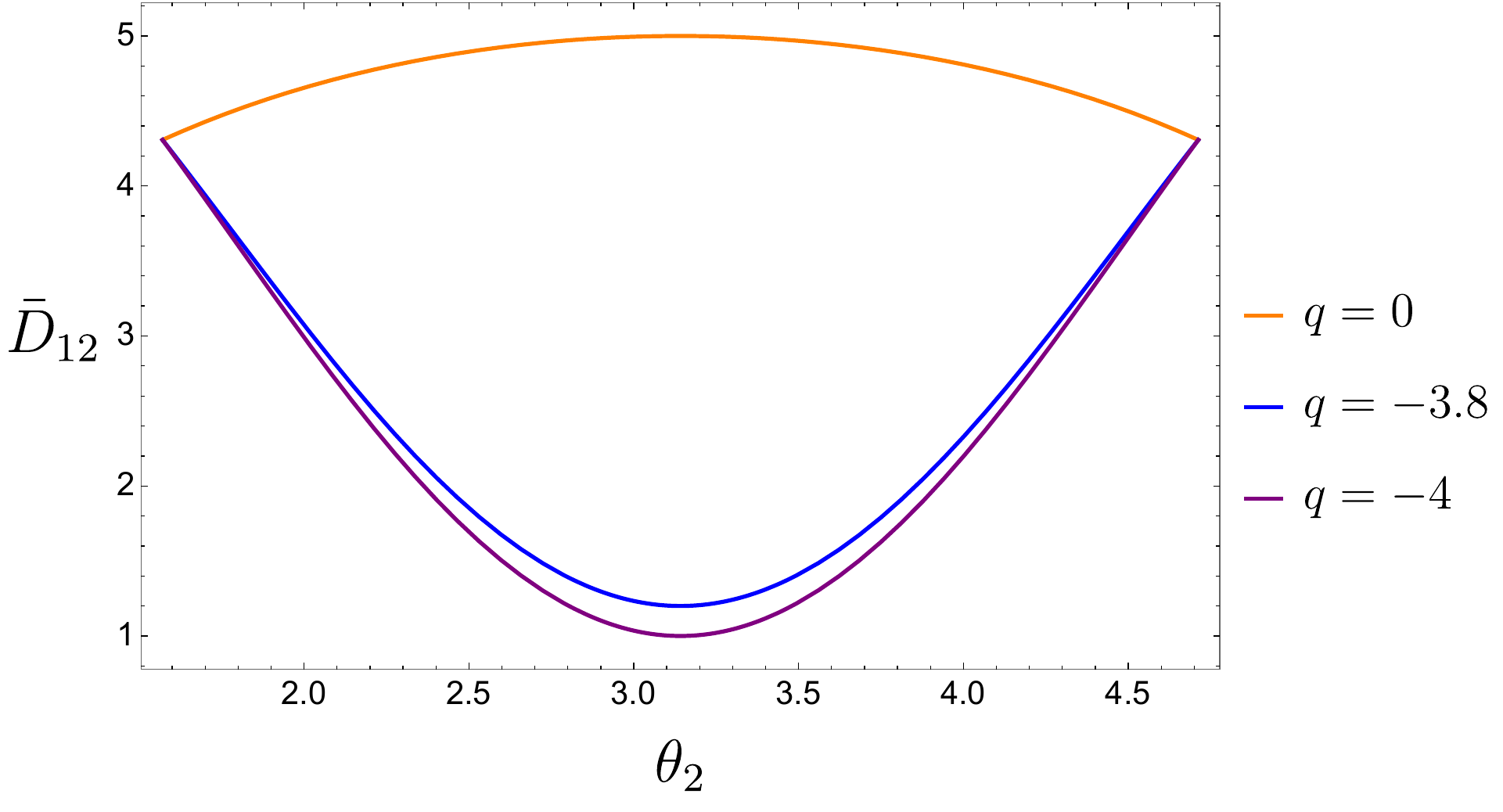}
   \includegraphics[width=3.2in]{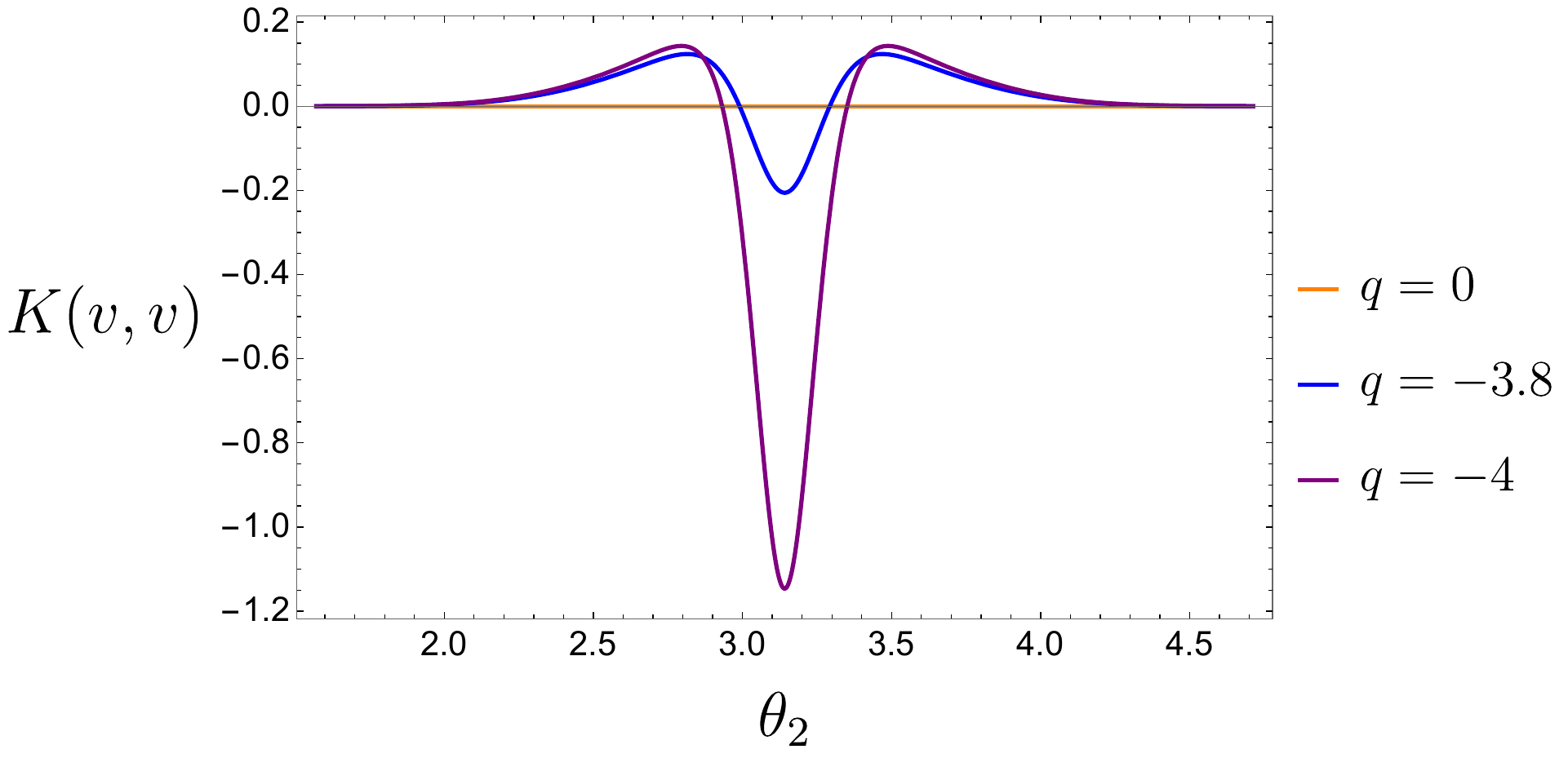}
      \caption{Left: The renormalized geodesic distance $\bar{D}_{12}$ \eqref{eq:barD12} as a function of the endpoint located at $\theta =\theta_2$ on the brane. In the case of a deformed dS brane with a negative value of $q$, there exists the possibility of an extremal point that is locally minimal. Right: The extrinsic curvature $K(\vecv, \vecv)$ as a function of $\theta_2$. The negative value of the extrinsic curvature $K(\vecv, \vecv)$ provides a sufficient condition for the extremal geodesic to be locally minimal. For both plots, we choose $\eta_b = 5$, $\theta_1 = 0 = t_1$.}
      \label{fig:dsdK}
\end{figure}

On the other hand, we can directly examine the variation of the geodesic length. The geodesic distance $D_{12}$ between one endpoint $(\eta_\infty, t_1, \theta_1)$ on the asymptotic boundary and the other endpoint $(f(\theta_2), t_2, \theta_2)$ on the EOW brane is given by eq.~\eqref{eq:Dab}. As $\eta_\infty \to \infty$, the regularized geodesic length is
\begin{equation}\label{eq:barD12}
    \bar{D}_{12}=\log(\cosh{f(\theta_2)} -\sinh{f(\theta_2)}\qty(\cosh{t_1}\cosh{t_2}\cos{(\theta_1-\theta_2)}-\sinh{t_1}\sinh{t_2}))\,. 
\end{equation}
We first consider the extremal condition, which requires that the first derivatives with respect to both $t_2$ and $\theta_2$ vanish. The extremization condition for the time direction simplifies to
\begin{equation}
    t_2=\arctanh\frac{\tanh t_1}{\cos(\theta_1-\theta_2)} \,,
\end{equation}
while the condition for the spatial direction generally yields a transcendental equation, which is difficult to solve analytically. To facilitate the calculation, we examine the example defined in eq.~\eqref{eq:dSdeformation} to illustrate the dependence of the geodesic length on $\theta_2$. For comparison, the corresponding $K(\vecv, \vecv)$ is shown in figure \ref{fig:dsdK}. It is evident that negative $K(\vecv, \vecv)$ provides a sufficient condition for a positive $L''(C_0)$, which indicates a stable extremal geodesic. This is particularly evident in the third case, where $q$ is negative and of a large magnitude.

\section{Conclusions and Discussions}\label{sec:dis}
In this paper, we explore the calculation of entanglement entropy in setups where a CFT is defined on one half of a two-dimensional de Sitter space, while a gravitational theory is defined on the other half across a timelike boundary. Considering that the subsystem resides within the CFT region, we initially applied the island formula. However, we discovered that no extremal configuration of the island minimizes the entropy functional, as discussed in section \ref{sec:island}. We found an extremal island solution that results in a local maximum, which is physically inadmissible. This situation contrasts with cases involving CFTs coupled to black holes in AdS spaces, which have been extensively studied in recent years. Nevertheless, in more general spacetimes, both extrinsic and intrinsic curvatures in higher-dimensional spaces play significant roles, as outlined in section \ref{sec:dif}. Consequently, these phenomena are not limited to specific de Sitter spacetimes but may occur in a wide range of gravitational settings.

To address issues related to the extremal island in dS space, section \ref{sec:doubleholography} explores the higher dimensional gravity dual through the framework of double holography. By computing the holographic correlation functions, specifically those of twist operators relevant to entanglement entropy calculations, in the AdS/BCFT setup involving the EOW brane, we found that the correction computation of the entanglement entropy is provided by a ``non-extremal island". This implies that the chosen island does not extremize the entropy functional but rather globally minimizes it. We further corroborated this approach through a direct holographic entanglement entropy calculation using the geodesic (denoted by $\Gamma_\mA$ in the AdS$_3$/CFT$_2$ setup). Utilizing this prescription, we computed the entanglement entropy in a Euclidean setup where a CFT on a two-dimensional hemisphere is coupled to a gravitational theory on the other hemisphere. We observed that for a small subsystem, the geodesic remains connected, resulting in the entanglement wedge excluding any portion of the gravitational region. As the size of the subsystem increases, a phase transition occurs, and the geodesic becomes disconnected. In contrast to standard scenarios where gravity resides in non-positive curvature spaces, we found that in our de Sitter setup, the geodesic always terminates at the boundary of the half dS space. Consequently, in this island phase, the entanglement wedge encompasses the entire gravitating region.

In the Lorentzian setup, where a CFT on one half of dS space is coupled to a gravitational theory on the other half, we can compute the entanglement entropy by analytically continuing the Euclidean results. We demonstrated that this method is effective in the early stages of time evolution. This finding was supported by explicitly computing the geodesic lengths in Lorentzian geometry according to the doubly holographic description. However, at late times, we observed that the endpoints of the geodesics on the EOW branes reach the future infinity of the timelike boundary. Beyond this critical point, no physically sensible geodesics exist in Lorentzian spacetime. Consequently, the analytically continued entropy becomes complex-valued, akin to the pseudo entropy \cite{Nakata:2020luh}, indicating that the geodesic corresponding to this entropy extends into a complexified coordinate direction. Additionally, complex entanglement entropy may arise from regulating horizon-like structures or quasi-normal modes \cite{Kawamoto:2023ade}. A more comprehensive understanding of the late-time behavior remains an open question. Although this paper focuses on two-dimensional examples for simplicity, we anticipate that our findings can be generalized to higher dimensions in a qualitatively similar manner, see \cite{Shaghoulian:2021cef} for previous investigations with higher dimensional dS space.

Compared to the results from the antipodal island and the non-extremal island suggested by double holography, the key distinguishing feature is that the boundary of the non-extremal island is not defined by the usual extremal surface. More importantly, the non-extremal island also suggests a different structure for the quantum state in de Sitter gravity when coupled with a bath system. To illustrate this, let us consider the simple model where
$\eta_b = \eta_\infty$ as an example. If we choose the subsystem $A$ to be the entire bath, \ie $\theta_A = \frac{\pi}{2}$, it is noteworthy that the corresponding entanglement entropy, as derived in eq.~\eqref{eq:SAtwophases} for the island phase, simply vanishes:  
	\begin{equation}
	S_{\mt{EE}}  (\text{entire bath})  = 	\tilde{S}_{\mathcal{A}}^{\mathrm{dis}} \big|_{\theta_A = \frac{\pi}{2}}  =0     \,. 
\end{equation}
This implies that the quantum state at any Cauchy slice of the total system, consisting of the half dS gravity and a non-gravitational bath, is a product state defined as 
\begin{equation}\label{eq:Hilbet_space}
	 | \Psi (t) \rangle  =    | \Psi_{\mathrm{half \,\,dS \,\,gravity}} \rangle  \otimes | \Psi_{\text{entire bath}} \rangle  \,.
\end{equation}
Moreover, we can imagine different setups by varying the relative sizes of the dS gravity and the dS bath. In all cases, we will find that the quantum state corresponding to the half dS gravity remains a pure state, regardless of its size. In other words, it is impossible to entangle either the global dS or any part of the dS gravity with a bath system. This result may or may not come as a surprise. A natural explanation is that the boundary dual of a half dS gravity, \ie defects in the boundary perspective does not contain any local degree of freedom \cite{Kawamoto:2023nki}. In other words, it has a vanishing boundary entropy\footnote{Instead, the result from the {\it incorrect} antipodal island indicates that the boundary defect is associated with a finite boundary entropy given by $S_{\rm bdy} = \frac{c}{6} \log \sqrt{ \frac{\mathcal{T}+L_{\mt{AdS}}}{\mathcal{T}-L_{\mt{AdS}}} } =\frac{c}{6} \eta_\infty$.}, \ie $S_{\mt{EE}} (\text{defects})  =0$. Another intriguing possibility is that the Hilbert space of the closed dS universe is trivial (\ie one-dimensional), implying that the quantum state in dS gravity is unique. For more recent discussion on the uniqueness of quantum states in closed universes, see, for example, \cite{Balasubramanian:2020xqf,Marolf:2020xie,McNamara:2020uza,Balasubramanian:2023xyd,Shaghoulian:2023odo,Usatyuk:2024mzs}. It would be interesting to explore further whether the pure state for dS gravity, $ | \Psi_{\mathrm{half \,\,dS \,\,gravity}} \rangle$ is unique.

Finally, it is also intriguing to explore quantum corrections to the non-extremal island contributions within the doubly holographic framework. In the leading classical gravity approximation, the standard area formula $\frac{\rm{Area}(\Gamma)}{4\GN}$ applies. Typically, in holographic entanglement entropy, gravity loop corrections are expressed as $\sum_{n=0}^{\infty}\mathcal{O}\left((\GN)^n\right)$. However, in the context of non-extremal islands, this conventional approach seems inadequate, as the leading contribution does not satisfy the equation of motion and does not represent the full saddle point. Consequently, the next leading contributions could surpass the order of $(\GN)^0$. Addressing this issue and gaining a deeper understanding of the implications of quantum corrections remains an important challenge for future research.

\acknowledgments
We would like to thank Vijay Balasubramanian, K. Narayan, Wei Song, Yu-ki Suzuki, Tomonori Ugajin, Zixia Wei and Yuzhen Zhang for useful comments and discussion.  We are also very grateful to Edgar Shaghoulian for bringing 
our attention to his relevant work \cite{Shaghoulian:2021cef}. This work is supported by MEXT KAKENHI Grant-in-Aid for Transformative Research Areas (A) through the ``Extreme Universe'' collaboration: Grant Number 21H05187. TT is also supported by Inamori Research Institute for Science and by JSPS Grant-in-Aid for Scientific Research (A) No.~21H04469. TK is supported by Grant-in-Aid for JSPS Fellows No. 23KJ1315. Work at VUB was supported by FWO-Vlaanderen project G012222N and by the VUB Research Council through the Strategic Research Program High-Energy Physics.

\appendix
\section{Islands in dS JT gravity}\label{sec:JT}

This paper's primary focus is on the issues pertaining to the antipodal island in dS gravity and its resolution. To this end, our result from holographic correlation functions suggests the non-extremal island, situated at the edge of the gravitational dS space. In order to align with the doubly holographic model in AdS$_3$, the two-dimensional dS gravity under consideration is selected to be the Liouville gravity. One might inquire whether the unphysical nature of the antipodal island is a consequence of the fact that induced dS$_2$ gravity is not dynamical. However, this appendix will demonstrate that it is, in fact, a universal feature of the cosmological horizon by considering dS JT gravity. Additionally, it is possible that there exists a physical extremal island whose boundary is given by the maximin surface. However, the extremal island must be situated in proximity to the black hole singularity.

\subsection{dS$_2$ black hole spacetime}
\begin{figure}[t]
	\centering
	\includegraphics[width=5.5in]{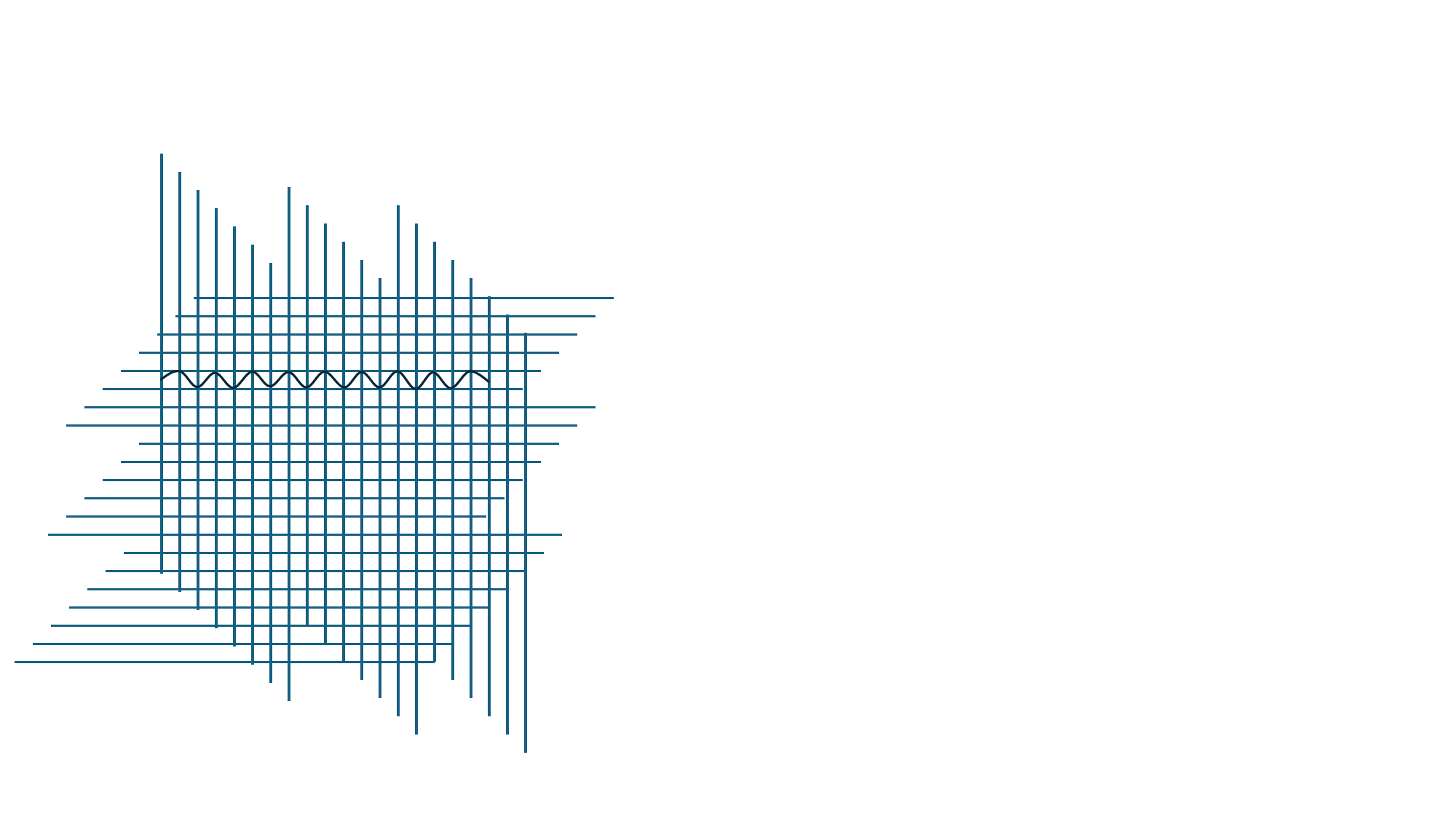}
	\caption{The Penrose diagram of the dS$_2$ black hole spacetime, where the dilaton profile is given by eq.~\eqref{eq:dilaton}.}\label{fig:dSBH}	
\end{figure}

As we have noted at the beginning of section \ref{sec:island}, there are two types of dS JT gravity \cite{Maldacena:2019cbz,Cotler:2019nbi,Sybesma:2020fxg,Kames-King:2021etp,Svesko:2022txo}, depending on whether there is a topological term originating from the higher dimensional extremal black hole. The island formula has been studied in dS JT gravity by many paper from different perspectives, see \eg \cite{Chen:2020tes,Hartman:2020khs,
Balasubramanian:2020xqf,Sybesma:2020fxg,Aguilar-Gutierrez:2021bns,Kames-King:2021etp,Aalsma:2021bit,Teresi:2021qff,Levine:2022wos,Aalsma:2022swk,Chang:2023gkt,Balasubramanian:2023xyd}. We start from considering the JT gravity derived from the Nariai limit of four dimensional Schwarzschild-de Sitter black hole: 
\begin{equation}\label{eq:JT03}
S_{\rm JT'}=\frac{\Phi_0}{16 \pi \GNN} \int d^2 x \sqrt{-g} R+\frac{1}{16 \pi \GNN} \int d^2 x \sqrt{-g}  \, \Phi \(R-\frac{2}{L^2}\) + \text{boundary terms} \,, 
\end{equation}
where the first $\Phi_0$ term with a proper boundary terms is topological for two dimensional spacetime but would be important for the following discussion. It is straightforward to get the equation of motion 
\begin{equation}
 R= \frac{2}{L^2} \,, \qquad  \left(g_{\mu \nu} \nabla^2-\nabla_\mu \nabla_\nu+ \frac{g_{\mu \nu} }{L^2}\right) \Phi=0 \,. 
\end{equation}
where the stress tensor of matter field is fixed as zero. For the solution with non-vanishing expectation values of the stress tensor, refer to \eg \cite{Balasubramanian:2020xqf}. The solution of dS$_2$ spacetime and dilaton field in conformal coordinates reads\footnote{The solutions in other coordinates can be derived from conformal coordinates by coordinate transformations. For example, the dilaton profile is given by 
\begin{equation}\label{eq:dilaton}
\Phi = \Phi_r \cos \theta \cosh t  = \Phi_r \, r = \Phi_r \frac{1+U V}{1- UV}  \,, 
\end{equation}
in global coordinates, static patch coordinates and Kruskal coordinates, respectively.}
\begin{equation}\label{eq:solutions}
  ds^2 = \frac{L^2}{\cos^2 T} \(  - dT^2 + d \theta^2 \) \,, \qquad \Phi (T,\theta) = \Phi_r \frac{\cos \theta}{\cos T}  \,.
\end{equation}
The corresponding Penrose diagram of dS$_2$ spacetime is shown in figure \ref{fig:dSBH}, where we have both a cosmological  horizon and a black hole horizon. This two-dimensional spacetime serves as a toy model for dS Schwarzschild black hole.

Although the additional topological term $\Phi_0$, which originates from the higher-dimensional extremal black hole is not dynamical, it nevertheless gives rise to the ground state entropy. Furthermore, this allows for the possibility of a negative dynamical dilaton field. As illustrated in eq.~\eqref{eq:solutions}, the dilaton value $\Phi$ assumes a negative value within the black hole region. In light of the constraints imposed by the effective Newtonian constant, it is necessary to require 
\begin{equation}\label{eq:dilatonzero}
 \Phi_0 + \Phi(T,\theta) \ge 0 \,. 
\end{equation}
In light of the assumption that the constant $\Phi_0$ is extremely large, the black hole singularity is frequently defined as $\Phi \to -\infty$, \ie 
\begin{equation}
\text{Black hole singularity:} \qquad T= \pm \frac{\pi}{2}, \qquad \theta \in (-\pi,-\frac{\pi}{2}) \cup (\frac{\pi}{2}, \pi) \,, 
\end{equation}
as shown in the Penrose diagram \ref{fig:dSBH}. However, more strictly speaking, the black hole singularity corresponds to the region with a vanishing dilaton $\Phi_0 + \Phi(T,\theta) \to 0$, \ie $\frac{1}{G^{(2)}_{\rm eff}} \to 0$. 

In the context of another dS JT gravity \eqref{eq:JT01}, derived from the dimensional reduction of AdS$_3$, the topological term with coefficient $\Phi_0$ is absent. Therefore, the only physical solution that remains is the dilaton profile with $\Phi \ge 0$. Consequently, only a cosmological horizon exists. It will become evident that the sign of the dilaton value is of pivotal importance in the subsequent analysis.

\subsection{Extremal island near the singularity}
In the following, let us first apply the island formula in dS$_2$ JT gravity by considering the black hole spacetime. For the purposes of this discussion, we will focus on one endpoint of the non-gravitational subsystem $\mA$ and assume that it is located at $(T_A, \theta_A)$. Similarly, it is assumed that the corresponding extremal surface is given by $(T_I, \theta_I)$. By employing conformal coordinates, we obtain the most general expression for the generalized entropy $S_{\rm gen}$. To wit, 
\begin{equation}
S_{\rm{gen}} =  \Phi_0 +  \Phi_r \frac{\cos \theta_I }{\cos T_I} +  \frac{c}{6}  \log \(  \frac{ 2\( \cos (T_A - T_I) - \cos(\theta_A -\theta_I) \)}{\epsilon^2 \cos T_A \cos T_I }  \) \,.\\ 
\end{equation}
The quantum extremal surface is fixed by solving the extremization equations, namely  
\begin{equation}
  \frac{\partial S_{\rm{gen}}  }{\partial T_I} = 0 = \frac{\partial S_{\rm{gen}}  }{\partial \theta_I} \,. 
\end{equation}
Due to the existence of the area term from the dilaton $\Phi$, the solutions becomes more complicated but have been explicitly shown in \cite{Hartman:2020khs}. Here, we are interested in examining the second derivative of the generalized entropy in order to check whether the extremal island solutions is physical. It is straightforward to obtain 
\begin{equation}
\begin{split}
   \frac{\partial^2 S_{\rm{gen}}  }{\partial \theta_I \partial \theta_I} &=
   - \frac{c}{24} \(   \frac{1}{\sin^2 \( \frac{1}{2} ( T_I - T_A + \theta_I - \theta _A ) \)  }  + \frac{1}{\sin^2 \( \frac{1}{2} ( T_I - T_A - \theta_I +  \theta_A ) \)  }   \) - \Phi_r \frac{\cos \theta_I}{\cos T_I}  \,,\\
   &= - \frac{c}{24} \(   \frac{1}{\sin^2 \( \frac{1}{2} ( T_I - T_A + \theta_I - \theta _A ) \)  }  + \frac{1}{\sin^2 \( \frac{1}{2} ( T_I - T_A - \theta_I +  \theta_A ) \)  }   \) - \Phi (T_I,\theta_I)
\end{split} 
\end{equation}
where the second term is nothing but the negative value of the dynamical dilaton $\Phi$. Obviously, we will always have 
\begin{equation}
  \frac{\partial^2 S_{\rm{gen}}  }{\partial \theta_I \partial \theta_I} <0, \qquad \text{with}\qquad \Phi \equiv \Phi_r \frac{\cos \theta}{\cos T} \ge  0  \,, 
\end{equation}
if the quantum extremal surface, \ie the boundary of extremal island is situated within the cosmological spacetime with $\Phi \ge 0$.
If we set $\Phi_r=0$, it reduces exactly to the antipodal island, which causes the issues discussed earlier. On the other hand, the same problems associated with the extremal island also apply to the JT gravity defined in eq. ~\eqref{eq:JT01} due to the constraint that its dilaton profile must be non-negative.

To put it another way. The quantum extremal surface could be locally minimal in the spatial direction (\eg with $ \frac{\partial^2 S_{\rm{gen}}  }{\partial \theta_I \partial \theta_I}  >0$) only if $\Phi (T_I, \theta_I) < 0$. So the extremal island must be in the black hole region. However, this quantum extremal surface behind the black hole horizon may not always exist, depending on the parameter regime. Combining the constraint \eqref{eq:dilatonzero} from the Newton constant and the condition for local minimization, one can find the following sequence inequalities:  
\begin{equation}
 - \Phi_0 < \Phi (T_I, \theta_I) < -\frac{c}{12}  \,.
\end{equation}
Obviously, this can only hold if $\Phi_0 > \frac{c}{12}$. From a more physical point of view, this is equivalent to the condition that the extremal island is not located behind the physical spacetime. Furthermore, it is worth noting that the extremal island must also be close to the black hole singularity ($\cos T = 0$) because the local minimization also requires 
\begin{equation}
\begin{split}
  \frac{1}{\cos T_I} &> \frac{c}{24 \Phi_r} \frac{1}{(-\cos \theta_I)} \(   \frac{1}{\sin^2 \( \frac{1}{2} ( T_I - T_A + \theta_I - \theta_A ) \)  }  + \frac{1}{\sin^2 \( \frac{1}{2} ( T_I - T_A - \theta_I +  \theta_A ) \)  }   \) \\  
  & > \frac{c}{12 \Phi_r} \gg 1 \,.  
\end{split}
\end{equation}
In summary, we can see that the same problems associated with the antipodal island also appear in dS JT gravity. This can only be avoided if the extremal island is placed close to the black hole singularity.


\section{One-point function with an AdS brane}\label{sec:appB}

In order to facilitate comparison with the dS brane case, we present the calculations of the one-point function by using the geodesic approximation with an AdS brane. The objective of this appendix is to demonstrate that the result obtained for the one-point function is equivalent to the holographic RT formula. 

\subsection{Geodesics approximation}
The bulk-to-bulk propagator \eqref{eq:bulk-to-bulk_Aropagator} for AdS$_{d+1}$ space is derived in terms of the hypergeometric function. We first focus on $\Delta \to \infty$ limit and derive the leading approximation, \ie the geodesic approximation \cite{Balasubramanian:1999zv,Louko:2000tp,Aparicio:2011zy,Balasubramanian:2012tu}. 
Note the integral representation of the hypergeometric function
 \begin{equation}
   \hypgeo{a}{b}{c}{Z}= \frac{\Gamma(c)}{\Gamma(b)\Gamma(c-b)}\int_0^1 d\tau \tau^{b-1}(1-\tau)^{b-c-1}(1-Z \tau)^{-a}
 \end{equation}
for $\Re(c)>\Re(b)>0,\; \abs{\arg(1-Z)}<\pi$. We can recast the bulk-to-bulk propagator $ G_{\mt{BB}}^\Delta$ and the bulk-to-boundary propagator $ K_{\text{Bb}}^\Delta$ in terms of the integral representation, \eg 
 \begin{equation}
     \begin{split}
         K_{\mt{Bb}}^\Delta \(\vec{x}; z', \vec{x}'\)=\lim _{z \rightarrow 0} \frac{1}{(z)^{\Delta}2^{\Delta}\pi^{\frac{d}{2}}}\frac{\Gamma(\Delta)\Gamma(\Delta-\frac{d}{2}+1)}{\Gamma(\Delta-\frac{d}{2})\Gamma(\frac{\Delta+1}{2})\Gamma(\frac{\Delta-d+1}{2})}\int_0^1 d\tau \tau^{\frac{\Delta-1}{2}}(1-\tau)^{\frac{\Delta-d-1}{2}}(1-\xi^2 \tau)^{-\frac{\Delta}{2}} \,. 
     \end{split}
 \end{equation}
In order to obtain the approximation at $\Delta \to \infty$ limit, we need to deal with the following integral 
 \begin{equation}
     I(\xi) = \int_0^1 d\tau \tau^{-\frac{1}{2}}\qty(1-\tau)^{-\frac{d+1}{2}}e^{- \Delta f(\tau)} \,, 
 \end{equation}
 with 
 \begin{equation}
 f(\tau)= \frac{1}{2}\( \log{\qty(1-\xi^2 \tau) - \log{\tau} - \log\qty(1-\tau)}   \)\,.
 \end{equation}
It is easy to find that there are two saddle points
 \begin{equation}
     \tau_*^{(1)} = \frac{1-\sqrt{1-\xi^2}}{\xi^2},\; \tau_*^{(2)} = \frac{1+\sqrt{1-\xi^2}}{\xi^2} \,, 
 \end{equation}
by solving $f'(\tau_*)=0$. Because of $0<\tau_*^{(1)}<1,\;\tau_*^{(2)} >1$, we only need to apply the saddle-point approximation to the first saddle point that corresponds to a local minimum with $f''(\tau_*^{(1)}) > 0$ and get 
 \begin{equation}
      I(\xi)\approx \xi^{-\Delta} \sqrt{\frac{2\pi}{\Delta}} \qty(\frac{1+\sqrt{1-\xi^2}}{\sqrt{1-\xi^2}})^{\frac{d}{2}}e^{-\Delta D(x;x')} \,. 
     \end{equation}
where $D(x,x')$ denotes the (dimensionless) geodesic distance \eqref{eq:geodesic_length}, namely 
\begin{equation}
D(x ; x') =  \, \ln \left(\frac{1+\sqrt{1-\xi^2}}{\xi}\right) \,, \qquad  \xi = \frac{1}{\cosh  \( D(x ; x')  \)} \,.
\end{equation}
Using Stirling's formula, the approximation of the additional constant coefficient can be derived as
\begin{equation}
    \begin{split}
     \frac{\Gamma(\Delta)\Gamma(\Delta-\frac{d}{2}+1)}{\Gamma(\Delta-\frac{d}{2})\Gamma(\frac{\Delta+1}{2})\Gamma(\frac{\Delta-d+1}{2})} \approx  \frac{1}{\sqrt{2\pi}}e^{1-\frac{d}{2}} e^{\frac{d+1}{2}\log{\Delta}+\qty(\Delta-\frac{d}{2})\log{2}+O\qty(\frac{1}{\Delta})} \,. 
\end{split}
\end{equation}
By combining these approximations, we can obtain 
     \begin{equation}
          \begin{split}
          G_{\mt{BB}}^\Delta(x;x')    
          &= \frac{1}{2\Delta-d}e^{-\Delta D(x;x')}e^{1-\frac{d}{2}}\qty(\frac{\Delta}{2\pi}\frac{1+\sqrt{1-\xi^2}}{\sqrt{1-\xi^2}})^{\frac{d}{2}} \,, 
          \\
          K_{\mt{Bb}}^\Delta(\vec{x}; z', \vec{x}')&=\lim _{z \rightarrow 0} e^{-\Delta\qty(D(x;x')+\log{z})}e^{1-\frac{d}{2}}\qty(\frac{\Delta}{2\pi}\frac{1+\sqrt{1-\xi^2}}{\sqrt{1-\xi^2}})^{\frac{d}{2}}\,, 
          \end{split}
     \end{equation}
where the subtraction $\log{z}$ with $z \to 0$ means that the bulk-to-boundary propagator is related to the renormalized geodesic length $\bar{D}(x,x_b)$. It can thus be concluded that the holographic one-point function of BCFT in the limit of large dimensions, as defined in eq.~\eqref{eq:Onepoint}, is given by the geodesic approximation:  
\begin{equation}\label{eq:onepointEOW}
    \ev{\mO(x)}_{\mt{BCFT}}  \approx \frac{1}{2\Delta-d}\int_{\mt{EOW}} d^d x_b \sqrt{g} e^{-\Delta D(x;x_b) }e^{1-\frac{d}{2}}\qty(\frac{\Delta}{2\pi}\frac{1+\sqrt{1-\xi^2}}{\sqrt{1-\xi^2}})^{\frac{d}{2}} \,, 
\end{equation}
where the integral is performed over the entire EOW brane.

\subsection{One-point function with an AdS bane}\label{app:AdSonepoint}

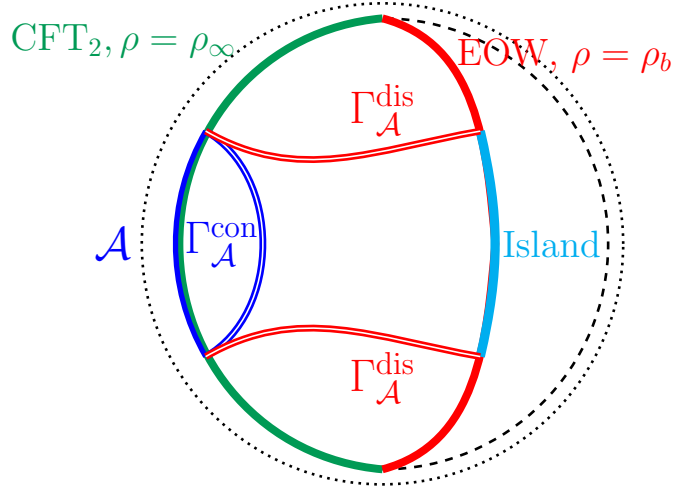
\begin{figure}[t]
\centering
    \begin{tikzpicture}
        \draw[ForestGreen,line width=3pt] (0,3) to [out=-175,in=90] (-2.7,0) to [out=-90,in=175] (0,-3);
        \draw[black, line width=1pt,dashed] (0,-3) arc(-90:90:3);
        \draw[black, line width=1pt,dotted] (3.2,0) arc(0:360:3.2);
        \draw[blue, line width=2pt] (-2.35,1.5) arc(150:210:3);

\draw[red, line width=3pt] (0,3) to [out=-15,in=95] (1.5,0) to [out=-95,in=15] (0,-3);
\draw[cyan, line width=3.5pt] (1.3,1.5) to [out=-77,in=77] (1.3,-1.5);
\draw[blue, line width=1pt,double] (-2.35,1.5)to[out=-30,in= 30](-2.35,-1.5);
         \draw[blue](-1.5,0)node[left]{\Large{$\Gamma_{\mathcal{A}}^{\mathrm{con}}$}};
          \draw[blue](-3.2,0)node[left]{\Large{$\mathcal{A}$}};
\draw[red, line width=1pt,double] (-2.35,1.5)to[out=-30,in= 190](1.3,1.5);
\draw[red, line width=1pt,double] (-2.35,-1.5)to[out=30,in= 170](1.3,-1.5);
 \draw[red](0,1.3)node[above]{\Large{$\Gamma_{\mathcal{A}}^{\mathrm{dis}}$}};
 \draw[red](0,-1.3)node[below]{\Large{$\Gamma_{\mathcal{A}}^{\mathrm{dis}}$}};
 \draw[cyan](1.45,0)node[right]{\large{Island}};
  \draw[ForestGreen](-1.8,2.7)node[left]{\large{CFT$_2,\rho=\rho_\infty$}};
  \draw[red](4,2.5)node[left]{\large{EOW, $\rho=\rho_b$}};
    \end{tikzpicture}
    \caption{The picture for the doubly holographic model with an AdS EOW brane.}
        \label{fig:AdSbrane}
\end{figure}

To compare with the dS brane setup discussed in the section \ref{sec:doubleholography}, we focus on the calculating holographic one-point function \eqref{eq:onepointEOW} with assuming that the EOW brane in the doubly holographic AdS$_3$ spacetime is an AdS$_2$ brane, as shown in figure \ref{fig:AdSbrane}. It is convenient to consider the AdS$_2$ slicing coordinate of AdS$_3$ space, \ie  
\begin{equation}
    ds^2= L_{\mt{AdS}}^2 \(    d\rho^2 + \cosh{\rho}^2 (dr^2+\cosh^2{r}d\rmtE^2)  \) \,, 
\end{equation}
where the conformal boundary locates at $\rho=\rho_\infty \to \pm \infty$ and each constant $\rho$-slice represents a AdS$_2$ space. 
We further that the AdS$_2$ brane is located at $\rho=\rho_b$ with a induced metric 
\begin{equation}
   ds^2 \bigg|_{\mt{EOW}} = L_{\mt{AdS}}^2  \cosh{\rho_b}^2 \(dr^2+\cosh^2{r}d\rmtE^2\)  \,.  
\end{equation}
The geodesic length between the boundary point $A=(\rho_\infty,\rmtE_A=\rmtE,r_A)$ and the AdS$_2$ brane at $I=(\rho_b,\rmtE_I,r_I)$ is expressed as 
\begin{equation}
    \cosh{D_{AI}} = \cosh\rho_\infty \cosh{\rho_b}\qty(\cosh r_A \cosh{r_I} \cos{\qty(t_{\mt{E}A}-\rmtE_I)} -\sinh{r_A}\sinh{r_I} )-\sinh{\rho_\infty}\sinh{\rho_b} \,. 
\end{equation}
The position of the extremal RT surface or the endpoints of the island is determined by the extremization conditions: 
\begin{equation}
    \frac{\partial D_{AI}}{\partial r_I}=0,\quad \frac{\partial D_{AI}}{\partial \rmtE_I}=0.
\end{equation}
The solution is simply given by 
\begin{equation}
    t_{\mt{E}I}=t_{\mt{E}A} \,, \quad  r_I=r_A \,. 
\end{equation}
In contrast to the dS brane case, it can be seen that the extremal point at $(t_{\mt{E}I}, r_I)$ is indeed a local minimum in all directions along the AdS EOW brane. It is straightforward to get 
\begin{equation}
    \frac{\partial^2 D_{AI} }{\partial r_I^2}= \frac{\cosh{\rho_\infty\cosh{\rho_b}}}{\abs{\sinh\qty(\rho_\infty-\rho_b)}}>0 \,, 
    \quad \frac{\partial^2 D_{AI} }{\partial \rmtE_I^2} = \frac{\cosh^2{r_A}\cosh{\rho_\infty\cosh{\rho_b}}}{\abs{\sinh\qty(\rho_\infty-\rho_b)}}>0 \,.
\end{equation}
In other words, the saddle point approximation can be applied once more in order to compute the integral on the AdS EOW brane, thus obtaining the one-point function. Moreover, this provides further evidence in support of the assertion that the conventional RT formula, which employs the area of the minimal surface anchored on the AdS EOW brane, can yield the same holographic entanglement entropy as that derived from the correlation functions of the twistor operator in BCFT.

\bibliographystyle{JHEP}
\bibliography{main}

\end{document}